\documentclass[usenatbib,usegraphicx,useAMS,a4paper]{mn2e}
\pdfoutput=1
\usepackage{graphicx}
\usepackage{subfig}
\usepackage{verbatim}
\usepackage[usenames,dvipsnames]{color}
\usepackage{booktabs}
\newcommand{\Eq}[1]{Eq.~\ref{#1}}
\newcommand{\Fig}[1]{Fig.~\ref{#1}}
\newcommand{\Tab}[1]{Table ~\ref{#1}}
\newcommand{\Sec}[1]{Sec.~\ref{#1}}

\def\ramses    {{\tt Ramses}}
\def\ramsesrt  {{\tt RamsesRT}}
\def\hi    {{\rm{H\textsc{i}}}}

\def\hei   {\rm{He\textsc{i}}}
\def\heii  {\rm{He\textsc{ii}}}
\def\hisub {\rm{H \scriptscriptstyle I}}
\def\hiisub {\rm{H \scriptscriptstyle II}}
\def\heisub {\rm{He \scriptscriptstyle I}}
\def\heiisub {\rm{He \scriptscriptstyle II}}
\def\heiiisub {\rm{He \scriptscriptstyle III}}
\def\nh {n_{\rm{H}}}
\def\snh {N_{\rm{H}}}
\def\nho {n_{\rm{H,0}}}
\def\nhe {n_{\rm{He}}}
\def\nhi {n_{ \hisub }}
\def\nhii {n_{ \hiisub }}
\def\nheii {n_{ \heiisub }}
\def\nheiii {n_{ \heiiisub }}
\def\ne {n_{\rm{e}}}
\def\xhi {x_{\rm{H \scriptscriptstyle I}}}
\def\xhii {x_{\rm{H \scriptscriptstyle II}}}
\def\xheii {x_{\rm{He \scriptscriptstyle II}}}
\def\xheiii {x_{\rm{He \scriptscriptstyle III}}}
\def\sighi {\sigma_{\rm{H \scriptscriptstyle I}}}

\def\cci {{\rm{cm}}^{-3}}
\def\cc {\rm{cm}^{3}}
\def\si {\rm{s}^{-1}}
\def\nhuv {n_{\rm{H}}^{UV}}

\def\recB {\alpha^{\rm{B}}_{\rm{H \scriptscriptstyle I}}}
\def\eps {\varepsilon}

\def\lya {{\rm{Ly}$\alpha$}}
\def\lyam {{\rm{Ly} \alpha}}
\def\fa{{f_{\alpha}}}
\def\LS {{\rm{erg \, s^{-1} \, kpc^{-2}}}}
\def\IS {{\rm{erg \, s^{-1} \, cm^{-2} \, arcsec^{-2}}}}
\def\nui {{\nu_{i \scriptscriptstyle 0}}}
\def\nuf {{\nu_{i \scriptscriptstyle 1}}}

\def\vsk{\vskip 0.2cm}

\def\FG {{FG10}}
\begin{document}

\title[Extended \lya{} emission]{Extended Lyman-alpha emission from
  cold accretion streams} \author[Rosdahl \& Blaizot]
{J.~Rosdahl\thanks{E-mail: joakim.rosdahl@univ-lyon1.fr}
  \thanks{Animations of our simulations can be found at
    http://www-obs.univ-lyon1.fr/labo/perso/joakim.rosdahl/LABs}
  and J.~Blaizot \\
  Universit\'e de Lyon, Lyon, F-69003, France ; \\
  Universit\'e Lyon 1, Observatoire de Lyon, 9 avenue Charles Andr\'e,
  Saint-Genis Laval,
  F-69230, France ; \\
  CNRS, UMR 5574, Centre de Recherche Astrophysique de Lyon.}

\maketitle
\begin{abstract}
  We investigate the observability of cold accretion streams at
  redshift 3 via Lyman-alpha (\lya{}) emission and the feasibility of
  cold accretion as the main driver of \lya{} blobs (LABs). We run
  cosmological zoom simulations focusing on 3 halos spanning almost
  two orders of magnitude in mass, roughly from $10^{11}$ to $10^{13}$
  solar masses. We use a version of the \ramses{} code that includes
  radiative transfer of ultraviolet (UV) photons, and we employ a
  refinement strategy that allows us to resolve accretion streams in
  their natural environment to an unprecedented level. For the first
  time in a simulation, we self-consistently model self-shielding in
  the cold streams from the cosmological UV background, which enables
  us to predict their temperatures, ionization states and \lya{}
  luminosities with improved accuracy. We find the efficiency of
  gravitational heating in cold streams in a $\sim10^{11}$ solar mass
  halo to be around 10-20\% throughout most of the halo but reaching
  much higher values close to the center. As a result most of the
  \lya{} luminosity comes from gas which is concentrated at the
  central 20\% of the halo radius, leading to \lya{} emission which is
  not extended. In more massive halos, of $\ga 10^{12}$ solar masses,
  cold accretion is complex and disrupted, and gravitational heating
  does not happen as a steady process. Ignoring the factors of \lya{}
  scattering, local UV enhancement, and SNe feedback, the cold `messy'
  accretion alone in these massive halos can produce LABs that largely
  agree with observations in terms of morphology, extent, and
  luminosity. Our simulations slightly and systematically over-predict
  LAB abundances, perhaps hinting that the interplay of these ignored
  factors may have a negative net effect on extent and luminosity.  We
  predict that a factor of a few increase in sensitivity from current
  observational limits should unambiguously reveal continuum-free
  accretion streams around massive galaxies at $z=3$.
\end{abstract}

\section{Introduction}
\label{Intro.sec}
The last decade has seen a shift in the way galaxies are thought to
have assembled. In the classic theory \citep{Rees:1977p4388,
  Silk:1977p4383,White:1978p4389}, galaxies collect their baryons via
so-called hot mode accretion where diffuse gas symmetrically falls
into dark matter (DM) halos and is shock-heated as it hits the gas
residing in them. Depending on the mass of the halo, the gas may or
may not eventually settle into the galaxy. However, it has become
increasingly apparent through theoretical work and simulations that at
high redshift ($z\ga 2$), galaxies get their baryons primarily via
accretion of relatively dense, cold ($10^4$ K) and pristine gas which
penetrates in the form of \textit{streams} through the diffuse
shock-heated medium \citep{Fardal:2001p3736, Birnboim:2003p3602,
  Keres:2005p3601, Dekel:2006p4450, Birnboim:2007p4448,
  Ocvirk:2008p2688, Dekel:2009p1318, Brooks:2009p3604,
  vandeVoort:2011p5673, FaucherGiguere:2011p5611,
  vandeVoort:2011p5669}. Simulations consistently show these streams
to exist and peak in activity around redshift 3, though it appears
that their widths are still dictated mostly by resolution.

The problem is that cold accretion streams have never been directly
observed, though we are starting to see some hints, both in emission
\citep{Rauch:2011p5439} and absorption \citep{Ribaudo:2011p5454}.

Is this lack of observational evidence consistent with the existence
of cold accretion streams? Do we not observe them because they're not
easily observable or simply because they don't exist?

\cite{FaucherGiguere:2011p3606} showed that the streams are hard to
detect directly via absorption due to their small covering factor and
surrounding galactic winds that overwhelm their signature.
\cite{Kimm:2011p4491} came to the same conclusion, adding that the low
metallicity in streams ($\leq 10^{-3}$ solar) further inhibits their
detection via metal line absorption. Even so,
\cite{Fumagalli:2011p2943} and \cite{vandeVoort:2011p5667} have argued
that a large fraction of observed metal-poor Lyman-limit systems
(LLSs) make up for indirect detections of cold streams.  Furthermore,
we may possibly have been directly observing the tips of these streams
during the last decade in the form of Lyman-alpha blobs (LABs).

\vsk LABs are extremely bright ($\ga 10^{43} \; \mathrm{erg \;
  s^{-1}}$) and extended ($\ga 30 \;\mathrm{kpc}$ in diameter) \lya{}
nebulae \citep[e.g.][]{Francis:1996p4544, Keel:1999p4529,
  Steidel:2000p2213, Matsuda:2004p3081, Palunas:2004p4550,
  Nilsson:2006p3525, Smith:2007p4610, Prescott:2009p3951,
  Yang:2010p3447, Erb:2011p5386}.  They have a slight tendency to be
filamentary in structure \citep[][hereafter M11]{Matsuda:2011p5426},
and often have short limbs protruding from the main body. They often
coincide with galactic sources that give hints about their physical
origin but the mechanism by which the emission becomes so strong and
extended is a matter of debate. A subset of LABs however have no
apparent coinciding galactic sources \citep[e.g.][]{Steidel:2000p2213,
  Weijmans:2010p5527, Erb:2011p5386}. Up until now about two hundred
LABs have been discovered, including about fifteen giant ones ($>100$
kpc). Smaller extended \lya{} emitters exist in large quantities over
a continuous range of sizes down to point sources. LABs appear to be
specific to the high-redshift Universe \citep{Keel:2009p4857} and most
of them have been detected at $2<z\la3$.

The physical nature of LABs is still a matter of debate, but by most
accounts they are powered by a combination of some or all of the
following processes: \textit{(a)} Cold stream accretion is a natural
explanation, where the fuel source is the dissipation of gravitational
potential, also termed gravitational heating
\citep[e.g.][]{Steidel:2000p2213, Haiman:2000p4632, Fardal:2001p3736,
  Dijkstra:2006p4697, Dijkstra:2009p3780}.  \textit{(b)}
Photo-fluorescence by near-lying sources, such as active galactic
nuclei (AGN) or starbursts \citep[e.g.][]{Haiman:2001p4742,
  Cantalupo:2005p4317, Kollmeier:2010p3256}, \textit{(c)} \lya{}
scattering, also fuelled by neighbouring star-forming regions
\citep[e.g.][]{Laursen:2007p4741, Zheng:2011p5486}. \textit{(d)}
Cooling radiation in galactic outflows, fuelled by AGN or supernovae
\citep[e.g.][]{Taniguchi:2000p4771, Ohyama:2003p4783, Mori:2004p4829}.

\cite{Furlanetto:2005p3744} used cosmological simulations to look at
the contributions of each of these processes, and found that
star-forming regions can in principle power all but the largest LABs
via photo-fluorescence and \lya{} scattering, but that cold accretion
alone cannot, except under very optimistic assumptions.  They however
pointed out that the \lya{} emissivity of their simulated gas is
highly uncertain due to the lack of modelling of self-shielding from
UV radiation: The self-shielding state of the gas affects both the
temperature and ionization state, which sensitively dictates the
\lya{} emissivity.  They also pointed out that the efficiency of
star-formation in powering LABs is very dependent on the presence of
dust. As pointed out by \cite{Cen:2011p5588}, massive galaxies tend to
have large dust content which makes them very efficiently transform
their UV (and \lya{}) output into infrared radiation. Thus it appears
problematic to associate the largest and most luminous LABs to
star-formation in the most massive halos in the Universe.

\subsection {Recent work on gravitationally driven \lya{} emission}
Notably, two recent simulation papers have studied gravitational
heating as the driver of LABs, but have reached conflicting
conclusions:

\citet[][hereafter G10]{Goerdt:2010p1237} analyze two suites of
cosmological adaptive mesh refinement (AMR) simulations. They assume
self-shielding in post-processing from the UV background in accretion
streams. Mock observations of halos of $\sim 4\ 10^{11}
\mathrm{M_{\sun}}$ at redshift 2.3 look similar to real LABs in
morphology and surface brightness profile, though the association of
LABs to halos of such low mass implies an unrealistically high LAB
abundance. A \lya{} luminosity function derived from their results is
not far off from a function derived from observations, though they
over-predict number densities somewhat, which implies the cooling
emission in their simulations is \textit{too} efficient. As pointed
out by \citet[][hereafter \FG{}]{FaucherGiguere:2010p5372} this
overestimate appears to be due to artificial photo-heating of stream
gas, which is not on-the-fly self-shielded from the UV background.

\FG{} analyze cosmological smoothed particle hydrodynamics simulations
to test different approaches and approximations. Based on radiative
transfer (RT) post-processing results, they apply on-the-fly
self-shielding by excluding UV photoionization from all gas denser
than $10^{-2}$ H atoms per $\cc$. Then they apply a \lya{} transfer
code to their output to model the scattering of \lya{} photons towards
the observer and obtain realistic mock observations.  According to
their results, which are in good agreement with
\cite{Furlanetto:2005p3744}, cooling radiation can in
\textit{principle} power LABs, provided one includes emission from gas
dense enough to be star-forming to some extent. They note that this
gas should be under the influence of feedback processes which
introduce a large uncertainty to the cooling emission.

Although G10 and \FG{} are not in good agreement on their conclusions,
they both agree with \cite{Furlanetto:2005p3744} on that proper
modelling of self-shielding from UV radiation is crucial to the
results.

\subsection {This work}
We have developed a radiation-hydrodynamics (RHD) version of the AMR
code \ramses{} \citep{Teyssier:2002p1740}, which puts us in a unique
position to continue the work of the aforementioned authors, to study
the emissivity of accretion streams in their natural environment at
high redshift in simulations that accurately and consistently model
self-shielding from the UV background. We also extend previous work by
simulating halos of larger masses, which are more likely to host LABs,
and by using an original refinement strategy which allows us to
describe cold streams with unprecedented resolution. The increased
resolution also allows us to accurately track the state of the gas up
to higher densities than in the previous works. The main motivations
of our work are: \textbf{(a)} Investigate whether gravitational
heating is capable and sufficient as a driver of observed
LABs. \textbf{(b)} Predict the observability of gravitationally
powered \lya{} emission from accretion streams at redshift 3.

\vsk The paper is structured as follows: Section 2 describes the
simulation code and the setup of our experiments.  Section 3 describes
the physical properties at redshift 3 of our simulated halos over a
range of masses. Section 4 presents our prediction of the \lya{}
emission from extended gas around galaxies and its observability. We
compare with observations of LABs.  In section 5 we discuss the
efficiency of gravitational heating as a source of extended \lya{}
emission and the contribution of cosmological UV fluorescence. We
discuss other factors that may affect the extended \lya{}
emission. Finally we conclude and discuss in section 6.

\section{Simulations} \label{Sim.sec}
\subsection{Code details}\label{Code.sec}
We run our simulations in \ramsesrt{}, a version of the AMR code
\ramses{} \citep{Teyssier:2002p1740} which we have modified to include
on-the-fly radiation-hydrodynamics describing the propagation in space
of UV photons and their interaction with gas via photoionization and
heating of hydrogen and helium.

The widely used \ramses{} code simulates the cosmological evolution
and interaction of dark matter, stellar populations and baryonic gas,
via gravity, hydrodynamics and radiative cooling. The gas evolution is
computed using a second order Godunov scheme for the Euler equations,
while trajectories of collisionless DM and stellar particles are
computed using a Particle-Mesh solver.

The \ramsesrt{} implementation and tests will be fully described in
Rosdahl et al. (2012, in preparation), and here, we only briefly
present the aspects of \ramsesrt{} which are most relevant to the
present work.

For the radiative transfer we use a moment-based method with the M1
closure relation, as described in \cite{Aubert:2008p1439}, the essence
of which is to turn rays of radiation into a fluid with a direction of
flow that corresponds to an average of rays over all angles. In
contrast to the usual ray-tracing codes currently on the market, this
gives the advantage that the computational load of RT does not scale
linearly - in fact hardly scales at all - with the number of radiative
sources in the simulation. This is a particular advantage here as we
simulate a spatially continuous source of radiation, which is hard to
do with a ray-tracing code. \ramsesrt{} takes full advantage of the AMR
structure of \ramses{} and photons are propagated through the same cells
that define the baryonic gas.

Our RT solver is explicit, which means the timestep length for the
propagation of photons is limited by the speed of light. This
typically makes the RT timestep three orders of magnitude shorter than
the hydrodynamical timestep. Since we're forced to apply a global RHD
timestep which is the minimum of the hydrodynamical step and the RT
one, we're faced with the rather horrifying prospect that \ramsesrt{}
simulations are slowed down by a factor of order one-thousand compared
to non-RT simulations. To get around this, we invoke the
\textit{Reduced Speed-of-Light Approximation} (RSLA) proposed by
\cite{Gnedin:2001p2858} \citep[see also discussion in
][]{Aubert:2008p1439}: The speed of light is reduced by a factor
$f_{c}$, bringing the RHD timestep closer to the normal \ramses{} one
and making \ramsesrt{} runnable in reasonable time.
In the simulations described here we use $f_{c} \sim 1/100$. To be
sure the choice of light speed is not affecting our results, we have
run analogs of our H1 simulation (see \Tab{tbl:sims}) with the light
speed changed by a factor of five in either direction,
i.e. $f_{c}=1/20$ and $f_{c}=1/500$. This has an insignificant effect
on the results, and we conclude it is an acceptable approximation for
our simulations.

In order to self-consistently evolve the UV field we implement
non-equilibrium gas cooling that keeps track of the abundances of all
ion species of hydrogen and helium. These abundances are stored in the
form of three ionization fractions, as passive scalars that are
advected with the gas,
\begin{eqnarray}\label{xion.eq}
  \xhii   & \equiv & \nhii   / \nh, \nonumber \\
  \xheii  & \equiv & \nheii  / \nhe, \\
  \xheiii & \equiv & \nheiii / \nhe, \nonumber 
\end{eqnarray}
where $n$ is number density. The non-equilibrium cooling
module evolves these ionization fractions along with photon fluxes and
temperature on a per-cell basis, with the timestep constraint that none
of these quantities changes by more than 10\% in a single timestep,
using sub-cycles when needed to fill the RHD timestep.

We have tested and verified \ramsesrt{} with the benchmark tests of
the `Cosmological radiative transfer comparison project'
\citep{Iliev:2006p1394, Iliev:2009p1494}, and the results will be
presented in Rosdahl et al. (2012, in prep.).

\subsection{Simulation setup}\label{Sec:simpars}
We run three cosmological zoom simulations, each targeting the
evolution until redshift 3 of a single halo and its large-scale
environment.  The initial conditions are generated using MPGRAFIC
\citep{Prunet:2008p5196}. We assume a $\Lambda$CDM Universe with
$\Omega_{\Lambda}=0.723$, $\Omega_m=0.277$, $\Omega_b=0.0459$, $h
\equiv H_0/100=0.702$ and $\sigma_8=0.817$, consistent with seven-year
WMAP results \citep{KomatsuEtal11}.  We assume hydrogen and helium
mass fractions $X=0.76$ and $Y=0.24$.

\setlength{\tabcolsep}{4.5pt}
\begin{table}
\begin{minipage}{0.5\textwidth}
  \caption{Simulation parameters}\label{tbl:sims}
  \begin{tabular}{p{0.6cm} p{0.95cm} p{1.1cm} p{0.9cm} p{0.9cm}
      p{0.8cm} p{1.0cm}}
    \toprule Name & Box size\footnote{Co-moving} [Mpc] & Halo
    mass\footnote{DM+baryons at $z=3$, (all the mass within the virial
      radius)} [$\mathrm{M_{\sun}}$] & DM res.\footnote{Optimal
      resolution} [$\mathrm{M_{\sun}}$] & Gas res.\footnote{ Optimal
      physical resolution (not co-moving) at $z=3$} [pc] &
    \(f_c\)\footnote{Reduced light-speed fraction, see
        \Sec{Code.sec}} &
    \(n_H^{UV}\) \footnote{
      Threshold for UV-emitting gas, see \Sec{Sec:simpars}} [$\cci$] \\
    \midrule H1 & $28.5$ & $2.9\ 10^{11}$ & $1.4\ 10^6$ & 217 &
    $1/100$ &
    $1\ 10^{-4}$ \\
    H2 & $28.5$ & $2.9\ 10^{12}$ & $1.1\ 10^7$ & 434 & $1/300$ &
    $3\ 10^{-4}$ \\
    H3 & $51.2$ & $1.3\ 10^{13}$ & $6.4\ 10^7$ & 780 & $1/300$ &
    $3\ 10^{-4}$ \\
    \bottomrule
  \end{tabular}
  \vspace{-5.3mm}
\end{minipage}
\end{table}

The masses of these three halos span almost two orders of magnitude,
the least massive halo ($\sim 3\ 10^{11} M_{\sun}$) roughly
corresponding to halos studied in G10 and \FG{}, and the more massive
halo simulations (up to mass $\sim 10^{13} M_{\sun}$) based on the
expectation that LABs are situated in over-dense regions of the
Universe \citep{Steidel:2000p2213,Prescott:2008p3849,Yang:2010p3447}.
The parameters for the individual simulations, named H1, H2 and H3 in
order of increasing halo mass, are listed in \Tab{tbl:sims}.

Each simulation has periodic boundaries and nested levels of
refinement in a zoom-region around the targeted halo, both in DM and
gas.

\vsk \textbf{On-the-fly refinement} is enforced inside the zoom
regions according to two criteria: The first is the traditional
`quasi-Lagrangian' criterion, where a cell is refined if it contains
more than 8 DM particles or the equivalent baryonic mass\footnote{A
  cell is refined if it contains a mass of baryons larger than $8 \;
  \Omega_b/\Omega_m \; m_{DM}$, where $\Omega_b$ and $\Omega_m$ are
  the cosmological mass fractions of baryons and matter, respectively,
  and $m_{DM}$ is the mass of the highest-resolution DM particles.  }.
This causes concentrations of mass to be refined to the maximum, but
will typically leave the resolution of cold flows many times less,
which is a problem when one is most interested in these flows. The
second refinement criterion, which is unique to this work, is applied
on the hydrogen neutral fraction gradient. According to it, two
adjacent cells at positions $i$ and $i+1$ are refined (up to the
maximum level of refinement), if
\begin{equation}
  2 \left|
    \frac{x^i_{\hisub}-x^{i+1}_{\hisub}}
    {x^i_{\hisub}+x^{i+1}_{\hisub}+x^{floor}_{\hisub}}
  \right| > \Delta x_{\hisub},
\end{equation}
where $x_{\hisub}=1-x_{\hiisub}$ and $x^{floor}_{\hisub}$ is a floor
on the neutral fraction under which the criterion becomes inactive. In
our simulations we typically use $\Delta x_{\hisub}=0.75$ and
$x^{floor}_{\hisub}=10^{-3}$ in order to resolve gas streams, though
we tweak these values a bit (even within the same simulation) to tread
the fine line of neither under-resolving the streams nor
over-resolving uninteresting regions. This enforces maximum refinement
in the cold streams, so while the optimal resolution in our
simulations is slightly less than in recent works, our resolution in
the streams is unprecedented in cosmological simulations.

\vsk \textbf{The cosmological UV background} is incorporated into our
simulations with an `outside-in' method, where it is propagated from
under-dense and transparent UV-emitting voids.  As such, our UV
background can be thought of as \textit{quasi-homogeneous}, as opposed
to the completely homogeneous and optically thin implementation
commonly used in cosmological codes that lack radiative transfer
\citep[e.g.][]{Cen:1992p2824,Katz:1996p2854, Rasera:2006p2855}.  The
reasons we chose this model are mainly twofold. First, it is only a
single step further than previous work on the subject. This allows us
to isolate the effect of self-shielding, and to interpret our results
in a well established theoretical framework. Second, the inside-out
method would require finely tuned star formation rates and UV escape
fractions for simulated galaxies to produce a `correct' UV background,
and this is a subject onto itself \citep[see
e.g.][]{WiseCen09,AubertTeyssier10}. Also, our simulations zoom in on
a relatively small volume with no star formation outside, which would
lead to a severe lack of external UV background radiation. We thus
postpone such a model to a future paper and instead demonstrate in
Sec. \ref{Sec:maxflu} that a local enhancement of UV radiation due to
star formation would not significantly change our conclusions.

In practice, we use a `void' density threshold $\nhuv$ such that all
gas cells lower in density are UV emitters, and we impose the
redshift-dependent UV background model from
\cite{FaucherGiguere:2009p1685} onto these cells under the valid
assumption that voids are optically thin. The radiative field is then
allowed to diffuse out towards denser regions. The idea is to have the
void threshold low enough that it doesn't include the potentially
shielded cold streams themselves, but high enough that radiation can
quickly reach them (a sensitive issue due to the reduced speed of
light).  We use void thresholds of $n_H^{UV}\ga 10^{-4} \;
\mathrm{\cci}$ in our simulations. Our results are not sensitive to
the fine-tuning of this as long as $10^{-4} \; \mathrm{\cci} \la
n_H^{UV} \la 10^{-2} \; \mathrm{\cci}$.

\vsk The spectral shape of the UV field is approximately taken into
account by using three (H{\sc i}-, He{\sc i}- and He{\sc ii}-ionizing)
packages of photons which are propagated independently (see
Appendix~\ref{App_UV.sec}). In this work we adopt the on-the-spot
approximation (OTSA), where any UV photon emitted from a recombination
is assumed to re-ionize a nearby atom (i.e. within the same grid cell)
-- in other words, the simulated gas does not emit UV photons due to
recombinations and case B recombination rates are used in computing
the gas cooling rate.

\vsk For the sake of simplicity, our simulations do not include SN
feedback or metals.  To prevent artificial fragmentation
\citep{Truelove:1997p3217} our simulations employ a polytropic
equation of state \citep{Dubois:2008p1848} as a subgrid recipe that
keeps the mostly unresolved multi-phase interstellar medium (ISM) from
collapsing and fragmenting. The recipe sets a density-dependent
temperature floor in every gas cell:
\begin{equation}\label{Polytrope.eq}
  T_{min} = T_0 \left(
    \frac{\nh}{\nho}\right)^{\gamma-1},
\end{equation}
where we've chosen the values $T_0=10^4$ K, $\nho=1$ $\cci{}$, and
$\gamma=1.6$. The value of $\nho$ also corresponds to the limit above
which gas is star-forming.

\vsk We identify halos in our simulation outputs with the AdaptaHOP
algorithm from \cite{Aubert:2004p4312} and \cite{Tweed:2009p4217},
where the virial radius of a halo, $R_{vir}$, is defined as the radius
where the average density is 200 times the critical density of the
Universe, and the halo center corresponds to the DM density maximum.

\subsection{Numerical issues}
We've established through convergence tests that resolution is
adequate in our simulations and that the chosen parameters of
light speed and UV emission threshold ($f_c$ and $\nhuv$) do not
affect our results noticeably. Three other issues should be noted:

\textbf{The gravitational potential} in our simulations is usually
dominated by DM particles, but it is resolved to the local cell
resolution.  With our strategy of optimally resolving gas streams
comes the danger that we may over-resolve the gravitational potential,
with few and far-between DM particles causing discreteness effects in
the potential, which may lead to artificial fragmentation and
complexity in the streams. This seems particularly ominous since we
find in our simulations that the streams indeed become fragmented and
complex around massive halos. To make sure this is not caused by an
over-resolved gravitational potential we have run analogues to our
simulations with smoothed potentials, which still reveal fragmented
streams. So while it is hard to tell whether or not this numerical
effect is nonexistent in our simulations, we can conclude that it is
not dominating our results and that the complex streams are physical
in nature.

\begin{figure*}\begin{center}
  \subfloat{\includegraphics[width=0.33\textwidth]
    {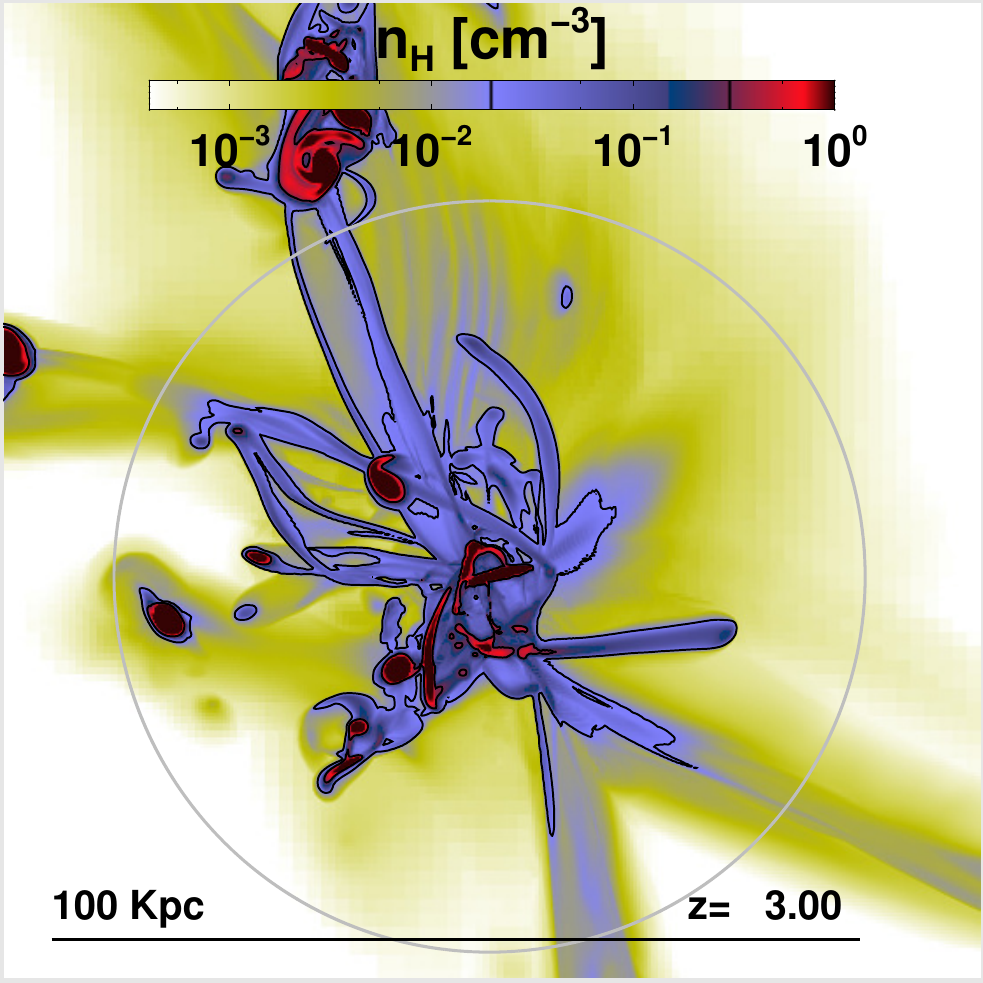}}\hspace{-1mm}
  \subfloat{\includegraphics[width=0.33\textwidth]
    {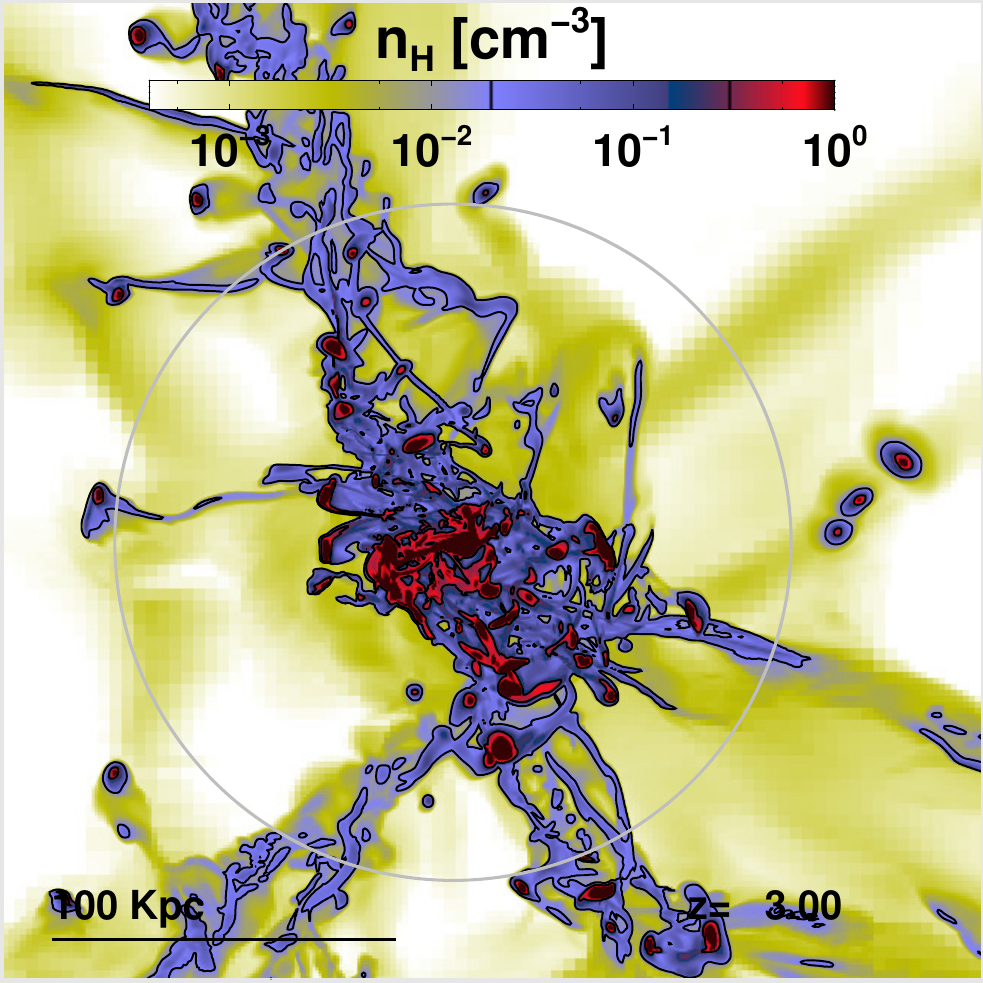}}\hspace{-1mm}
  \subfloat{\includegraphics[width=0.33\textwidth]
    {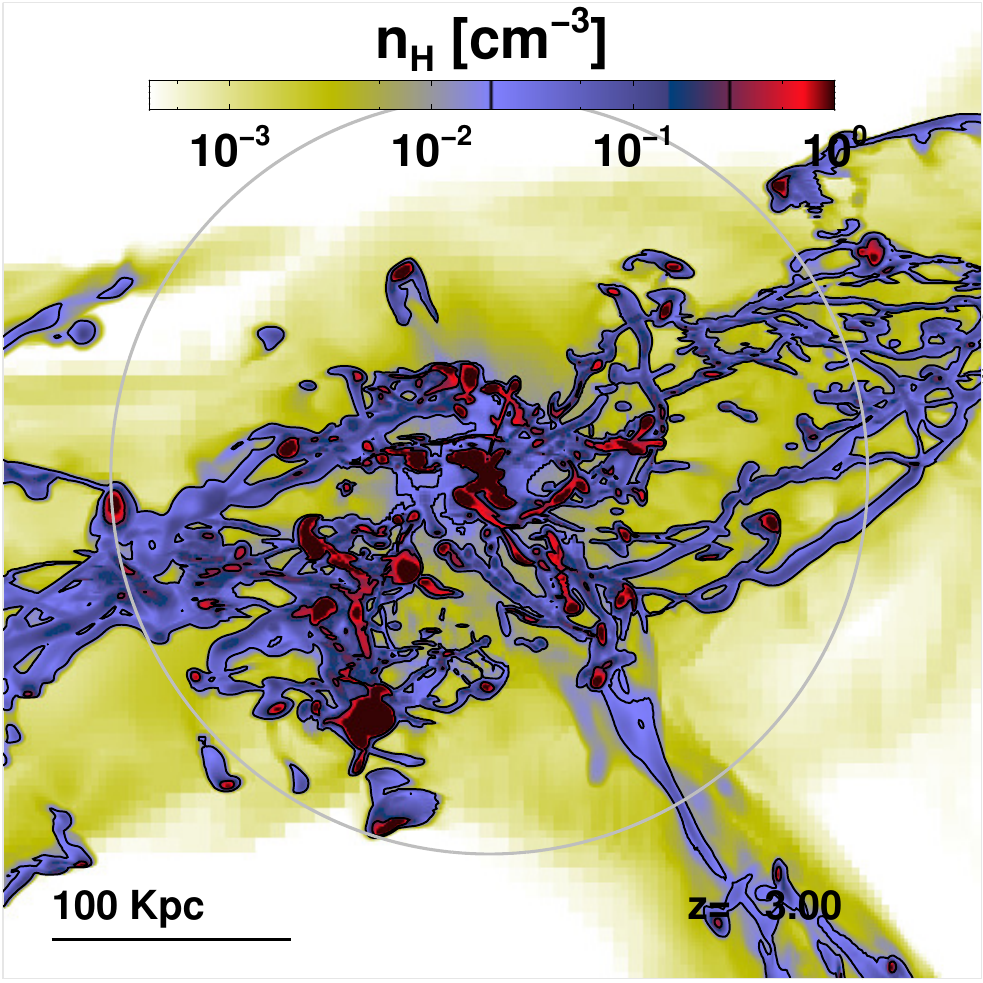}}\hspace{-1mm}
  \vspace{-3.9mm}

  \subfloat{\includegraphics[width=0.33\textwidth]
    {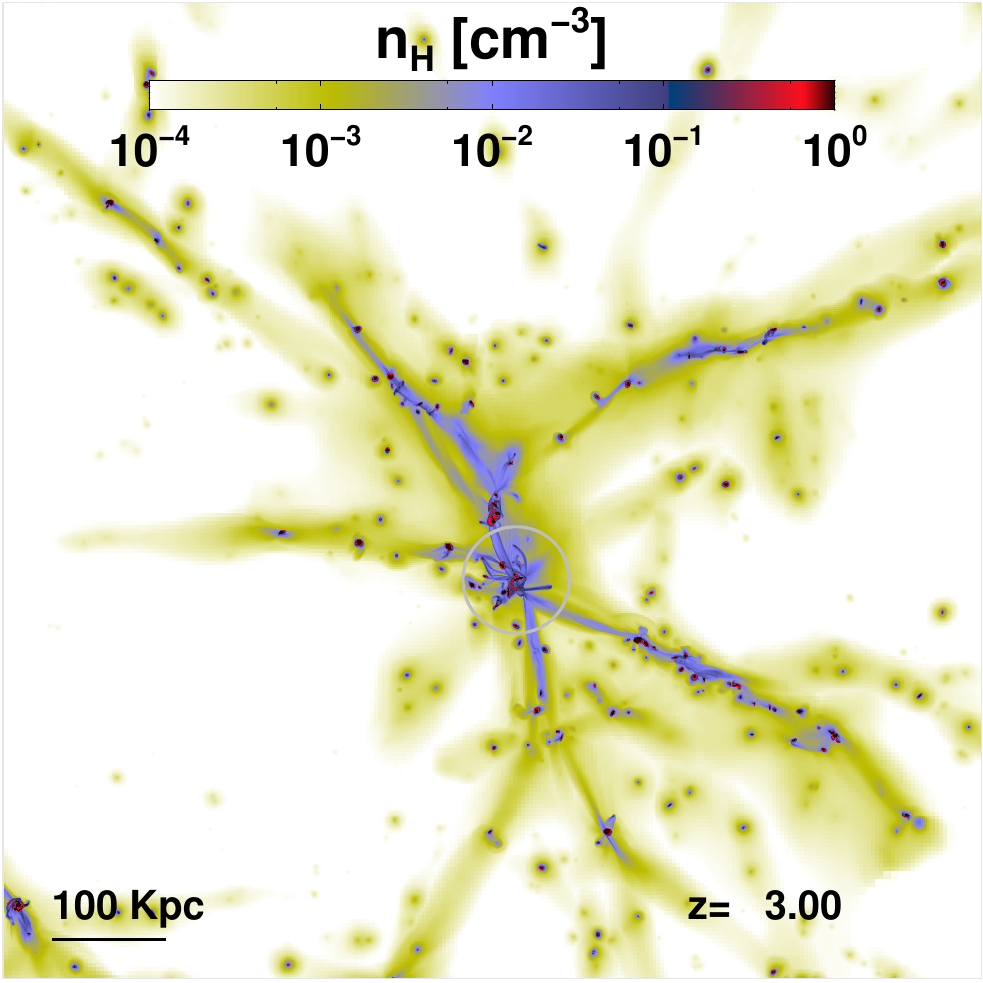}}\hspace{-1mm}
  \subfloat{\includegraphics[width=0.33\textwidth]
    {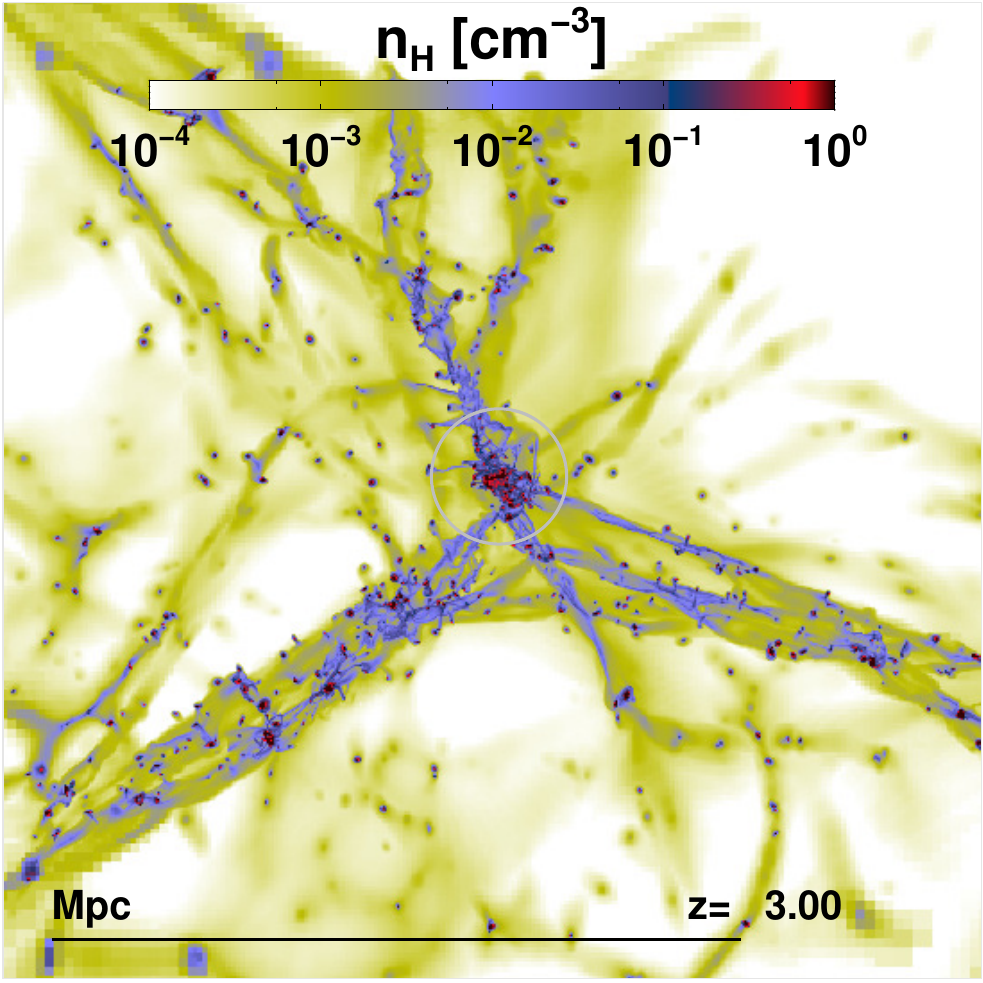}}\hspace{-1mm}
  \subfloat{\includegraphics[width=0.33\textwidth]
    {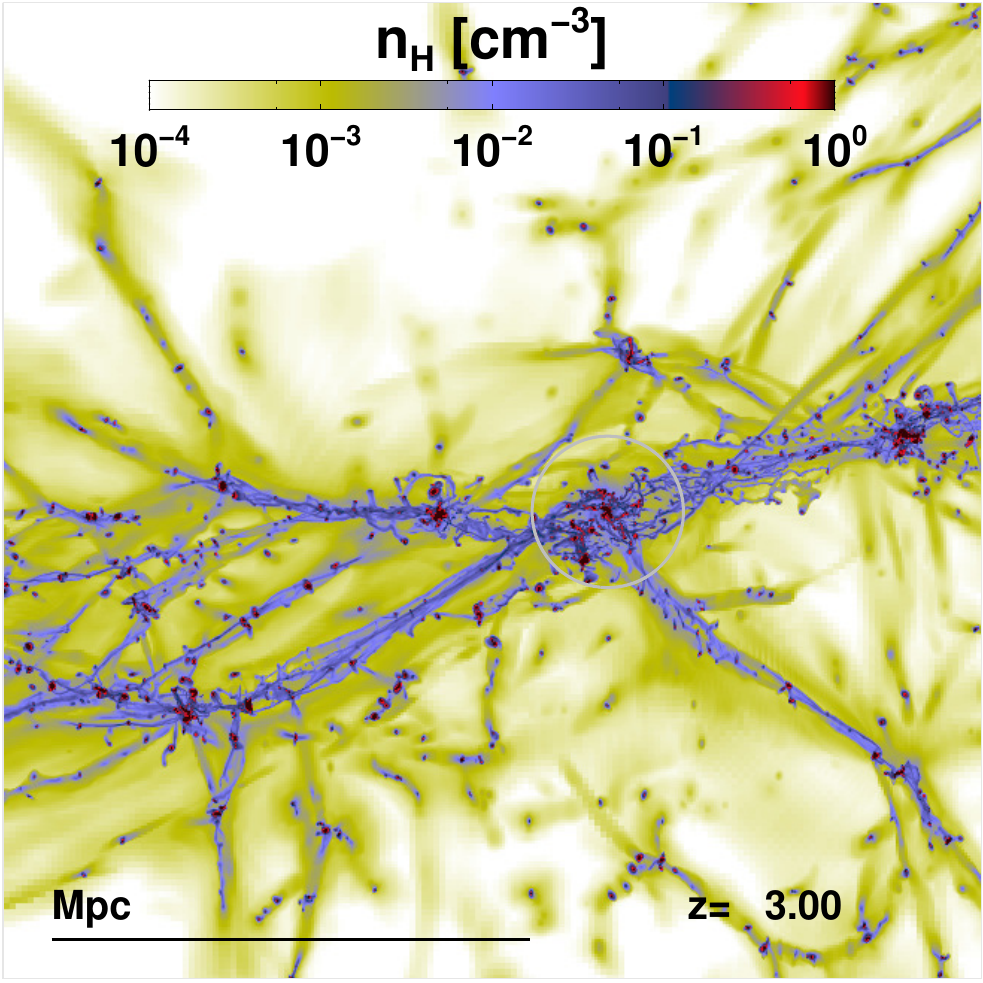}}\hspace{-1mm}
  \vspace{-3.9mm}

  \subfloat[H1]{\includegraphics[width=0.33\textwidth]
    {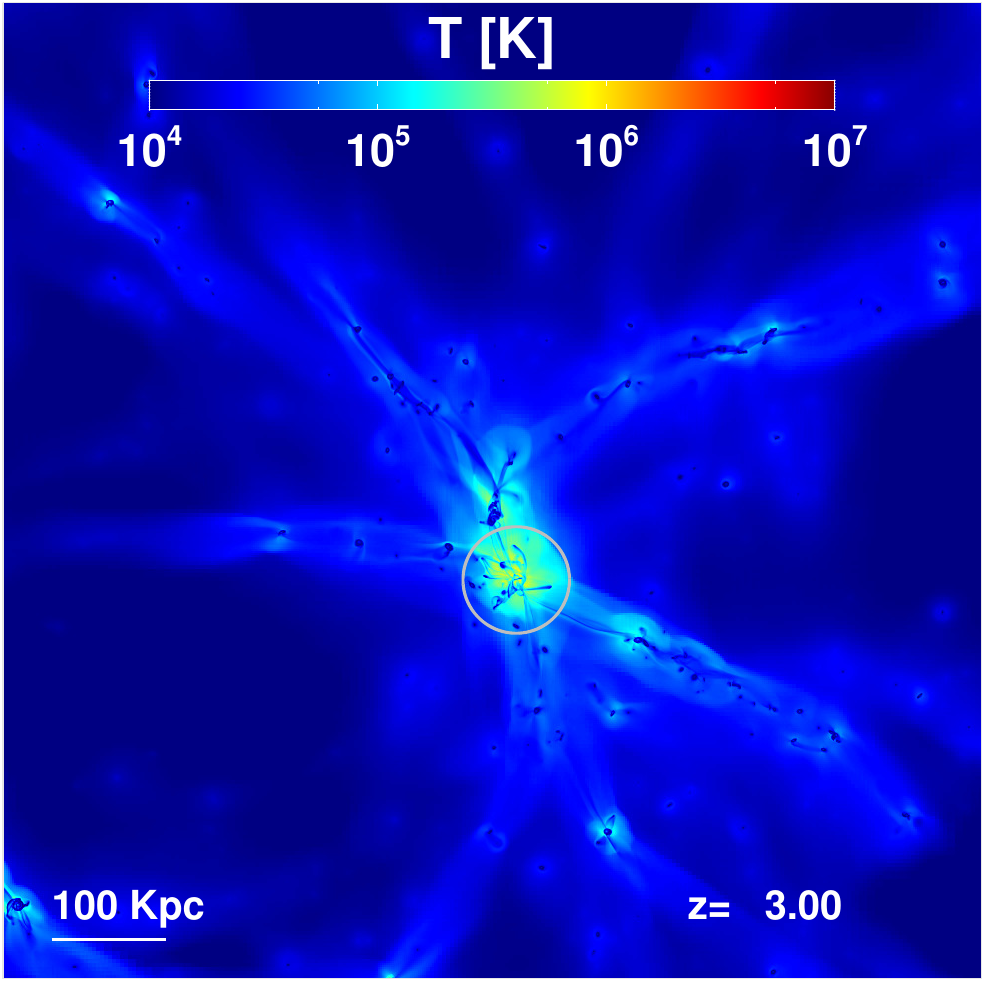}}\hspace{-1mm}
  \subfloat[H2]{\includegraphics[width=0.33\textwidth]
    {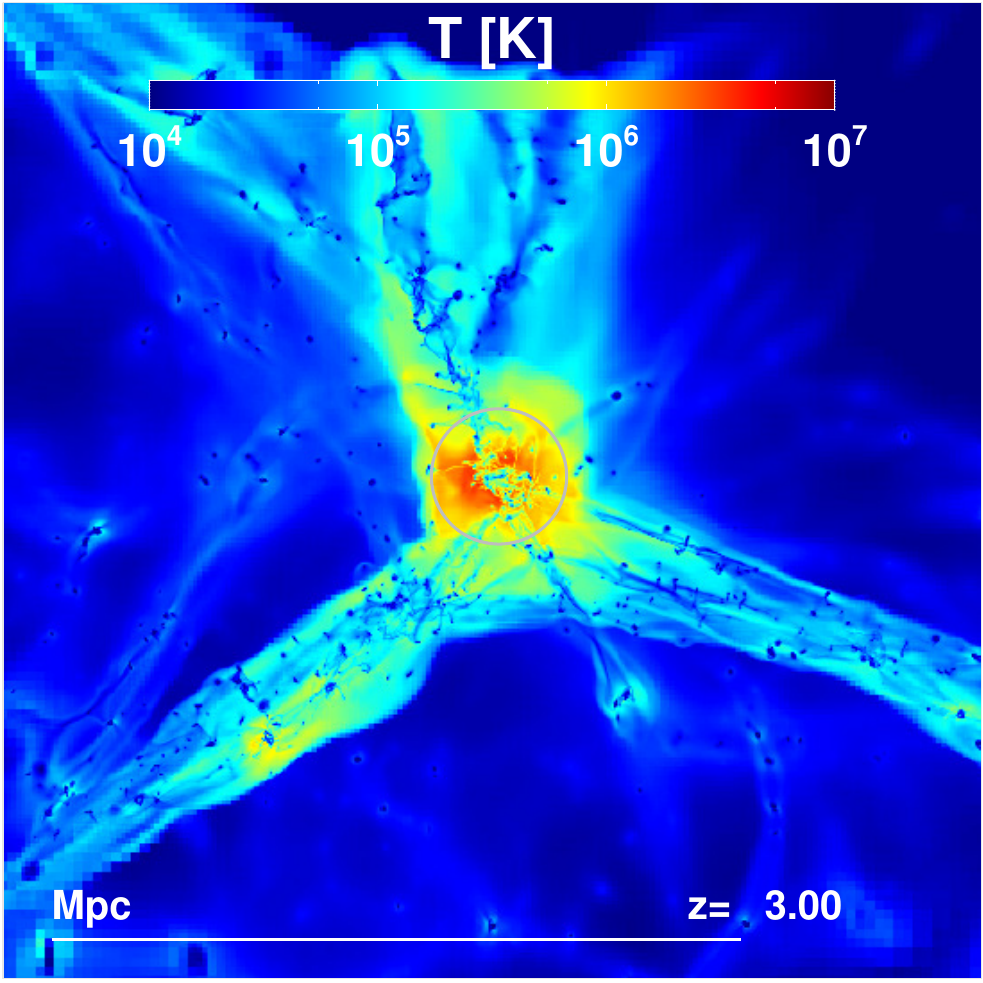}}\hspace{-1mm}
  \subfloat[H3]{\includegraphics[width=0.33\textwidth]
    {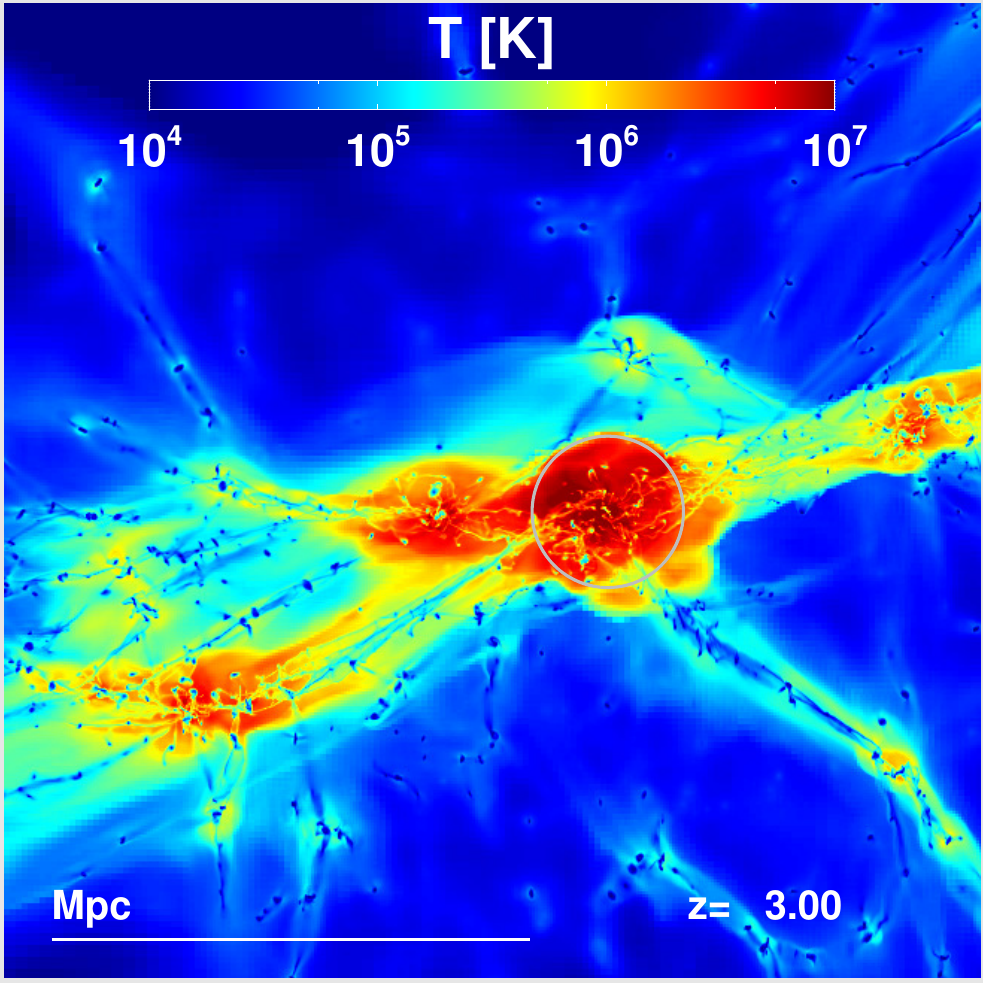}}\hspace{-1mm}
  \vspace{-1.3mm}

  \caption[]{\label{Halos.fig}Redshift 3 maps of the three targeted
    halos in simulations H1, H2 and H3 from left to right (increasing
    halo mass). Grey circles indicate virial radii of the halos; $46$,
    $98$ and $158$ kpc for the H1, H2 and H3 halos
    respectively. \textbf{Top row:} Number density maxima along the
    line of sight, with contours marking $0.02$ and $0.3$ $\cci$ as
    indicated in the color bars, corresponding to our definition of
    streams. \textbf{Middle row:} The same but zoomed out to show the
    large-scale environment. \textbf{Bottom row:} Mass weighted
    temperature maps, on the same scale as the middle row.}
  \vspace{-3.3mm}
\end{center}\end{figure*}

\textbf{Operator splitting} is a widely-used method of decomposing
unwieldy differential equations into separate parts that can be solved
independently and in sequence \citep[e.g.][]{Toro99,
  1992nrfa.book.....P}. \ramsesrt{} employs this method to split the
radiative-hydrodynamics equations into \textit{(i)} advection of gas
and photons between cells and \textit{(ii)} chemical reactions within
the cells (radiative cooling and photo-heating). The advection part is
first solved and then cooling, using the advection result as the
initial state and subcycling when needed. Gas normally exists in a
competition between advective/gravitational heating and radiative
cooling, where the temperature `adjusts' to a value where these
processes cancel each other out. However, when the cooling time is
shorter than the advection time, operator splitting may artificially
give cooling the upper hand, leading to a slight underestimate in the
temperature.  Normally this is no big deal, but considering how
sensitive \lya{} emissivity is to gas temperature (see
\Fig{LyaEmi.fig}), this can result in a severe underestimate of \lya{}
emissivity in the gas. We've verified that this is indeed the case in
our simulations.  To prevent this from affecting our results, we
restart the simulations at $z=3$ with the global timestep reduced by
orders of magnitude, to make sure it is everywhere shorter than the
local cooling-time, and run until we reach convergence in \lya{}
luminosity (this takes a few-thousand fine-cell time-steps).

\textbf{Cell merging:} With the bookkeeping on ionization states, and
due to the fact that cell de-refinement takes place just before
outputs are written in \ramses{}, special care must be taken on cell
merging. Applying the traditional method of giving a merged cell a
children-averaged ionization state can sometimes result in a
combination of temperature and ionized state which causes it to
outshine whole galaxies in \lya{} emissivity (see discussion in
\Sec{emission.sec}). To prevent this we enforce a photoionization
equilibrium (PIE) ionization state to merged cells, assuming the
children-averaged values of gas density, pressure, and UV flux.

\section{Physical properties of 3 halos}
In this section, we first review the qualitative properties of our
three simulated halos, and define the different phases of the
intra-halo gas. We then describe in detail the impact of
self-shielding on the ionization and thermal states of cold streams,
and discuss the validity of an approximate treatment of self-shielding
introduced by FG10.
\subsection{Basic halo properties}
Gas density maps of the three targeted halos at redshift 3 are shown
in \Fig{Halos.fig}, in close-ups of the halos and zoom-outs to show
their environments. Also shown are zoom-out maps of temperature. The
halos display a tendency with increasing mass towards more intense,
complex and fragmented accretion, and larger and hotter domains of
shock-heated intergalactic medium (IGM).

The least massive halo (H1, left) has narrow (down to $\sim 1$ kpc in
diameter) and unperturbed accretion streams and tidal tails stretching
from the central galaxy, which can be seen in red at the center of the
halo, edge-on and slightly inclined from the horizontal. It is about
to undergo a major merger with another halo five times less massive,
situated just outside $R_{vir}$ and coming in from the north. Two
parallel accretion streams bridge the merging halos. Another accretion
stream extends towards a factor 100 smaller merging halo, also at the
edge of $R_{vir}$, but towards the line of sight (LOS), seen as a
moon-shaped clump to the south-west. To the south and south-east are
two relatively thick and diffuse accretion streams and another one
even more diffuse to the west.  Other structures in the map are
orbiting satellites and tidal tails.

The intermediate mass halo (H2, middle) is more a group of orbiting
galaxies than a single galaxy. On the large scale there is a network
of filaments mixed with galaxies of varying masses, with at least 6
large scale streams extending towards the central halo. Movies show
that the accretion here is notably more spiral than around the H1
halo, with the streams starting to curve around the center of gravity
already well outside $R_{vir}$.  Inside the halo we see plenty of
streams and tidal tails, but much more disrupted and messy than in the
H1 halo, as a result of stronger and more frequent interactions with
other streams and galaxies.

This tendency continues with the most massive halo (H3, right), where
we find even more disrupted streams, to the point that many of them
seem to be completely obliterated close to the halo center. The H3
halo has just undergone a major merger, which makes the accretion
activity particularly violent at this point in time.

\vsk To facilitate our analysis, we apply the following categorization
to divide the gas into phases, as shown in the temperature-density
phase diagram in \Fig{H1_PH.fig} (note that the categorization is
specific to this paper and does not apply in general):

\begin{figure}\begin{center}
  \includegraphics[width=.5\textwidth]{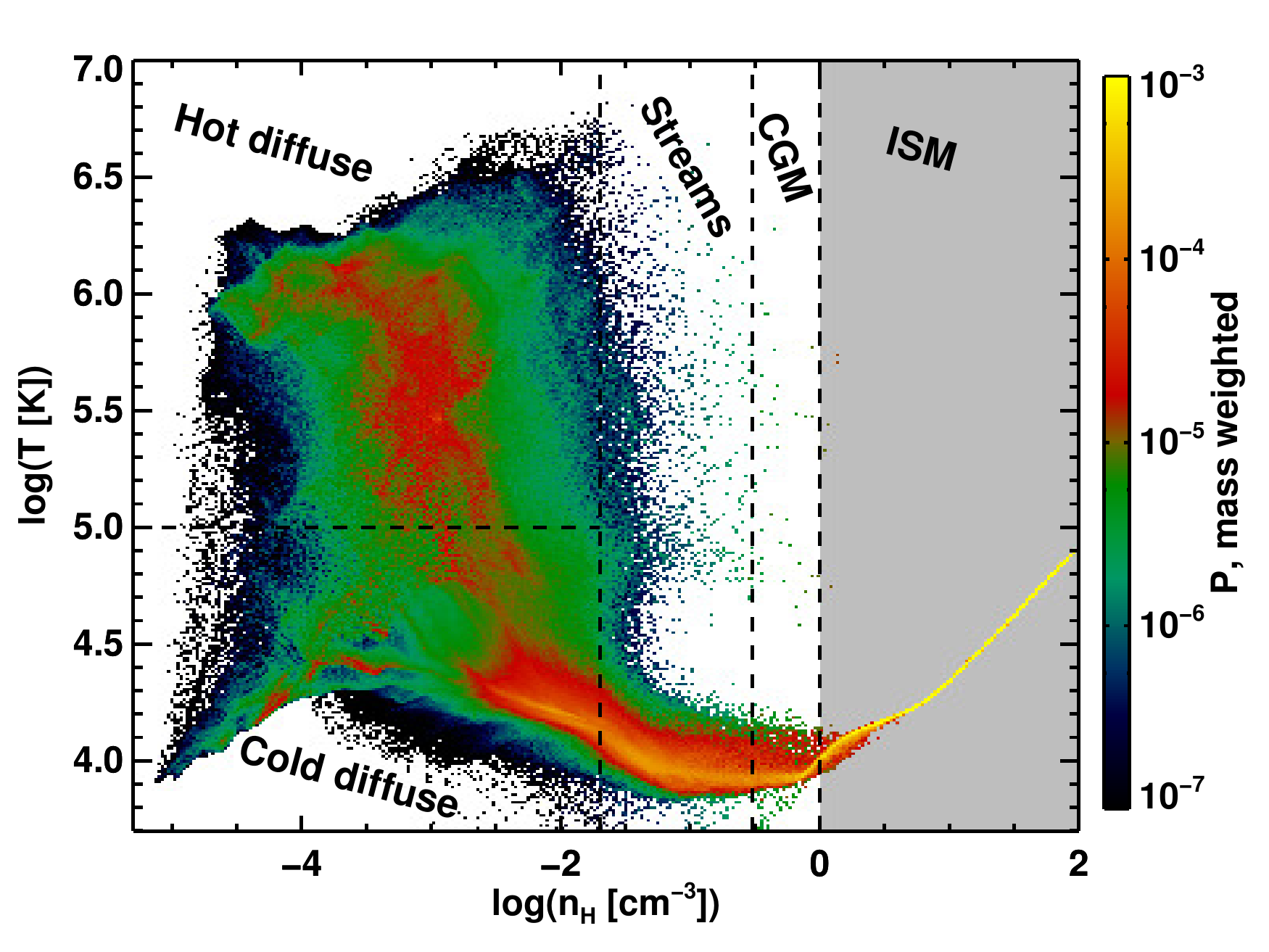}
  \caption[]{\label{H1_PH.fig}Phase diagram of the H1 halo, showing
    our density-dependent definitions of streams, CGM and ISM. The ISM
    is shaded to indicate that we always ignore ISM gas when adding up
    \lya{} emissivities. The color scale represents mass weighted
    probability per temperature-density bin.}
\end{center}\end{figure}

\vsk \textbf{The star-forming ISM} is all gas denser than $1$ $\cci$
and as discussed in \Sec{Sec:simpars} we apply a temperature floor in
the form of a density-dependent polytrope to keep this gas from
artificially fragmenting, which manifests itself in the constrained
temperature-density relation in the shaded area of the diagram. Our
simulations lack the ingredients to accurately model \lya{} emission
from the ISM (multiphase resolution, dust, \lya{} scattering) and our
reaction to that is to simply ignore the \lya{} emission from there in
our analysis. The shaded color of the ISM region in \Fig{H1_PH.fig}
should remind the reader of this and that this work is about modelling
the \lya{} emission coming from galactic environments and not the
galaxies themselves.  The ISM gas density threshold is resolution
dependent and reflects the density at which further collapse of gas --
i.e. the Jeans length -- is no longer resolved. At our chosen density
threshold, assuming minimum temperatures of $10^4$ K, the Jeans length
is resolved by approximately 10, 5, and 2.5 cell widths in the H1, H2
and H3 simulations respectively. It should be noted that our density
threshold is almost an order of magnitude above what has typically
been used in recent similar works (e.g. G10, FG10).

\textbf{The CGM} is gas with number densities between $0.3$ and $1$
$\cci$. Ideally these densities form membrane interfaces between the
ISM and their more diffuse environment. The lower density limit
corresponds to the inner contours in the density maps of
\Fig{Halos.fig} (top row), and from those maps it can be seen that the
CGM gas is indeed mostly constrained to galaxies (in red). In the
phase diagram we find that most of the CGM gas is cooled down to the
temperature floor of $\sim 10^4$ Kelvin where radiative cooling
basically stops (metals can cool gas further but we don't include
those).  Although CGM gas in our simulations is not directly affected
by the polytropic equation of state, one may expect gas at densities
$\ga 0.1 \ \cci$ to be multiphase and star-forming
\citep[][]{Schaye:2004p5757}. This cannot happen in our simulations
because they don't describe cooling below $\sim 10^4$ K, and this
temperature floor provides artificial pressure support for dense
gas. This implies a potentially high error in our predictions for the
\lya{} luminosities of halos, resulting from an overestimated CGM
contribution. Thus, while we in general include the CGM gas in our
analysis of \lya{} luminosities, we also consider at some points the
effect of excluding it, to get a grip on how sensitive our results are
to the density thresholds applied. In summary we find that GCM gas
typically provides a $~40\%$ of the \lya{} luminosities of our
simulated halos, but that in terms of \lya{} extent it is less
substantial.

\textbf{Streams} are defined in this work as gas with densities
between $0.02$ and $0.3$ $\cci$. These limits correspond to the
density contours in \Fig{Halos.fig} (top row) and from those we can
see that these densities indeed correspond to thin filamentary
structures.  Much like the CGM most of the stream gas is found at the
bottom of the temperature curve at $\sim 10^4 $ K though we do see an
increase in temperature in the more diffuse stream gas due to a
combination of photo/gravitational heating and inefficient cooling
(because of the low densities). Gas at sub-stream densities turns out
to be negligible in terms of \lya{} emissivity and thus not very
important to our results so we crudely split what remains into two
categories.

\setlength{\tabcolsep}{4.5pt}
\begin{table}
  \caption{Halo sizes and mass budgets (\% of mass within $R_{vir}$).}
  \label{tbl:mass}
  \begin{tabular}{llllllll}
    \toprule
    \multicolumn{1}{c}{Halo} & \multicolumn{1}{c}{$R_{vir}$} 
    & \multicolumn{1}{c}{DM} &
    \multicolumn{1}{c}{Stars} &  \multicolumn{4}{l}{Gas}   \\
    &  &    &     & ISM & CGM & Streams   & Hot \\
    \midrule
    H1 & 46 kpc  & 82\% & 8\% &  10\%   &     &      &      \\ \cline{5-8}
    &         &      &     &  73\%   &3\%  & 8\%  & 8\%  \\ 
    H2 & 98 kpc  & 81\% & 7\% &  12\%   &     &      &      \\ \cline{5-8}
    &         &      &     &  60\%   & 9\% & 16\% & 13\% \\  
    H3 & 158 kpc & 82\% & 5\% &  13\%   &     &      &      \\ \cline{5-8}
    &         &      &     &  58\%   & 6\% & 12\% & 23\%  \\
    \bottomrule
  \end{tabular}
\end{table}

\begin{figure*}\begin{center}
  \renewcommand*{\thesubfigure}{}
  \subfloat{\includegraphics[width=0.33\textwidth]
    {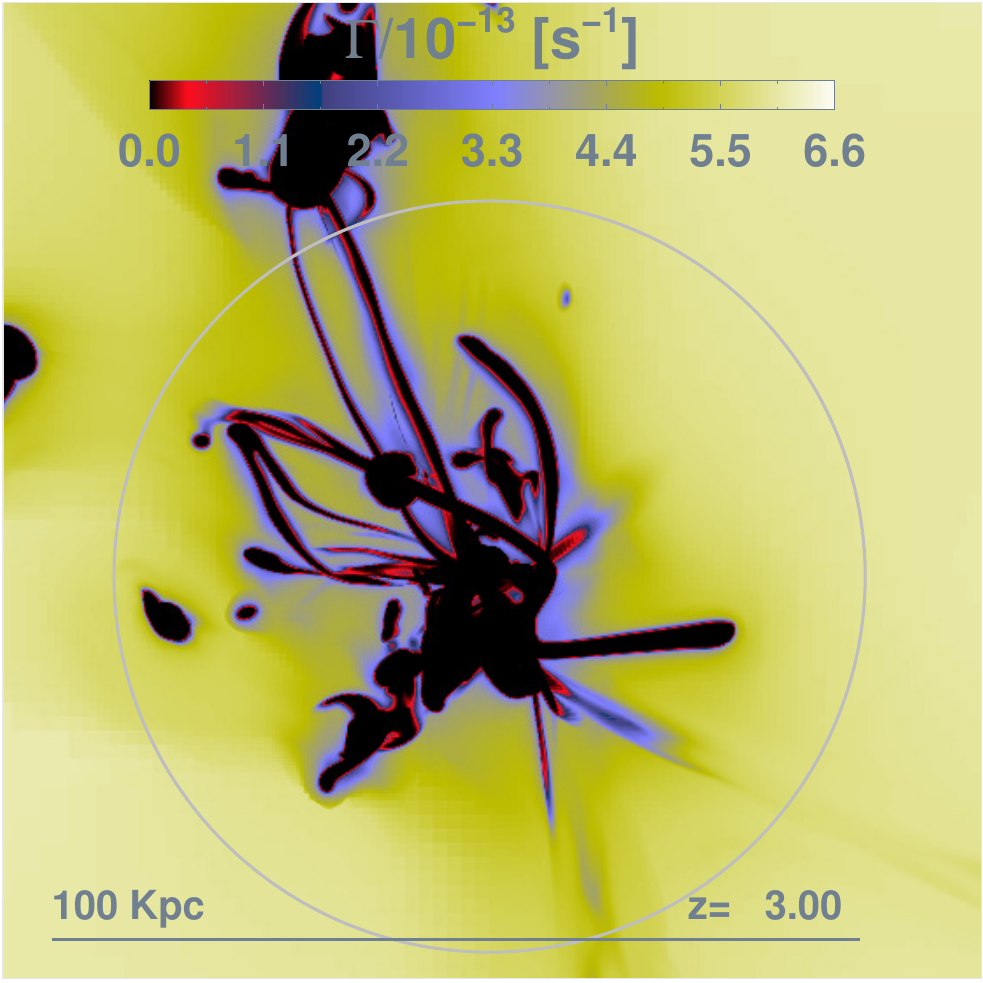}}\hspace{-1mm}
  \subfloat{\includegraphics[width=0.33\textwidth]
    {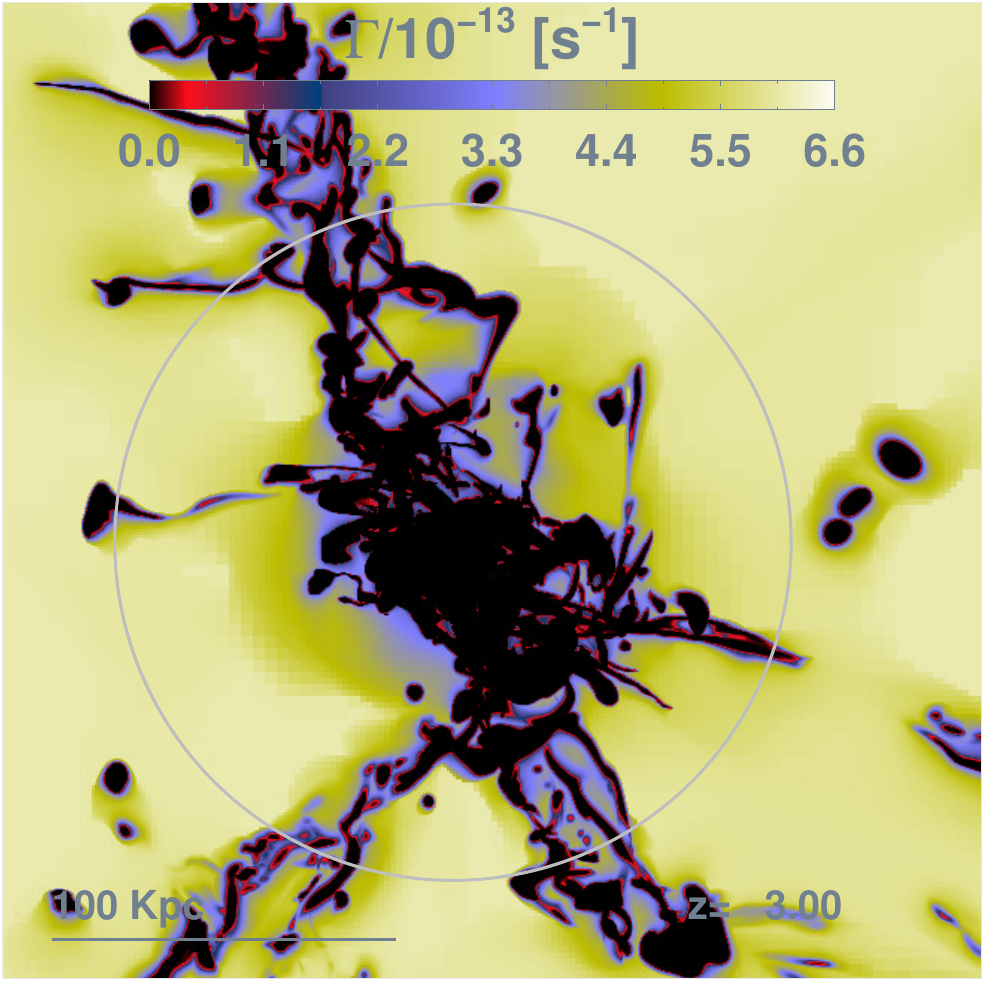}}\hspace{-1mm}
  \subfloat{\includegraphics[width=0.33\textwidth]
    {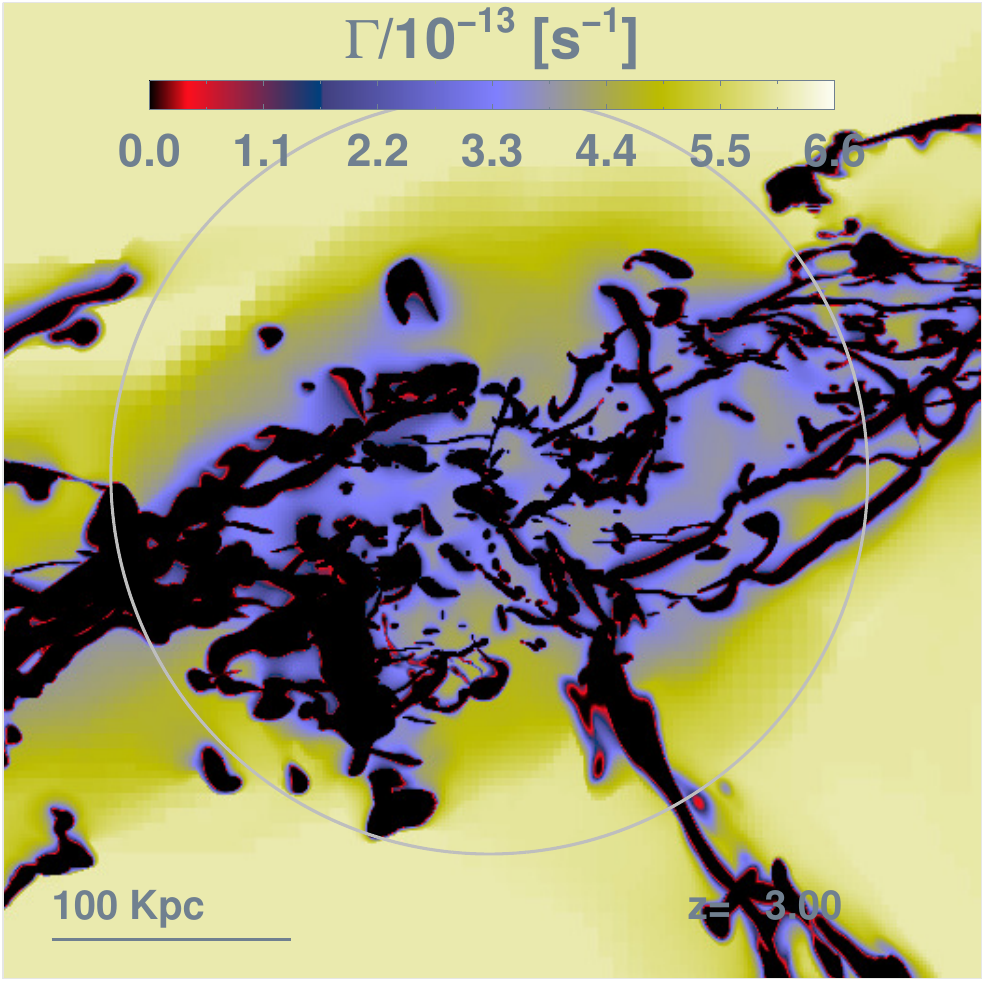}}\hspace{-1mm}
  \vspace{-1mm}

  \hspace{-2mm}
  \subfloat{\includegraphics[width=0.424\textwidth]
    {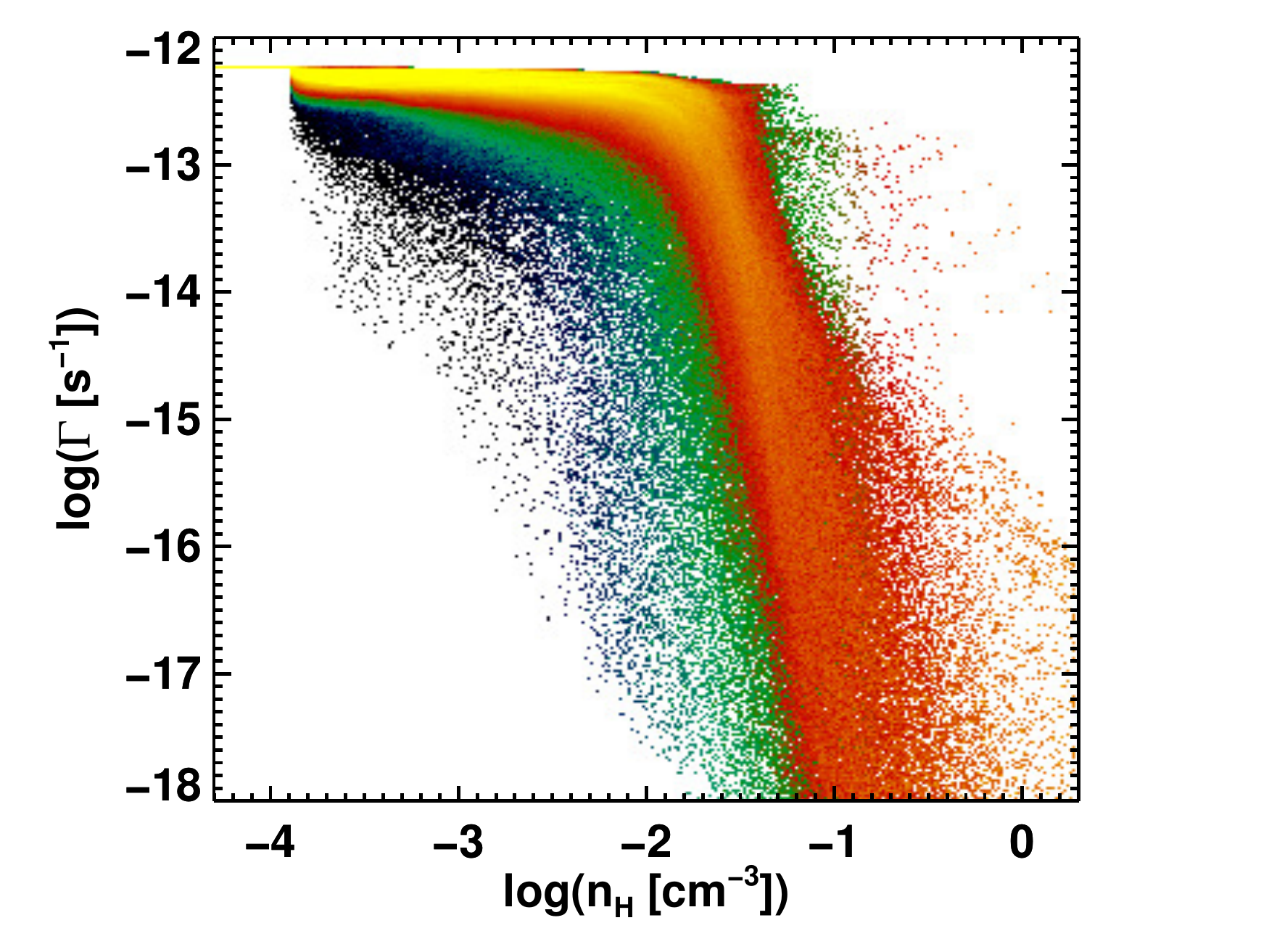}}\hspace{-25.3mm}
  \subfloat{\includegraphics[width=0.424\textwidth]
    {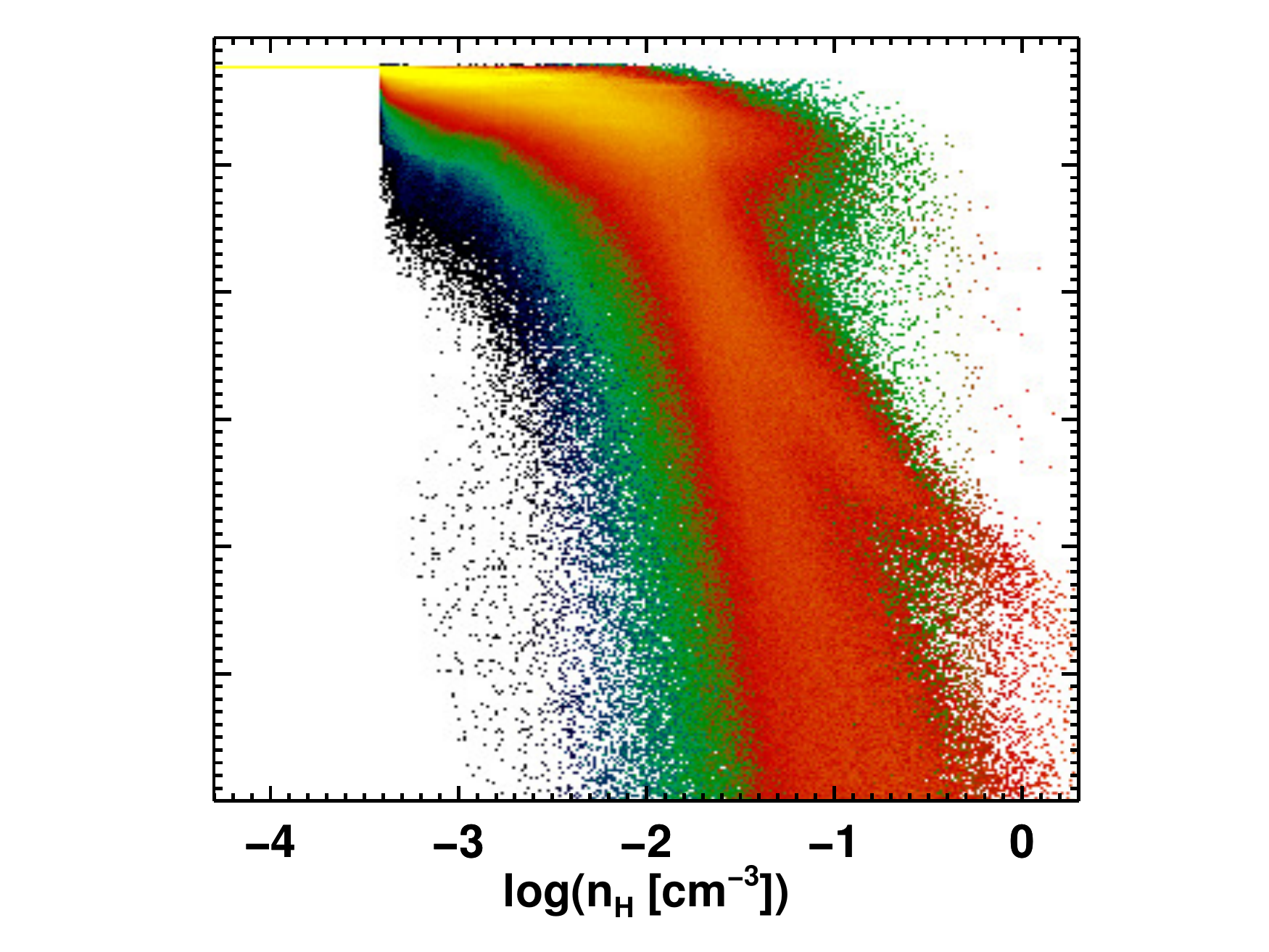}}\hspace{-25.3mm}
  \subfloat{\includegraphics[width=0.424\textwidth]
    {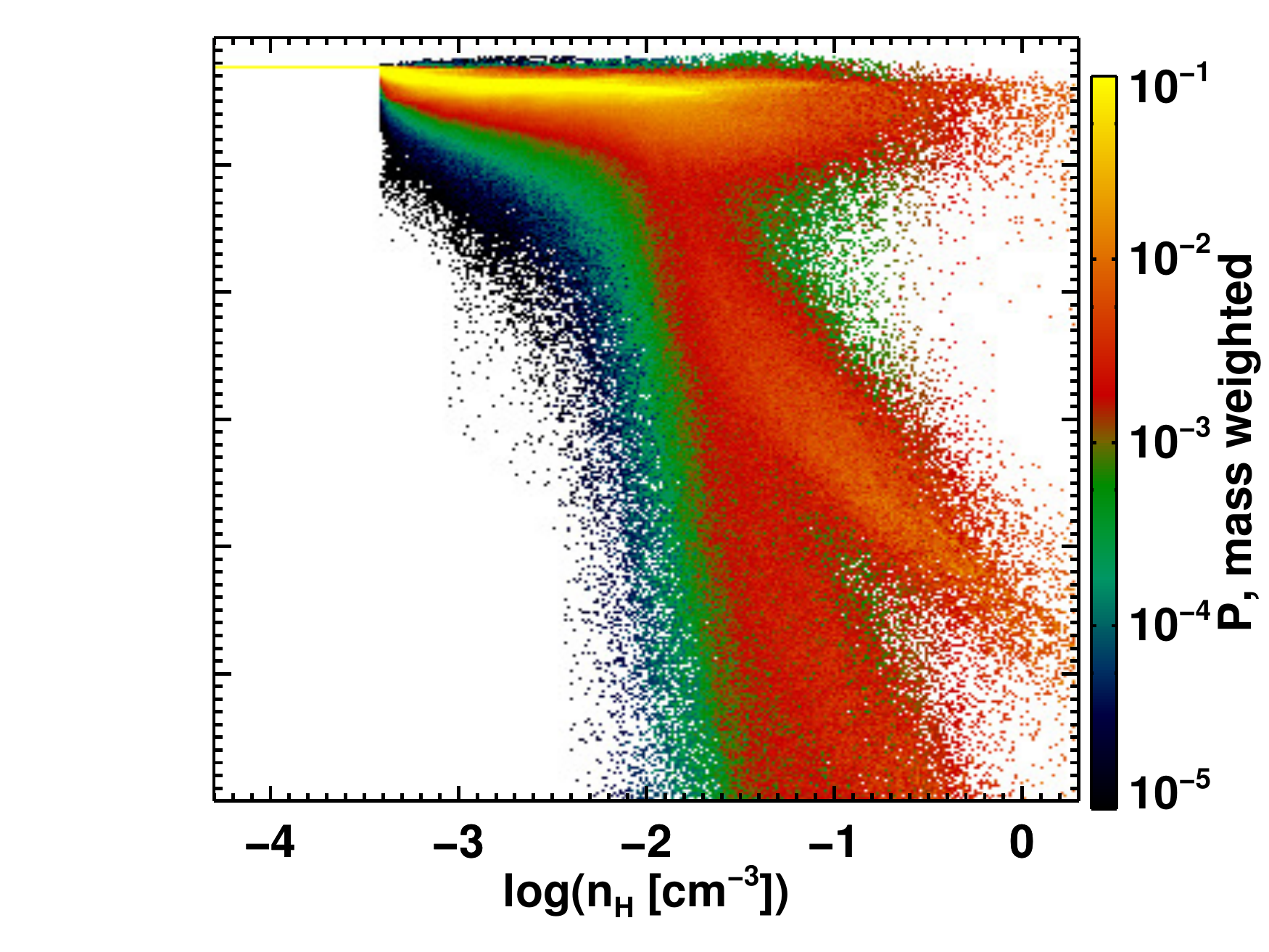}}
  \vspace{-2mm}
  \caption[]{\label{SS.fig}Self-shielding at redshift 3 in the three
    halos, from left to right, H1, H2 and H3. \textbf{Top row:} Maps
    of projected minima along the LOS of the hydrogen photoionization
    rate. The scale is non-logarithmic and in units of $10^{-13} \;
    \mathrm{s^{-1}}$. \textbf{Bottom row}: Phase diagrams of hydrogen
    photoionization rate versus density. The color scale represents
    mass-weighted probability per $\Gamma-\nh$ bin over the plotted
    $\Gamma$ range.}
\end{center}\end{figure*}

\textbf{Hot diffuse} gas has been shock heated above $10^5$
Kelvin. As seen in the temperature maps in \Fig{Halos.fig} (bottom
row) this gas exists in abundance within the virial radii of the
halos, but there also seems to be weaker heating around the
large-scale accretion streams (and actually not so weak in the large
streams around the H3 halo). Shock heating gets decidedly stronger
with increasing halo mass, with gas reaching $\sim 2\ 10^7$ K in H2
and $\sim 6\ 10^7$ K in H3. Also, increasingly dense gas exists above
$10^5 $ K in the more massive halos; CGM in H2 and ISM in H3.

\textbf{Cold diffuse} gas is partly gas which is slowly condensing
towards the streams and the CGM and partly cosmological gas that has
not interacted with the halos at all and is being cooled down by the
cosmological expansion.

\begin{figure*}\begin{center}
  \subfloat{\includegraphics[width=0.33\textwidth]
    {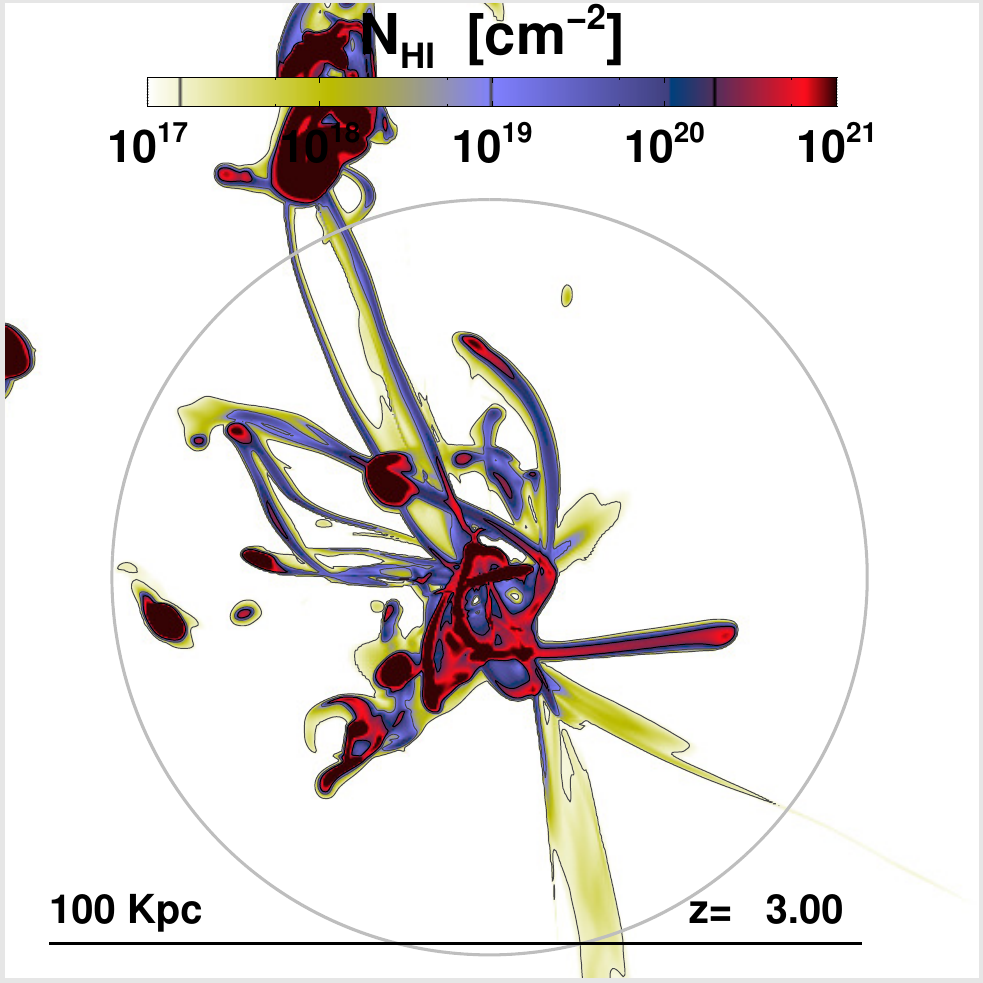}}\hspace{-1mm}
  \vspace{-1.5mm}
  \subfloat{\includegraphics[width=0.33\textwidth]
    {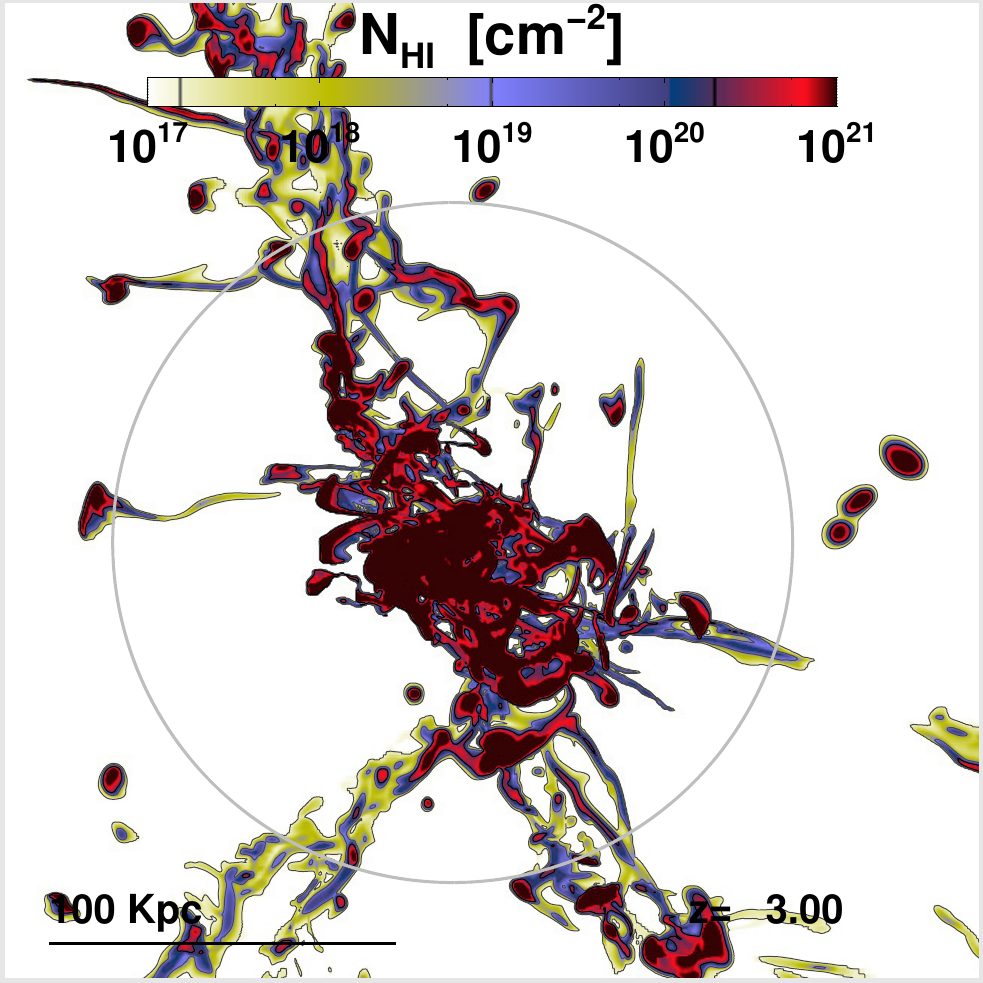}}\hspace{-1mm}
  \subfloat{\includegraphics[width=0.33\textwidth]
    {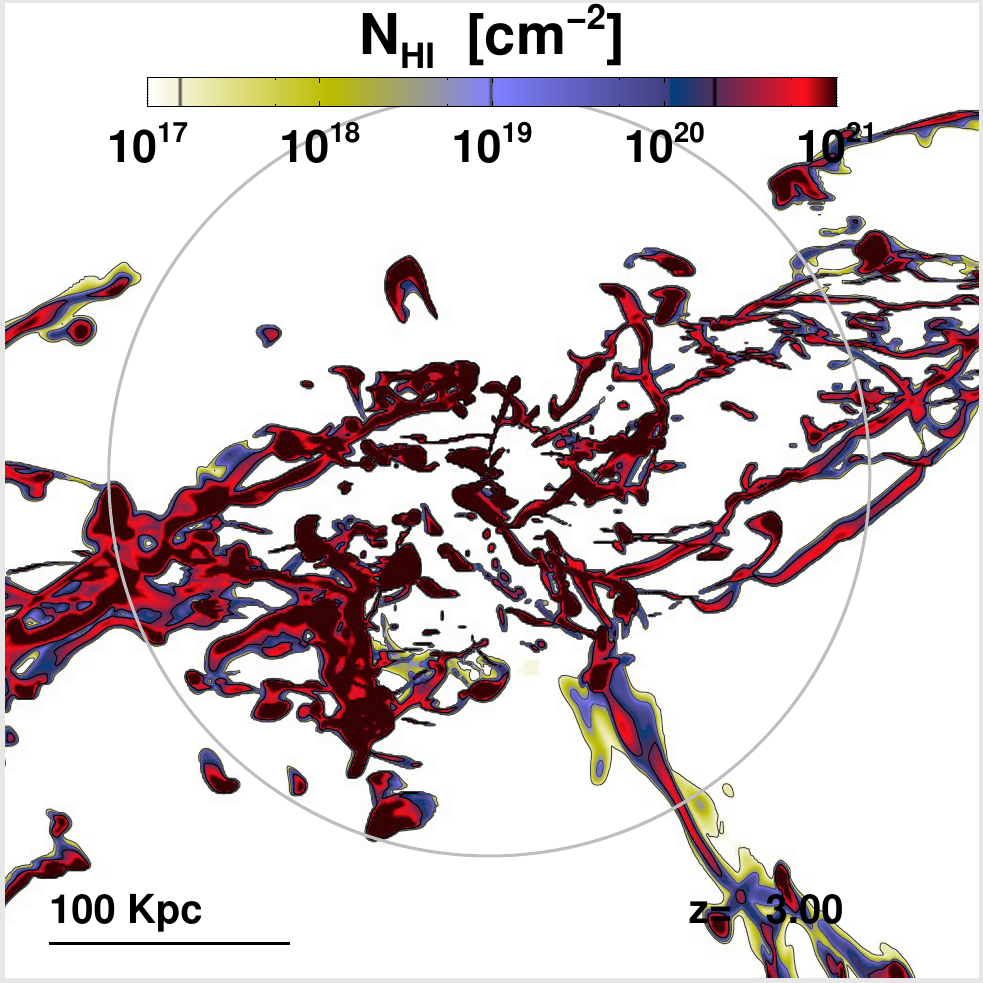}}\hspace{-1mm}
  \renewcommand*{\thesubfigure}{}

  \hspace{-3mm}
  \subfloat{\includegraphics[width=0.405\textwidth]
    {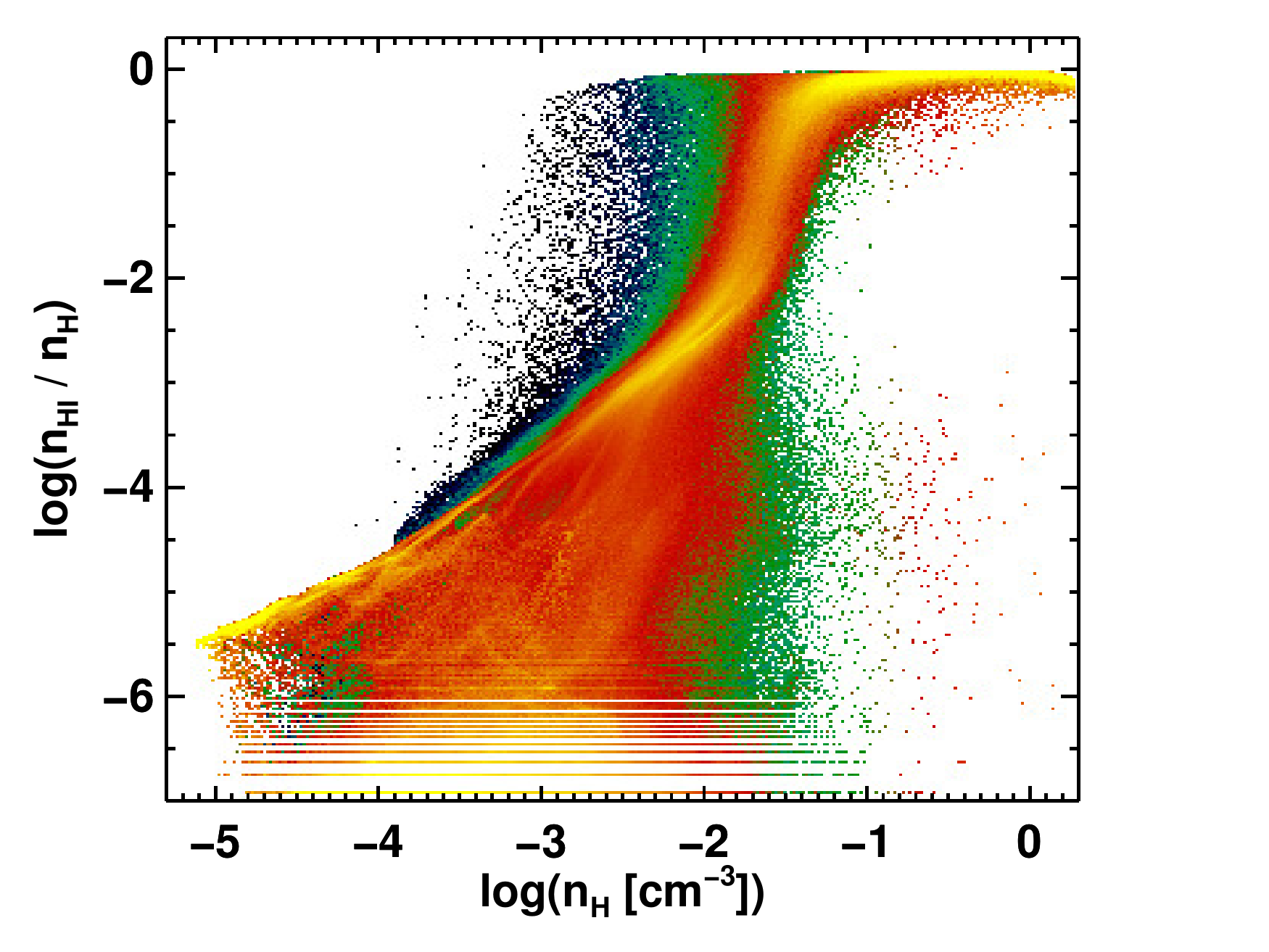}}\hspace{-21.5mm}
  \subfloat{\includegraphics[width=0.405\textwidth]
    {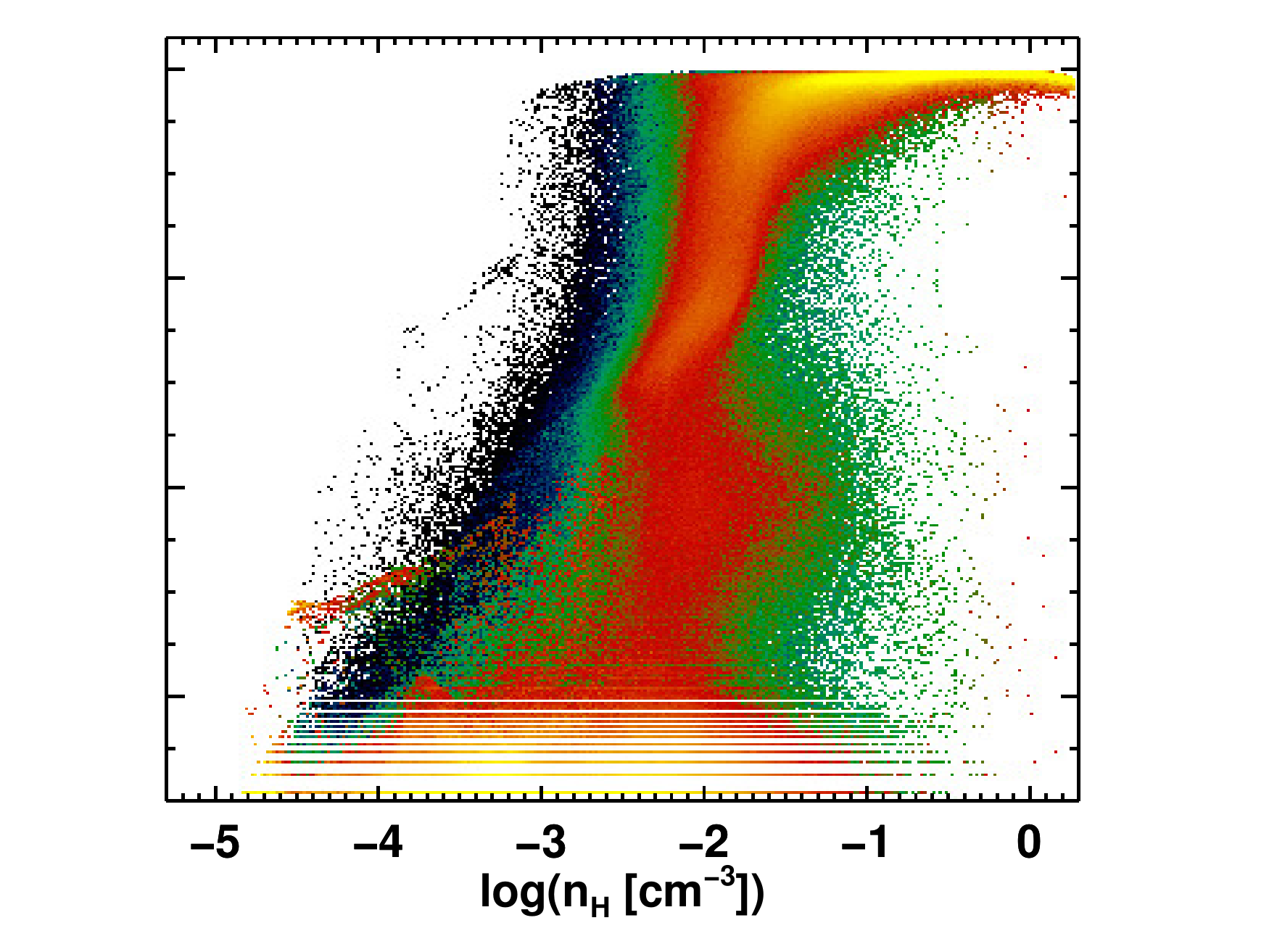}}\hspace{-21.5mm}
  \subfloat{\includegraphics[width=0.405\textwidth]
    {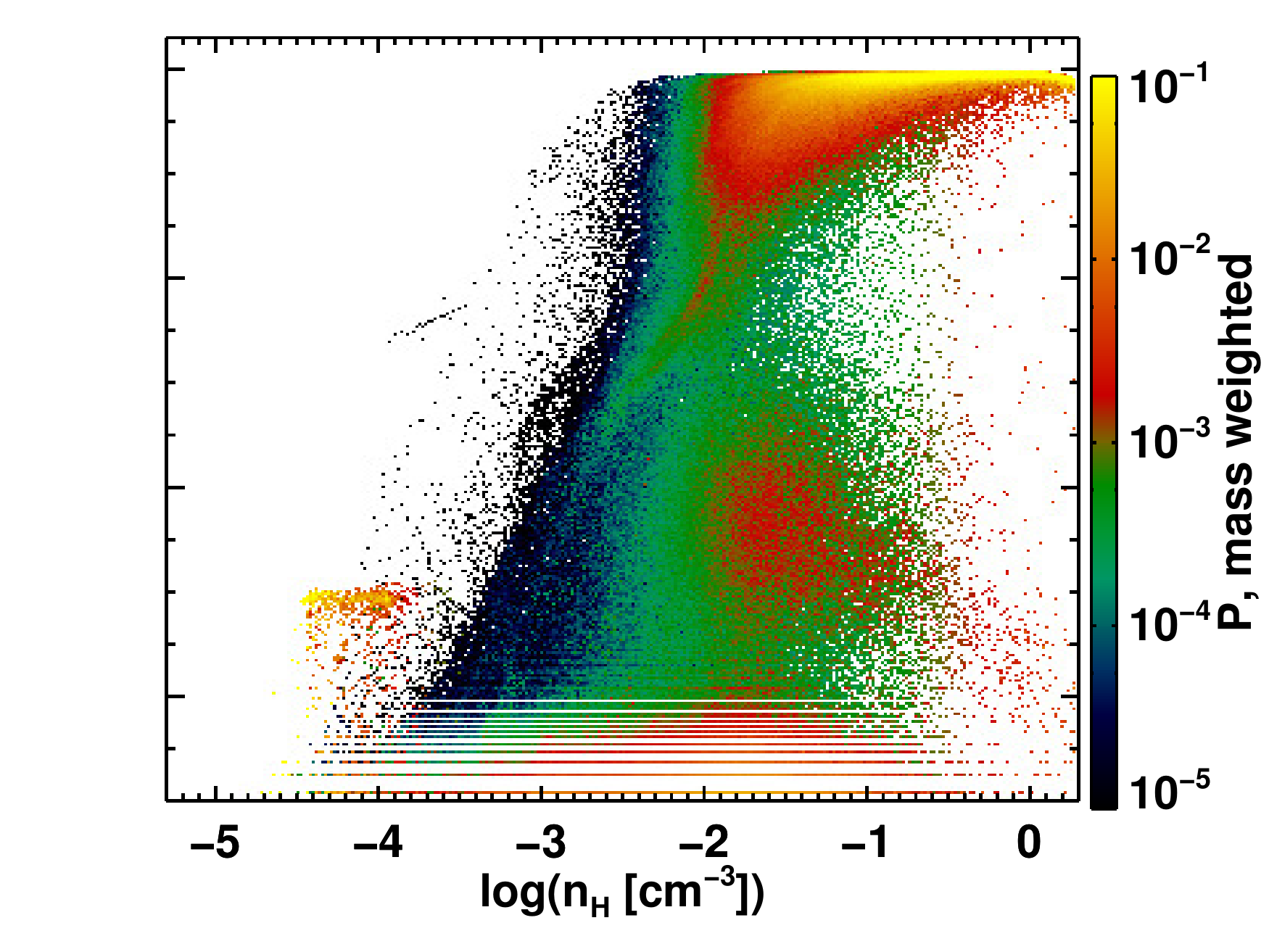}}\hspace{-1mm}
  \vspace{-2mm}
  \caption[]{\label{xHI.fig}Neutral hydrogen at redshift 3 (H1, H2, H3
    halos from left to right). \textbf{Top row:} Maps of projected
    \hi{} column density. Contours correspond to the lower limits for
    DLAs ($\snh = 2\ 10^{20} \; \mathrm{cm^{-2}}$), SLLSs ($10^{19} \;
    \mathrm{cm^{-2}}$) and LLSs ($1.6\ 10^{17} \; \mathrm{cm^{-2}}$).
    \textbf{Bottom row:} Phase diagrams of neutral hydrogen fraction,
    $\xhi\equiv \nhi / \nh$, versus density. The color scale
    represents mass-weighted probability per $\xhi-\nh$ bin over the
    plotted $\xhi$ range.  The quantization-like horizontal lines at
    the bottom of all diagrams are due to numerical precision of the
    cell variable $\xhii$ in \ramsesrt{}, which is roughly seven
    digits. }
\end{center}\end{figure*}

\vsk
The sizes and mass budgets of our three targeted halos are listed in
\Tab{tbl:mass}. Each of the halo masses consists of roughly $80\%$
dark matter and $20\%$ baryons. The stellar/gas ratio decreases with
halo mass, going from roughly one-to-one in H1 to about a one-to-three
in H3. The gas mass is primarily in the ISM, going from $73\%$ in H1
to $58\%$ in H3. The hot gas fraction clearly increases with halo
mass, going from $8\%$ in H1 to $23\%$ in H3 and correspondingly the
cold fraction decreases, going from $8\%$ to less than
$1\%$. Interestingly the stream fraction peaks in the intermediate
mass halo at $16\%$, with half and two-thirds of that in H1 and H3
respectively. The low fraction in H1 can be explained by the smooth
accretion that efficiently moves the gas straight into the ISM (hence
a high ISM fraction), whereas H3 streams are disrupted to the point of
obliteration when they approach the halo center (hence the low ISM
fraction and large fraction of hot gas).

\subsection{On-the-fly self-shielding}
The transfer of UV photons gives us the opportunity to study the
extent of self-shielding in gas clumps and streams. We quantify the
local UV field intensity in terms of the hydrogen photoionization rate
$\Gamma$, which expresses the average number of photoionization events
per hydrogen atom per unit time (see Appendix
\ref{App_PHrate.sec}). In the UV model we use, the unattenuated
photoionization rate at redshift 3 is $\Gamma=6.1\ 10^{-13} \;
\mathrm{s^{-1}}$ (see \Fig{FG2.fig}), and shielded regions should have
$\Gamma \rightarrow 0 \; \mathrm{s^{-1}}$.

\Fig{SS.fig} shows the UV attenuation in the three targeted halos at
redshift 3.  The top row contains non-logarithmic maps of projected
\textit{minima} of the photoionization rate along the LOS. The light
color on the edges of the maps corresponds to the unattenuated
value. Towards the centers of the halos the UV field becomes
increasingly attenuated due to photo-absorption of the gas and in the
densest streams and clumps we see $\sim 100\%$ attenuation.  The
diffuse streams, with densities $\la 0.02$ $\cci$, are not
self-shielded. Gas at the centers of the H1 and H2 halos is
efficiently shielded but at the center of the H3 halo gas is thermally
ionized and thus optically thin.

The bottom row of \Fig{SS.fig} shows logarithmic phase diagrams of the
hydrogen photoionization rate $\Gamma$ versus density for the same
three halos. The most diffuse gas is UV emitting and has corresponding
horizontal lines in the diagrams up to the $n_H^{UV}$-threshold. Above
this threshold there is an immediate spread in the photoionization
rate in all three halos, ranging from unattenuated UV to about half
attenuated.  Gas in the H1 halo is mostly self-shielded at $n_H \ga 2\
10^{-2}$~$\cci$.  In more massive halos, the advent of thermal
ionization in dense gas makes the situation more complex, and gas at
$n_H \ga 2\ 10^{-2}$~$\cci$ exists in two phases, either self-shielded
as in H1 or optically thin.  The bifurcation in the diagram around
$\Gamma\sim 10^{-17}-10^{-15} \; \mathrm{s^{-1}}$, $n_H \sim
0.1-1$~$\cci$, which grows more conspicuous with increasing halo mass is
an effect of the ionization fronts becoming under-resolved at high
densities, where the mean free path becomes comparable or shorter than
the cell sizes. This feature does not affect our results.

\Fig{xHI.fig} shows maps of \hi{} column densities and phase diagrams
of neutral fraction $\xhi\equiv \nhi / \nh$ vs. density. We find the
CGM and ISM regions correspond mostly to damped \lya{} absorbers
(DLAs, $\snh > 2\ 10^{20} \; \mathrm{cm^{-2}}$) and the streams to
Lyman limit systems (LLSs, $\snh = 1.6 \ 10^{17}- 10^{19} \;
\mathrm{cm^{-2}}$) and even super Lyman limit systems (SLLSs,
$\snh=10^{19}-2\ 10^{20} \; \mathrm{cm^{-2}}$), according to the
definitions found in e.g. \cite{Fumagalli:2011p2943}.  The column
densities are likely over-estimated where they are highest due to lack
of locally enhanced UV from star-formation. We see in the phase
diagrams an abrupt transition of cold gas from ionized to neutral
states, at about $5\ 10^{-2}$~$\cci$ in all halos. This generic result
is consistent with early expectations from \citet{Schaye01} and recent
numerical estimates
\cite[e.g.][]{Kollmeier:2010p3256,FaucherGiguere:2010p5372,AubertTeyssier10}.
The dense ($n_H \sim 0.1$~$\cci$) and ionized ($\xhi \la 10^{-4}$)
cells which become increasingly abundant with halo mass correspond to
hot shock-heated gas, which is thermally ionized and optically thin.

\subsection{A self-shielding approximation}\label{shieldapp.sec}
\begin{figure}\begin{center}
  \hspace{-13mm}
  \vspace{-2mm}
  \includegraphics[width=.405\textwidth]{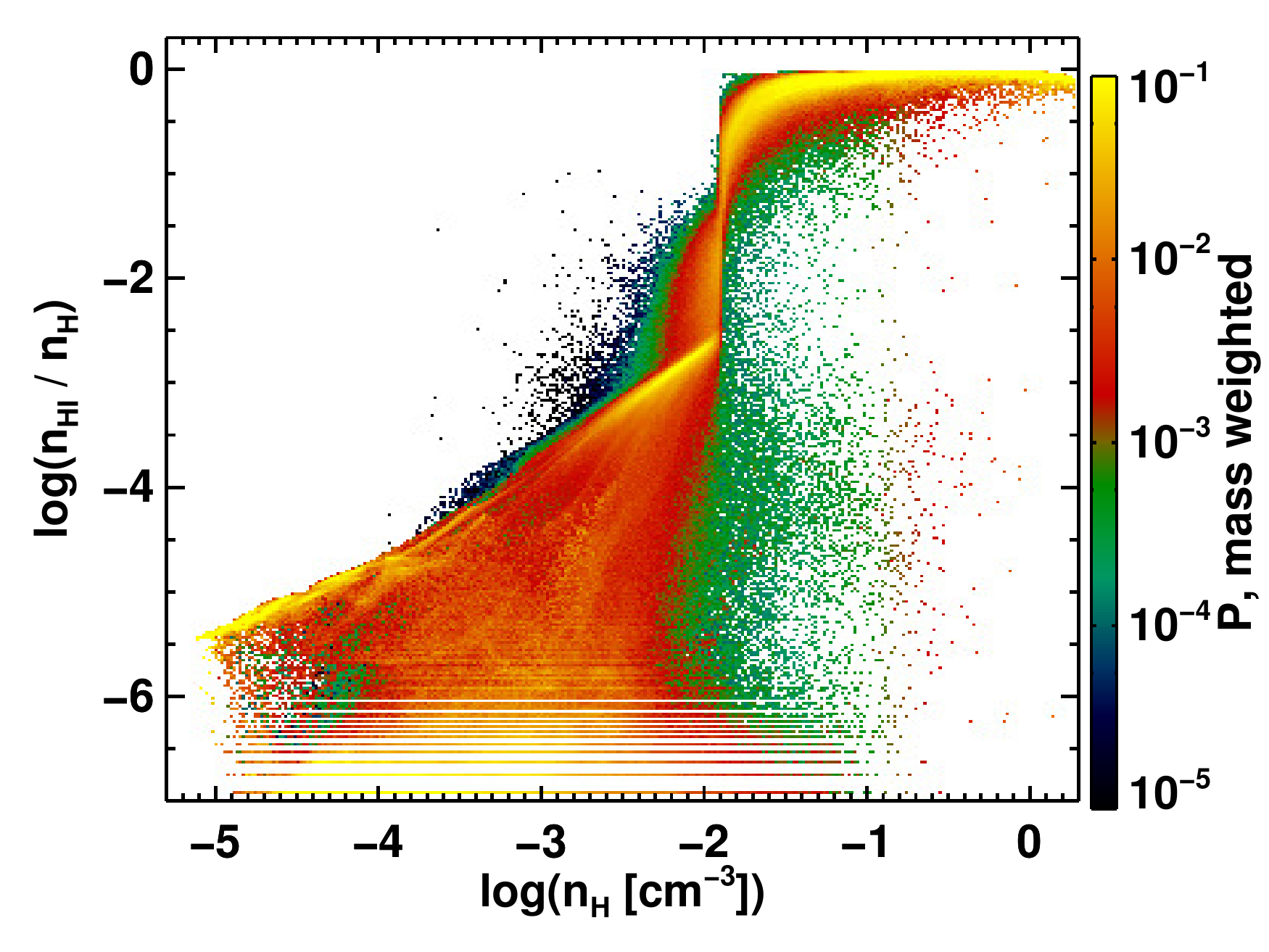}
  \caption[]{\label{H103_xHI.fig}Phase diagram of neutral fraction
    versus density at redshift 3, in a version of the H1 simulation
    where a self-shielding approximation is used (instead of RHD) of
    applying full-strength UV background at densities below
    $10^{-2}$~$\cci$ and zero strength above.}
  \vspace{-2mm}
\end{center}\end{figure}

\FG{} applied a self-shielding approximation in their simulations,
where a UV field is applied homogeneously to gas but with a cutoff at
an assumed self-shielding density threshold of $10^{-2}$~$\cci$. We
have run an analogue to our H1 simulation using the same
self-shielding approximation instead of radiative
transfer. \Fig{H103_xHI.fig} shows the neutral fraction versus density
phase diagram at $z=3$ in this simulation. Apart from a much more
discrete jump from ionized to neutral, the diagram is similar to the
RHD counterpart (\Fig{xHI.fig}, bottom left), and we find $50\%$
neutral fraction at half the density of the RHD counterpart, or at
$0.025$~$\cci$. In terms of getting right the ionization state of gas
at redshift 3 it thus appears that this non-RT self-shielding
approximation holds fairly well. One might perhaps consider moving the
self-shielding threshold a factor of two towards higher density, but
one should be careful not to move it higher than that to avoid
over-predicting \lya{} emissivities due to photo-heating and
photo-fluorescence. The approximation inaccurately describes UV
attenuation in more massive halos, where much of the gas is thermally
ionized and thus UV transparent. This doesn't matter however, since an
absence/presence of the UV background in gas which is already so
ionized has a negligible effect on its \lya{} emissivity (which is
dictated by collisional ionization equilibrium (CIE) heating/cooling.

\clearpage
\section{Predicted \lya{} luminosities}
\subsection{Computing the gas \lya{} emission} \label{emission.sec}

In most astrophysical contexts, an electron in the excited level 2P of
the hydrogen atom will practically instantly relax to the ground state
(1S) via the emission of a \lya{} photon. There are two channels to
produce such excited atoms, and hence to produce \lya{}
radiation\footnote{To be exhaustive, there is a third one, which is
  absorption of photons with energies in the range 10.2 - 13.6 eV,
  which will excite the electron to any level $\geq 2$, which will in
  turn cascade down and sometimes produce a \lya{} photon. This
  process is likely sub-dominant in the regime that we are
  investigating \citep{Furlanetto:2005p3744,Kollmeier:2010p3256}, and
  requires \lya{} radiative transfer, which we postpone to a future
  paper.}:

\vsk \textbf{Collisional:} A collision with a free electron excites
the H-atom, which may release a \lya{} photon when it relaxes back to
the ground state. The collisional emissivity is approximated with
\begin{equation}\label{LyEmColl.eq}
\eps_{coll}=C_{\lyam}(T) \; \ne \; \nhi \; \epsilon_{\lyam},
\end{equation}
where $\ne$ and $\nhi$ are number densities of electrons and neutral
hydrogen, respectively, and $C_{\lyam}(T)$ is the rate of
collisionally induced 1S-to-2P level transitions. An expression for
this rate is given by G10, fitting results from
\cite{Callaway:1987p2948}. It is always less than the hydrogen
collisional excitation cooling rate, $\Lambda_{coll}^{HI}$, used in
the code \citep[from][]{Maselli:2003p4122}, since cooling also takes
into account excitations to atomic states other than 2P (the most
likely of which is the non-\lya{} releasing 2S state). The ratio of
$C_{\lyam} / \Lambda_{coll}^{HI}$ goes from $71\%$ at $10^4$ K to
$57\%$ at $5\ 10^4$ K.

\vsk \textbf{Recombinative:} A free electron combines with a proton at
any level ($\geq 2$), and may cascade down to the 2P level.  The
recombinative \lya{} emissivity of this process is given by
\begin{equation}\label{LyEmRec.eq}
\eps_{rec}=0.68 \; \recB(T) \; \ne \; \nhii \;
\epsilon_{\lyam},
\end{equation}
where the $0.68$-factor is the average number of \lya{} photons
produced per case B recombination \citep[from][]{2006agna.book.....O}
and $\recB(T)$ is the case B recombinations rate, i.e. counting
recombinations to all levels except directly to the ground one.  We
use the expression from \cite{Hui:1997p2465}.

\vsk Unless otherwise specified, the \lya{} emissivities calculated in
this paper are:
\begin{equation}\label{LyEm.eq}
  \eps=\eps_{coll} + \eps_{rec}.
\end{equation}

\begin{figure}\begin{center}
  \includegraphics[width=0.5\textwidth]{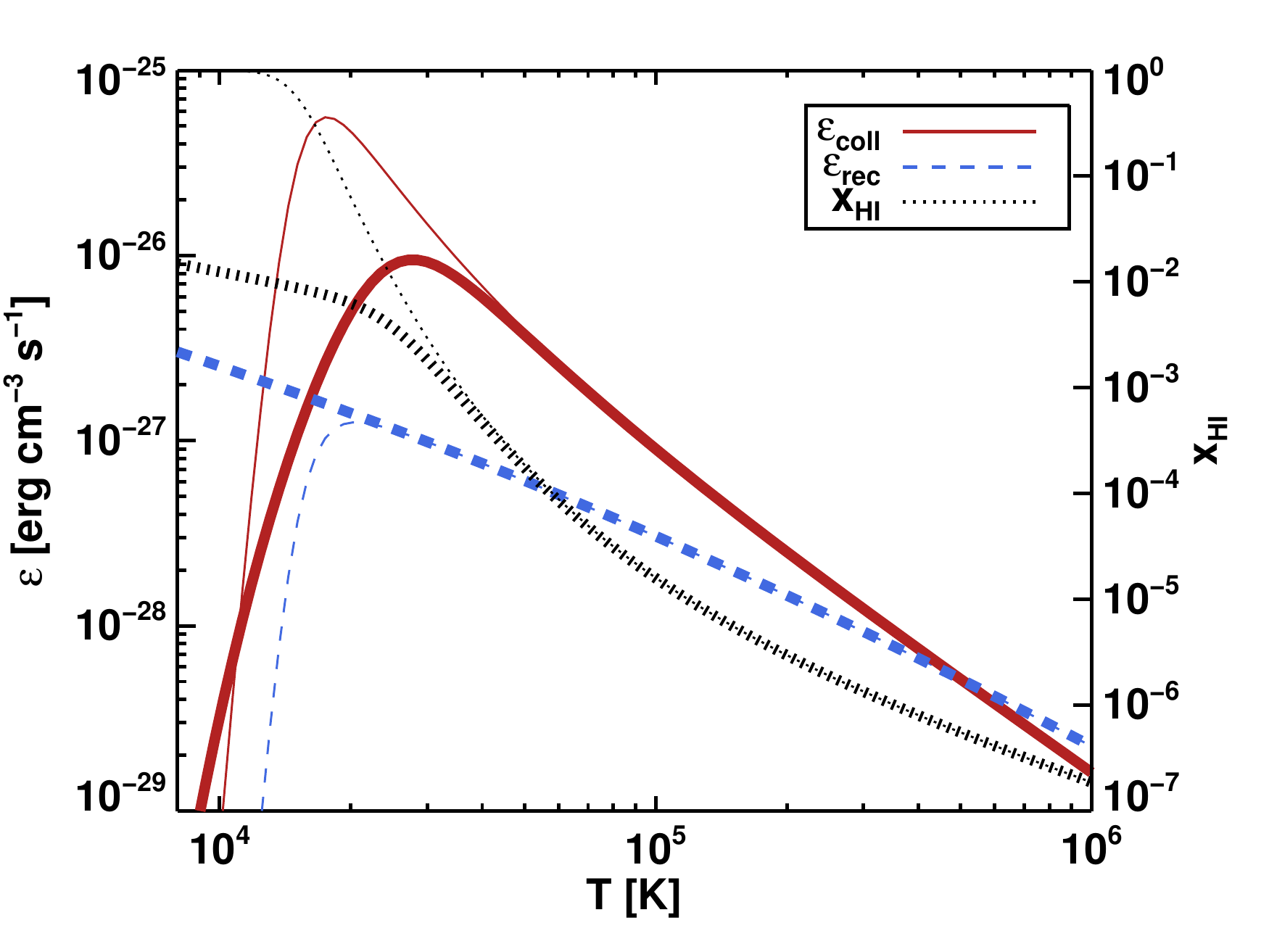}
  \caption[]{\label{LyaEmi.fig}\lya{} emissivity of gas at number
    density $\nh=3\ 10^{-2} \; \cci$, which is close to the lower
    limit for accretion streams in our simulations. Thick curves show
    gas exposed to a UV field with $\Gamma = 6.1\ 10^{-13} \; \si$,
    corresponding to redshift 3. Thin curves show UV shielded gas
    ($\Gamma=0 \; s^{-1}$). Blue dashed curves show recombinative
    \lya{} emissivity, red solid curves show collisional \lya{}
    emissivity and black dotted curves show the approximate neutral
    hydrogen fraction in the gas, assuming equilibrium between
    photoionization, collisional ionization and recombinations.}
\end{center}\end{figure}

Figure \ref{LyaEmi.fig} shows the collisional and recombinative \lya{}
emissivities of gas at typical stream density, assuming the gas is UV
exposed (thick curves) and self-shielded (thin). Also shown in dotted
black curves is the neutral fraction of the gas, approximately
extracted using a simplified model of hydrogen-only and PIE/CIE
equilibrium.

The plot illustrates that it is crucial to be \textit{consistent} in
following the gas state of $(T, \xhi, \Gamma)$ in the simulation code,
since independently changing one of those factors without considering
the effect on the others can have a dramatic effect on $\eps$. If, for
example, self-shielding is assumed in post-processing and the neutral
fraction changed accordingly without considering the change in
temperature, the $\eps_{coll}$ estimate can increase by almost an
order of magnitude. This is unphysical -- what really happens if gas
suddenly becomes UV shielded is that the temperature drops somewhat
due to lack of photo-heating, the end-result being a slightly lowered
value of $\eps$.

The accuracy of the $(T, \xhi, \Gamma)$-state is secondary to
consistency, because if the code handles things properly, $\eps$
should simply reflect the work put into the gas by the UV background
and gas advection. In the limit that the UV energy input is negligible
compared to gravitational heating, accurate modelling of the UV
background isn't really crucial in the context of \lya{} emissivity,
and applying e.g. a sensible shielding approximation like the one
discussed in \Sec{shieldapp.sec} should be OK. This breaks down when
UV photo-fluorescence becomes non-negligible.

\subsection{Intrinsic luminosities}
\begin{figure*}\begin{center}
  \subfloat{\includegraphics[width=0.33\textwidth]
    {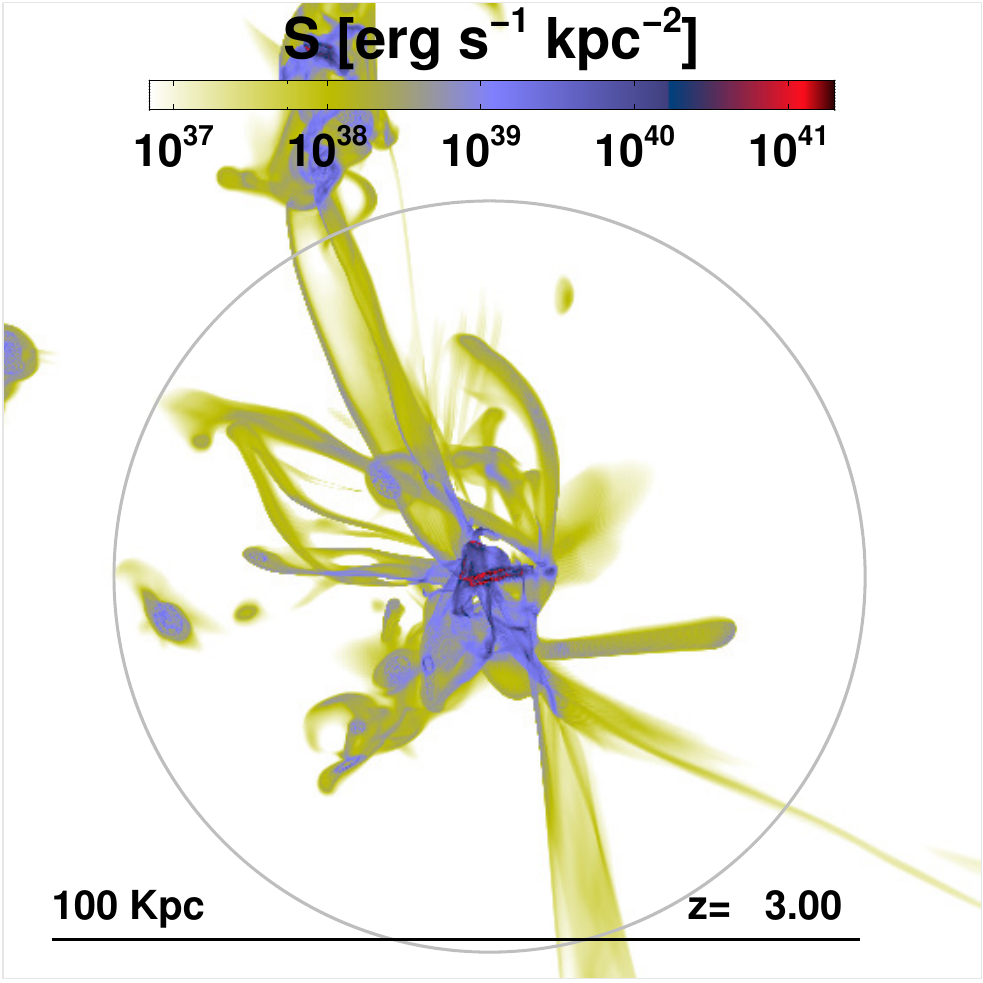}}\hspace{-1mm}
  \subfloat{\includegraphics[width=0.33\textwidth]
    {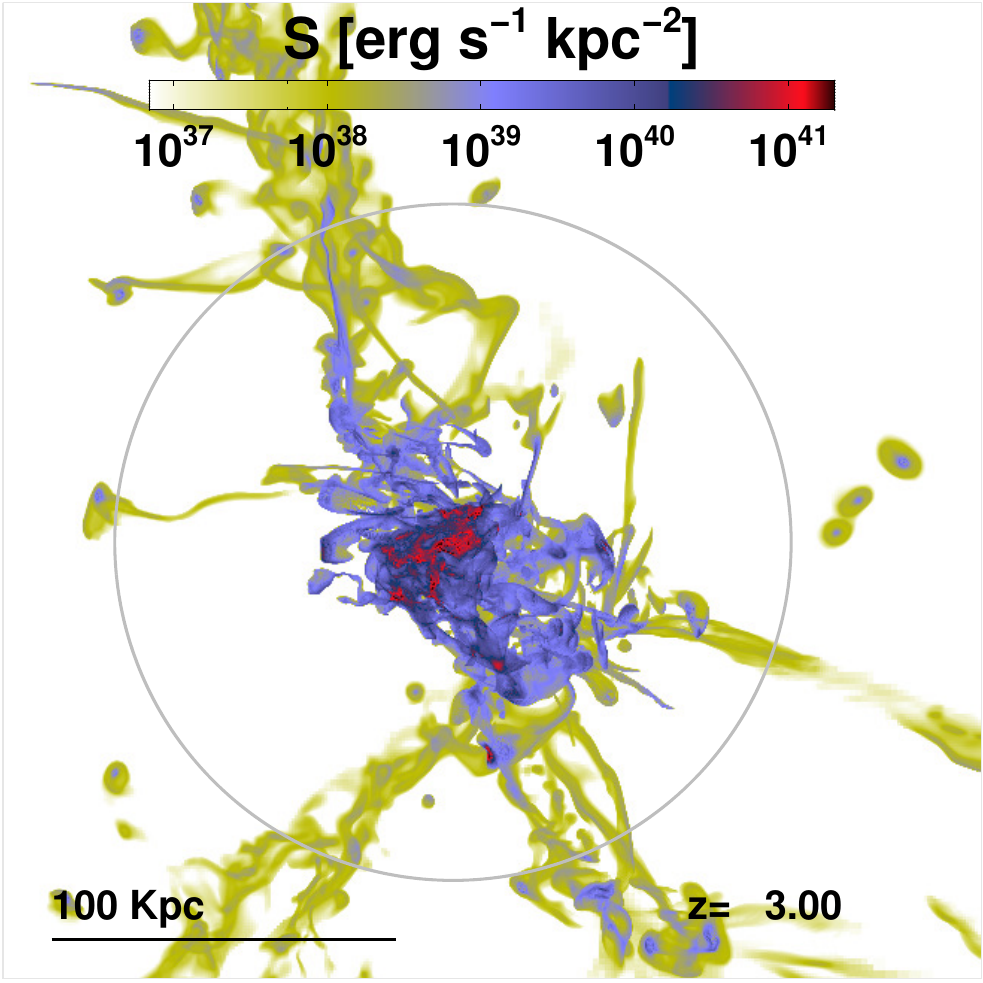}}\hspace{-1mm}
  \subfloat{\includegraphics[width=0.33\textwidth]
    {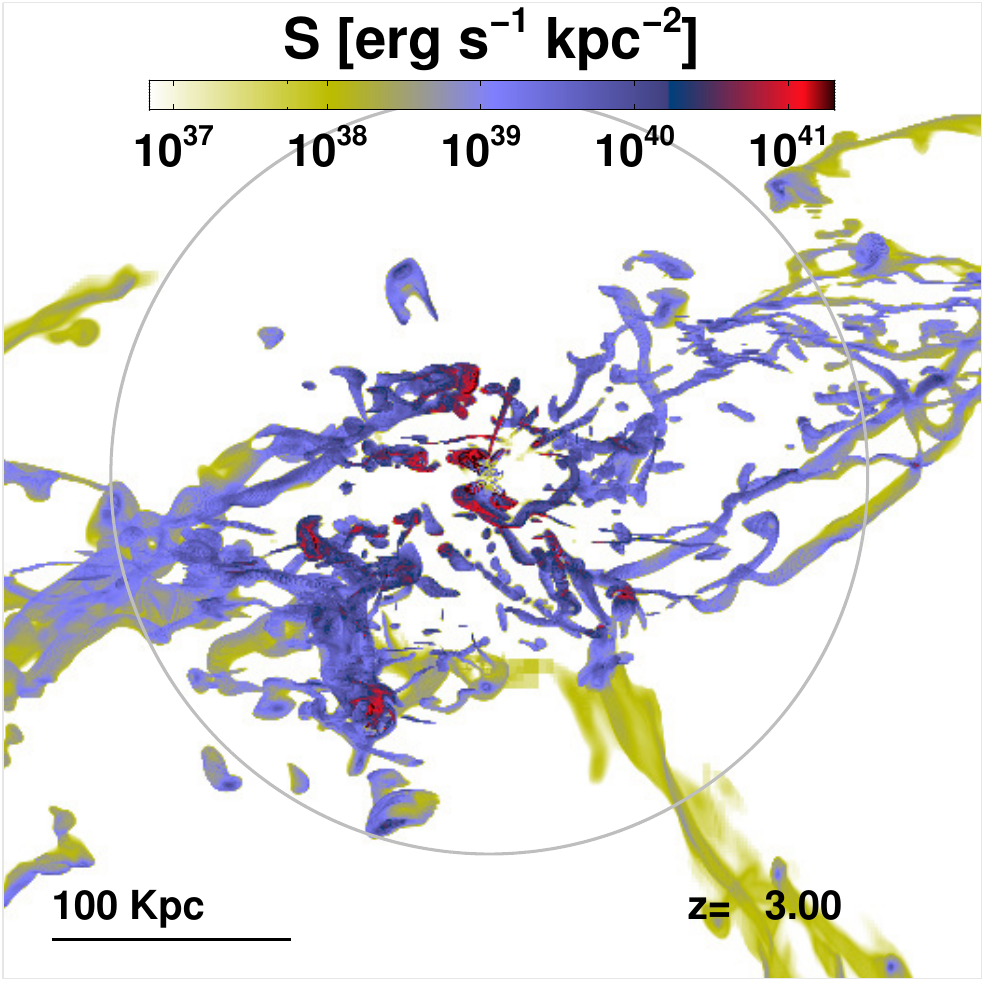}}\hspace{-1mm}
  \vspace{-6.mm}

  \subfloat{\includegraphics[width=0.379\textwidth]
    {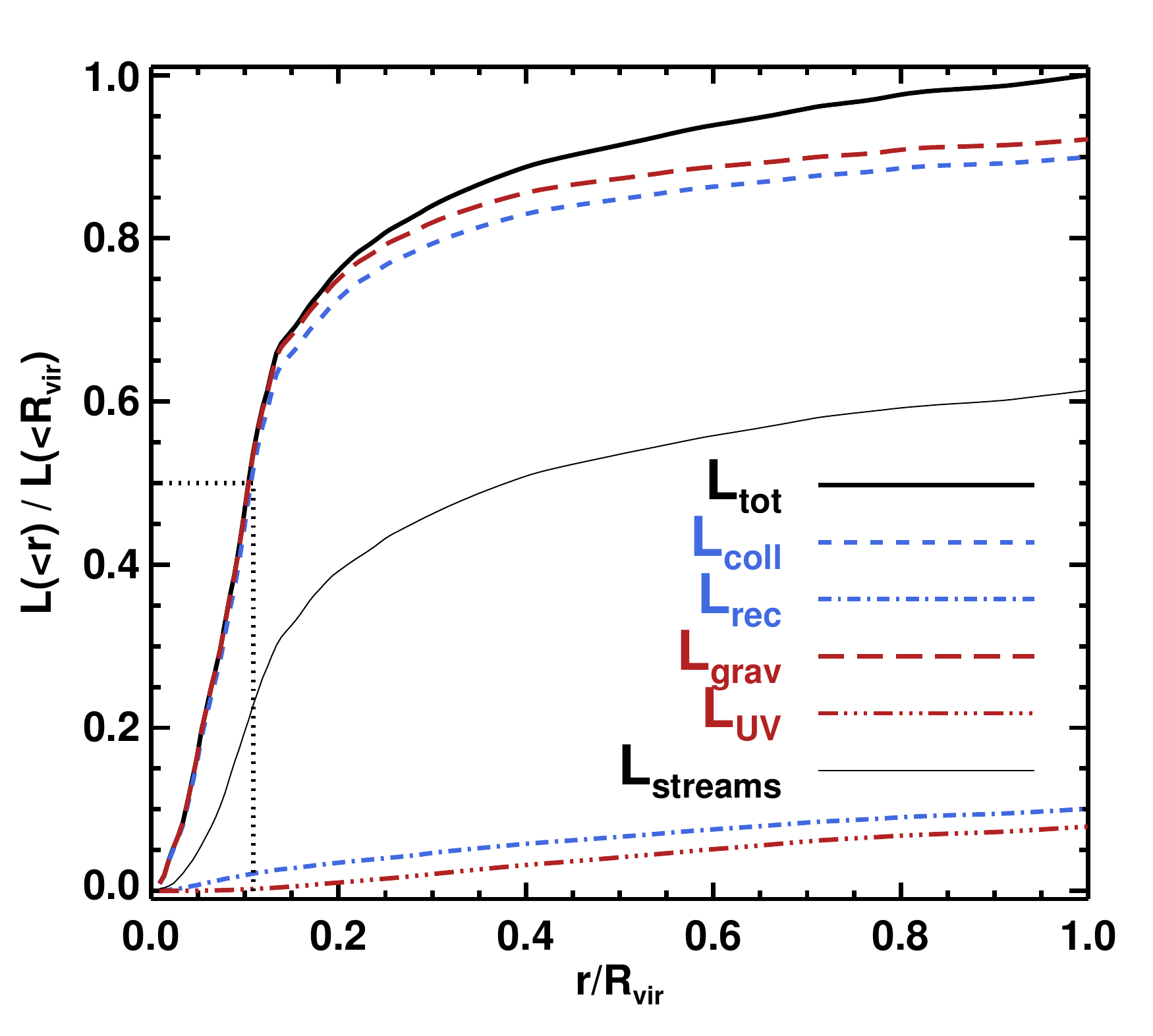}}\hspace{-13.8mm}
  \subfloat{\includegraphics[width=0.379\textwidth]
    {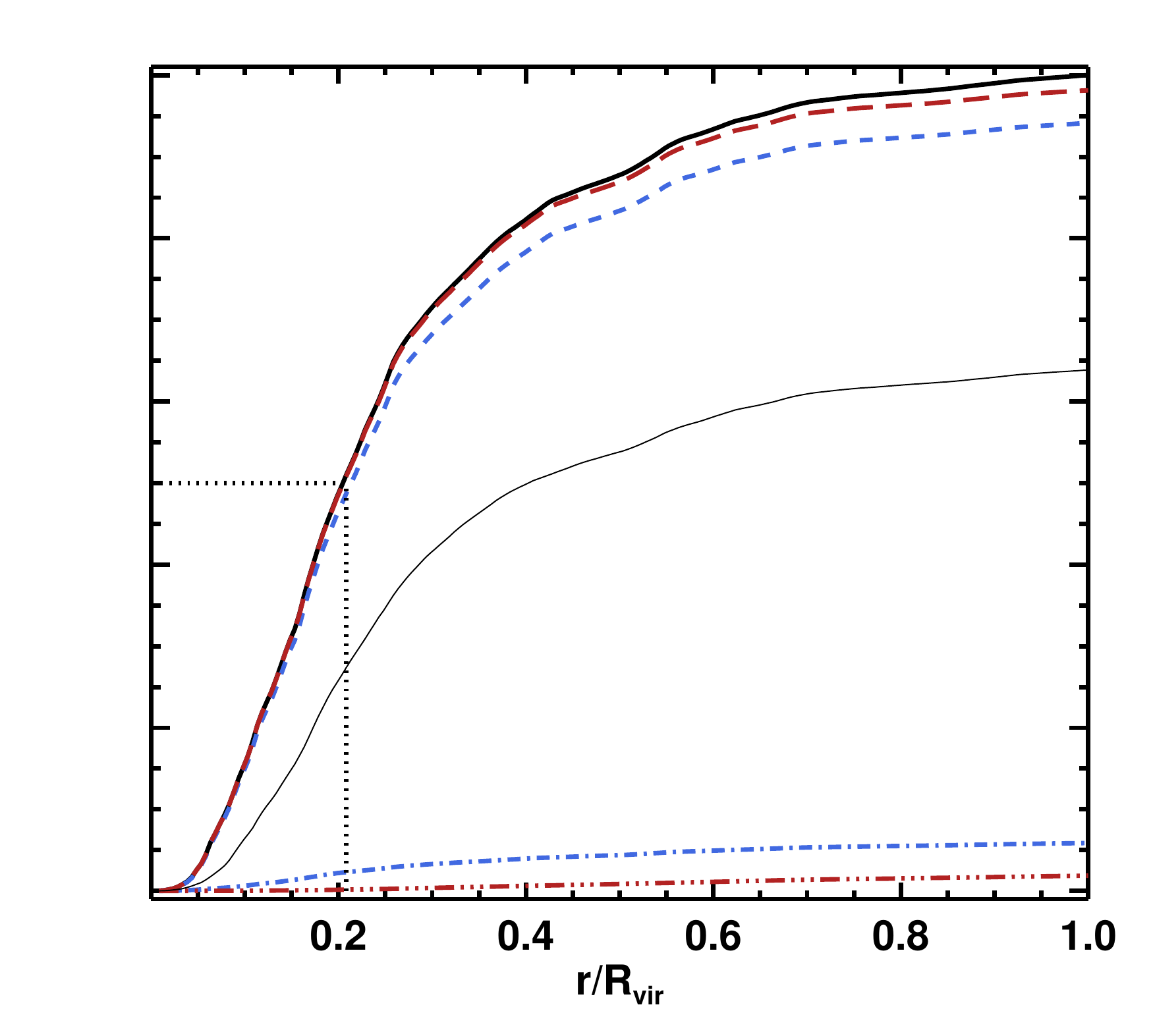}}\hspace{-13.8mm}
  \subfloat{\includegraphics[width=0.379\textwidth]
    {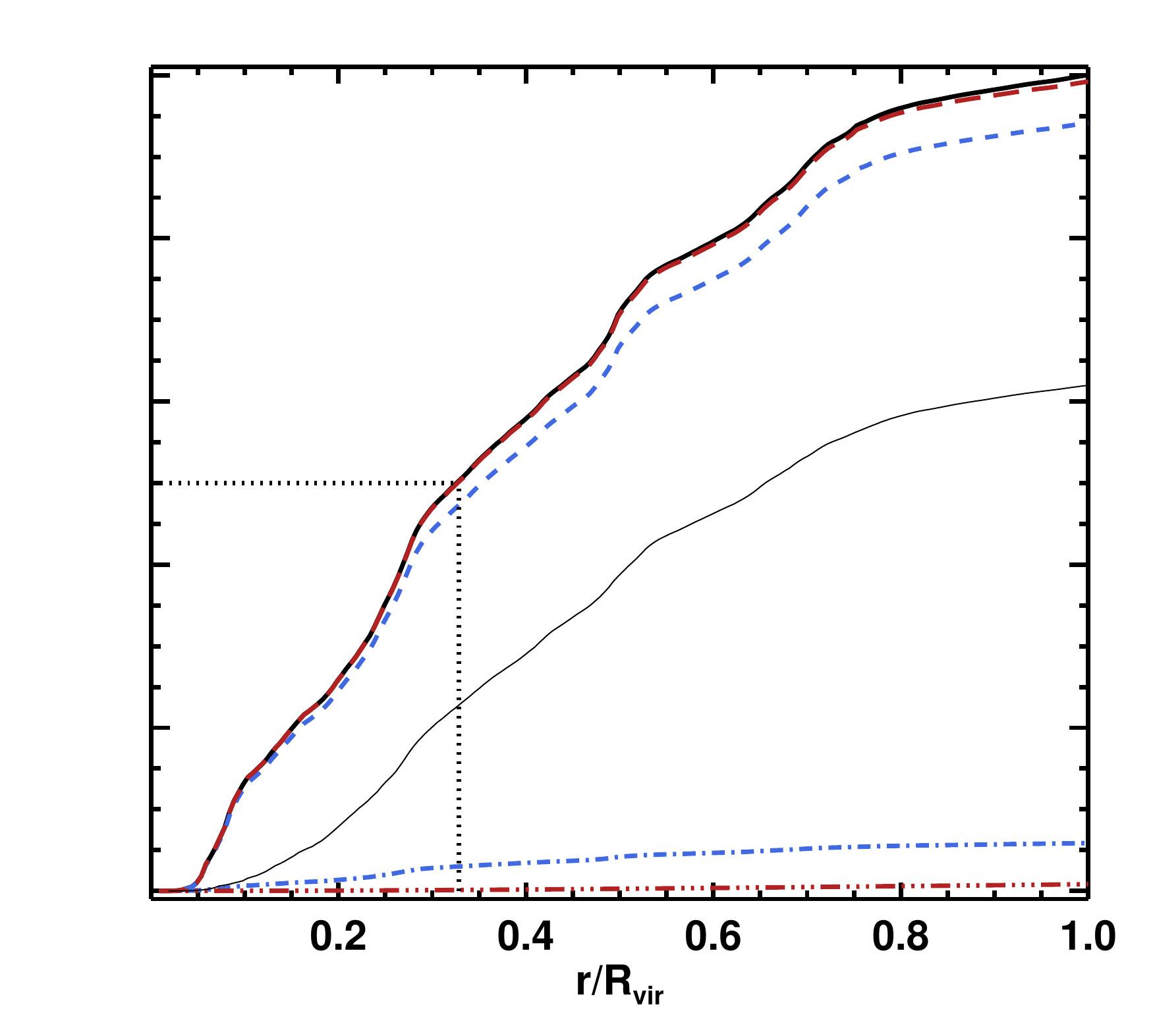}}\hspace{-4mm}
  \vspace{-3.mm}
  \caption[]{\label{SLE.fig} \textbf{Top row:} Rest-frame \lya{}
    surface brightness maps of the targeted halos (H1, H2 and H3
    simulations, from left to right). \textbf{Bottom row:} Radially
    (3D) cumulative \lya{} luminosities of the same halos. The thick
    black curves show the total luminosity. The dotted lines denote at
    what radius $50\%$ of the total luminosity is accounted for. The
    blue curves show how the total luminosity is split between
    collisional and recombinative channels, given in
    Eqs. \ref{LyEmColl.eq} and \ref{LyEmRec.eq} respectively. The red
    curves show how the total is split between the contributions of
    the UV background and gravitational dissipation, discussed in
    \Sec{contr.sec}.  The thin black curves show the contribution of
    sub-CGM density gas (i.e. mostly streams) to the total
    luminosities.}
\end{center}\end{figure*}
\Fig{SLE.fig}, top row, shows maps of the rest-frame \lya{} surface
brightness $S$ of the three targeted halos, which is calculated by
integrating the \lya{} emissivity (Eq. \ref{LyEm.eq}) along the
LOS. We don't take absorption or scattering into account: These
factors would certainly diminish the brightest spots associated with
CGM regions, but we don't expect them to affect the more diffuse
streams much (see discussion in \Sec{obs.sec}). The surface brightness
is concentrated around CGM regions in all three halos, with $S\approx
10^{40}-10^{41} \; \LS$, and the brightness in streams is typically
lower by one or two orders of magnitude.

The bottom row in \Fig{SLE.fig} shows radially cumulative \lya{}
luminosities for the halos, i.e. fraction of the total luminosity
within a given radius (black solid curves). Streams and the diffuse
medium consistently contribute about 60\% of the total luminosity, as
indicated by the thin black curves.

It is evident from both the maps and plots that there is a trend of
more extended emission with increasing halo mass. In the H1 halo, half
of the total luminosity comes from the central $16\%$ of the virial
radius (dotted lines), while in H2 this radius is $20\%$ and $33\%$ in
H3. Partly this is because the more massive halos consist of
increasing quantities of orbiting galaxies so the surface brightness
is just following the increased spread of CGM regions, as can be seen
from red dots of surface emissivity in the maps and from corresponding
steps in cumulative surface brightness in the plots. That is not the
whole story though: The streams become more efficient \lya{} emitters
with increasing halo mass.

As seen from the blue curves in the luminosity plots,
electron-hydrogen collisions dominate the total luminosity, and
recombinations are borderline negligible, as should be expected
outside ISM regions. This dominance increases with halo mass, with
recombinations contributing $10\%$ to the total in the H1 halo and
only about $5\%$ in H2 and H3. The red curves in Fig. \ref{SLE.fig}
will be discussed in Sec. \ref{Sec:What}.

\begin{figure}\begin{center}
  \includegraphics[width=0.5\textwidth]{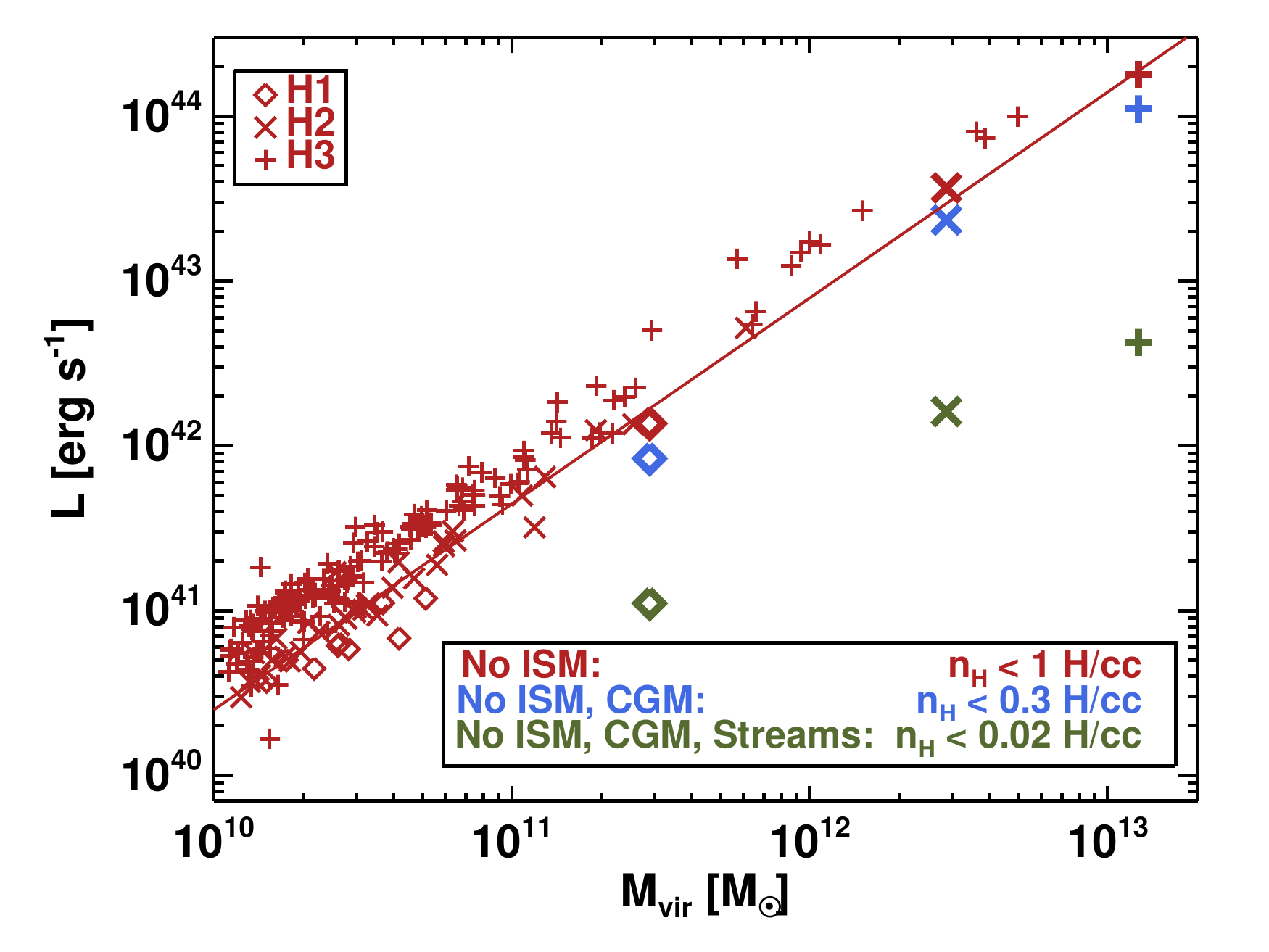}
  \caption[]{\label{HLL.fig}Total halo luminosities versus halo mass
    for the three simulations. The halos are extracted from the
    zoom-in volumes, and we exclude sub-halos. The three targeted
    halos are indicated by large thick symbols. For those halos we
    also show luminosities excluding different phases of the gas. The
    red solid line indicates a power-law with exponent $1.25$.}
\end{center}\end{figure}

\vsk In \Fig{HLL.fig} we plot total halo luminosities versus halo mass
(which is defined, as in \Tab{tbl:sims}, as the total mass of all dark
matter and baryons within the virial radius). From each of the three
simulations we extract all halos from within the zoom-in volume and
integrate \Eq{LyEm.eq} over their virial radii, excluding ISM gas. The
halos roughly line up into a power law indicated by a red solid line,
with exponent $1.25$. There is a systematic tendency for halos in more
massive simulations to be more luminous for a given mass, which is
presumably an environment effect since the cosmological over-density
of the zoom-in regions increases between the H1, H2 and H3 simulations
respectively.

Large thick symbols mark the three main halos targeted in our
simulations. For those we also plot luminosities excluding consecutive
phases of gas. Blue symbols show luminosities when ignoring the ISM
and the CGM, and green symbols show what happens if we also ignore the
stream densities. The CGM accounts for about 40\% of the total
luminosity in all three targeted halos (see also \Fig{SLE.fig}) and
the streams account for most of the rest, or $50-60\%$, with
sub-stream densities accounting for $8\%$, $4\%$ and $2\%$ in the
targeted halos of H1, H2 and H3 respectively.
 
\begin{figure}\begin{center}
  \includegraphics[width=.5\textwidth]{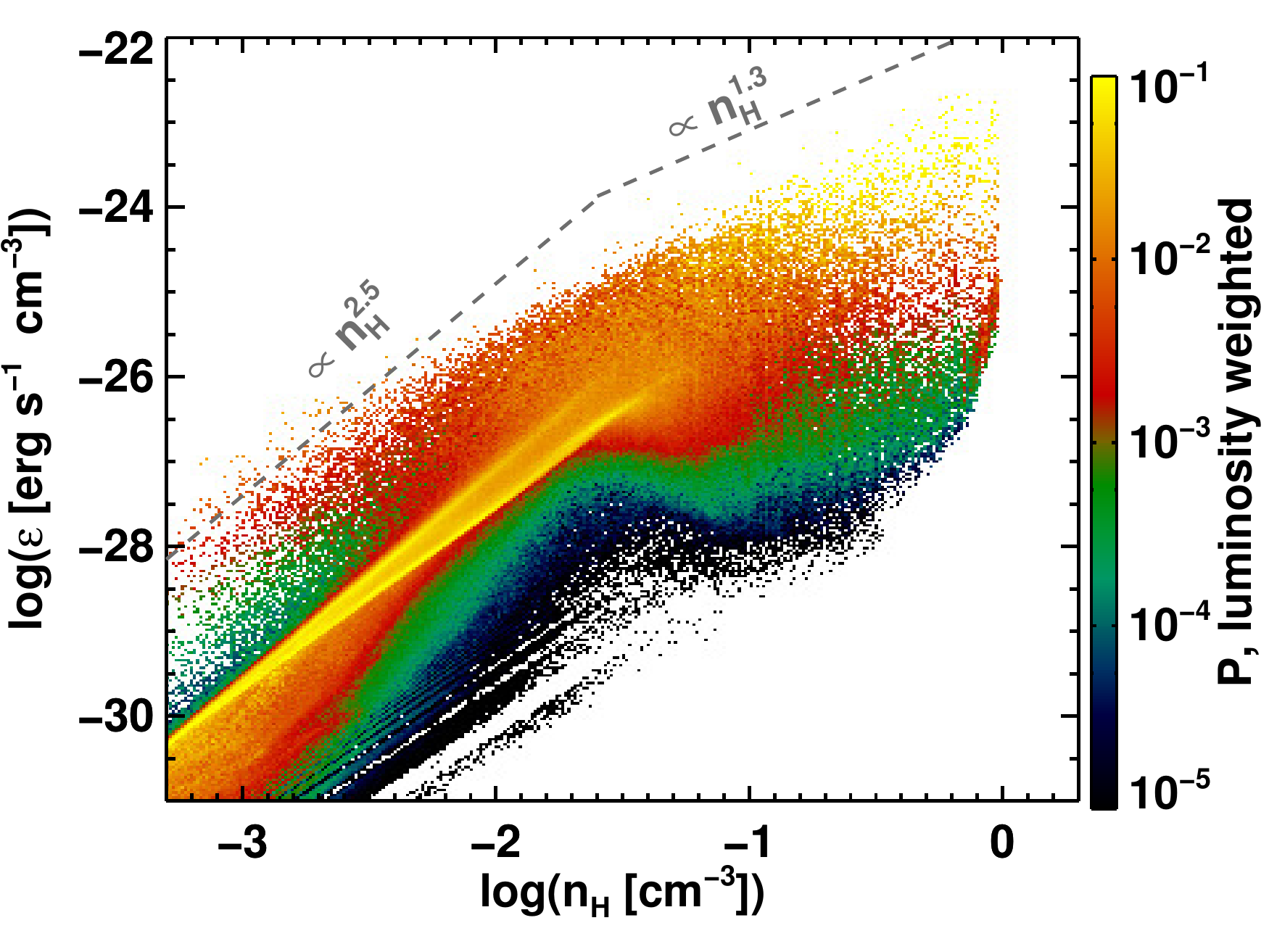}
  \caption[]{\label{PH_LE.fig}Phase diagram showing \lya{} emissivity
    of gas versus density at z=3 in the H1 halo. Each density bin is
    \lya{} luminosity weighted independently. The grey dashed lines
    show power laws that approximately fit the data. The color scale
    represents \lya{} luminosity weighted probability per $\eps-\nh$
    bin over the plotted $\eps$ range.}
\end{center}\end{figure}

As can be read directly from Eqs. (\ref{LyEmColl.eq}) and
(\ref{LyEmRec.eq}), the \lya{} emissivity of gas in principle scales
with density squared, though temperature and ionization state have
their influence as well. \Fig{PH_LE.fig} shows a luminosity weighted
phase diagram of \lya{} emissivity of gas versus density in the H1
halo (the H2 and H3 diagrams are similar). Over-plotted on the diagram
in dashed grey lines are two power laws that the gas emissivity
approximately follows, with a knee between $10^{-2}$ and
$10^{-1}$~$\cci$. The knee roughly corresponds to where the gas
becomes self-shielding and the change in slope is caused by the
corresponding transition in temperature and ionization state. Below
the knee the gas emissivity is split in two ridges with slightly
different slopes. The upper one has power index $\approx 2.5$ and is
dominated by collisional emission (Eq. \ref{LyEmColl.eq}), whereas the
lower one has power index $\approx 2.2$ and is dominated by
recombinations (Eq. \ref{LyEmRec.eq}). The emissivity above the knee
is completely dominated by collisions. The stronger than $2$ power law
below the knee stems from the increasing abundance of neutral atoms
with density and a temperature that tends towards peak \lya{}
emissivity, whereas the less than $2$ power law above it results from
the decreasing relative abundance of electrons with density.

\subsection{Observational properties}\label{obs.sec}
\begin{figure*}\begin{center}
  \subfloat{\includegraphics[width=0.33\textwidth]
    {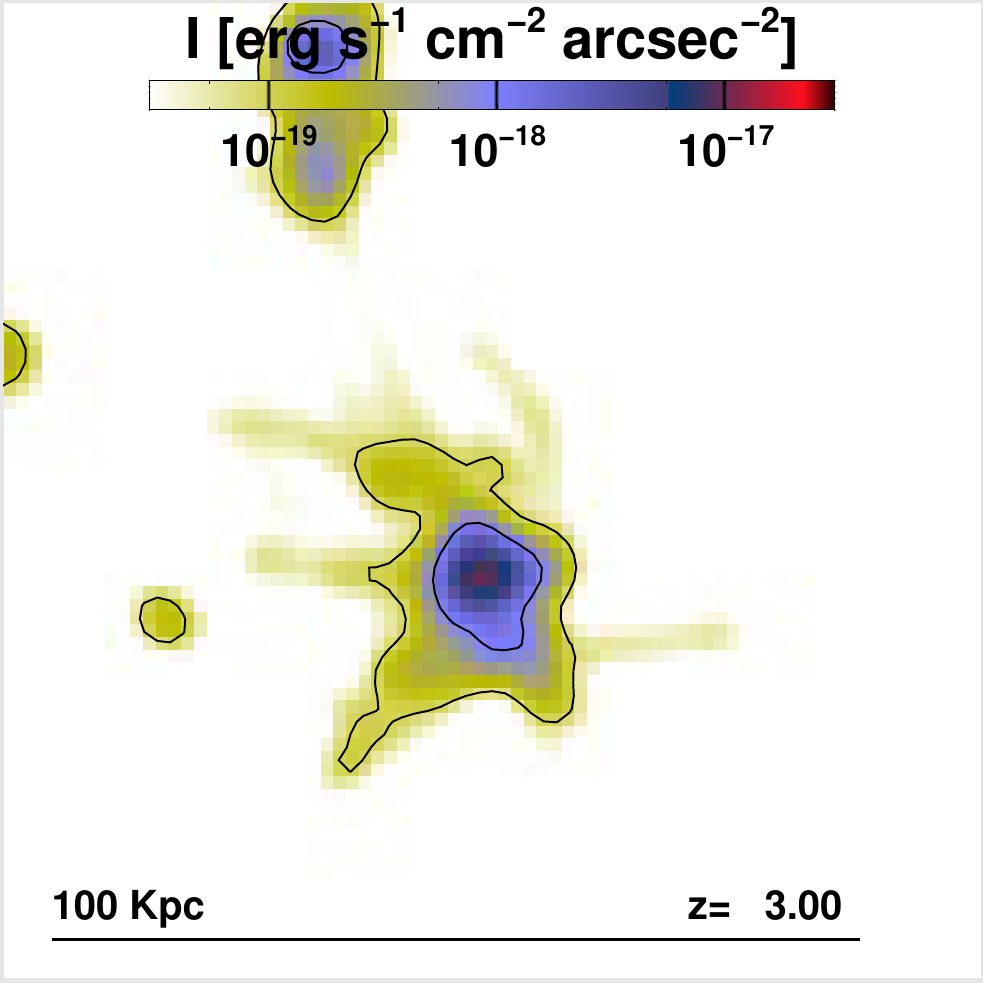}}\hspace{-1mm}
  \subfloat{\includegraphics[width=0.33\textwidth]
    {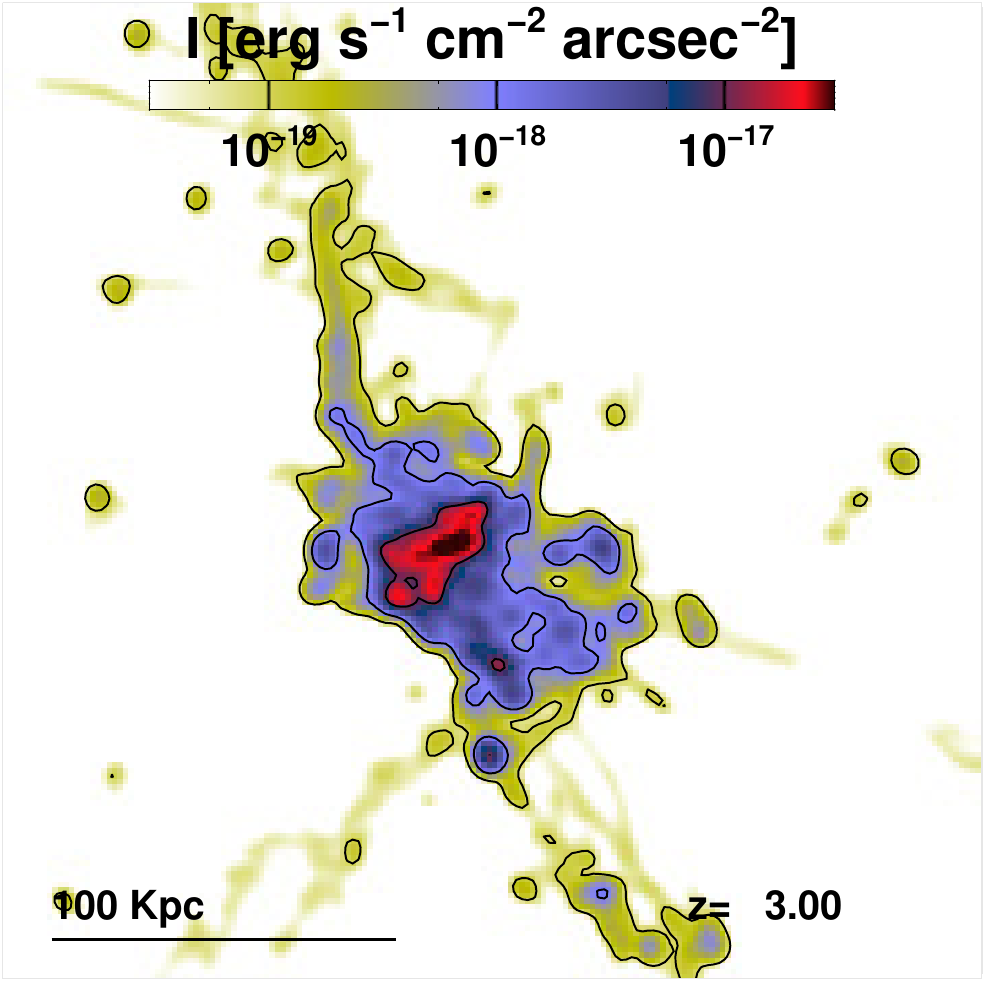}}\hspace{-1mm}
  \subfloat{\includegraphics[width=0.33\textwidth]
    {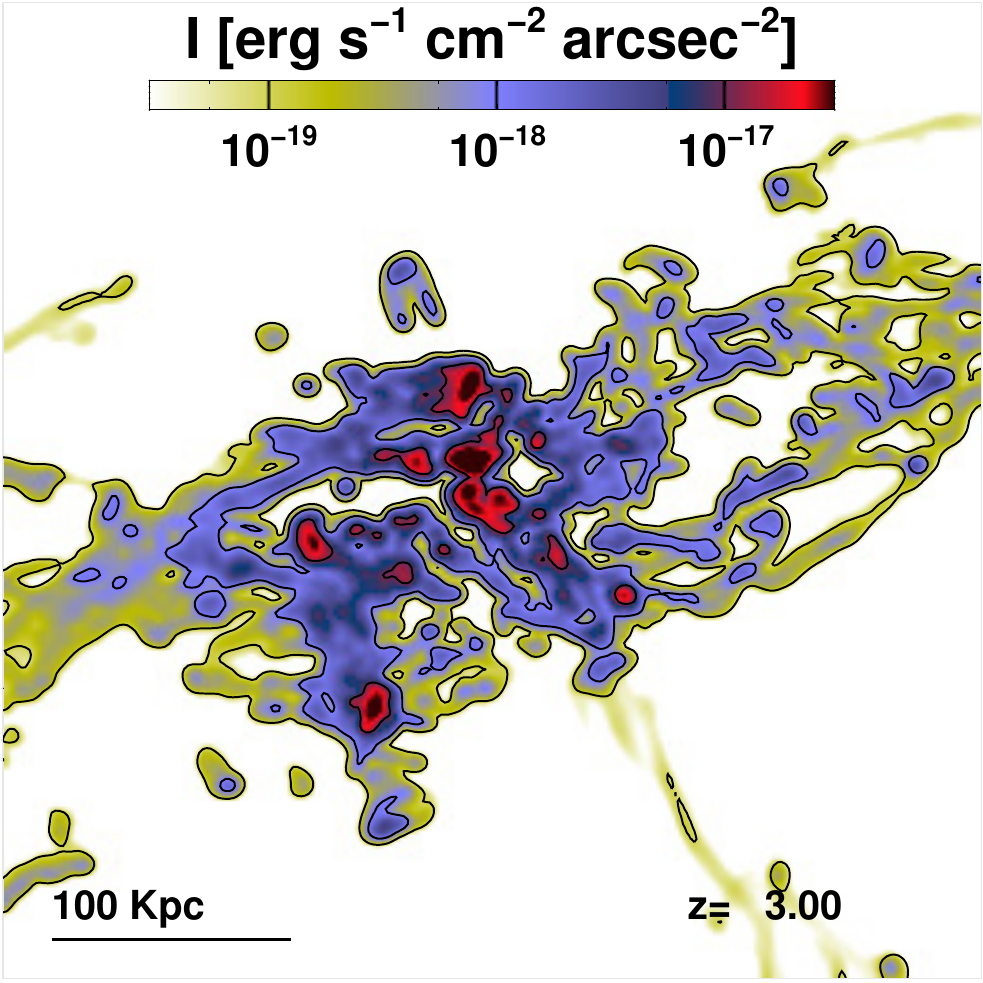}}\hspace{-1mm}
  \vspace{-3.9mm}

  \subfloat{\includegraphics[width=0.33\textwidth]
    {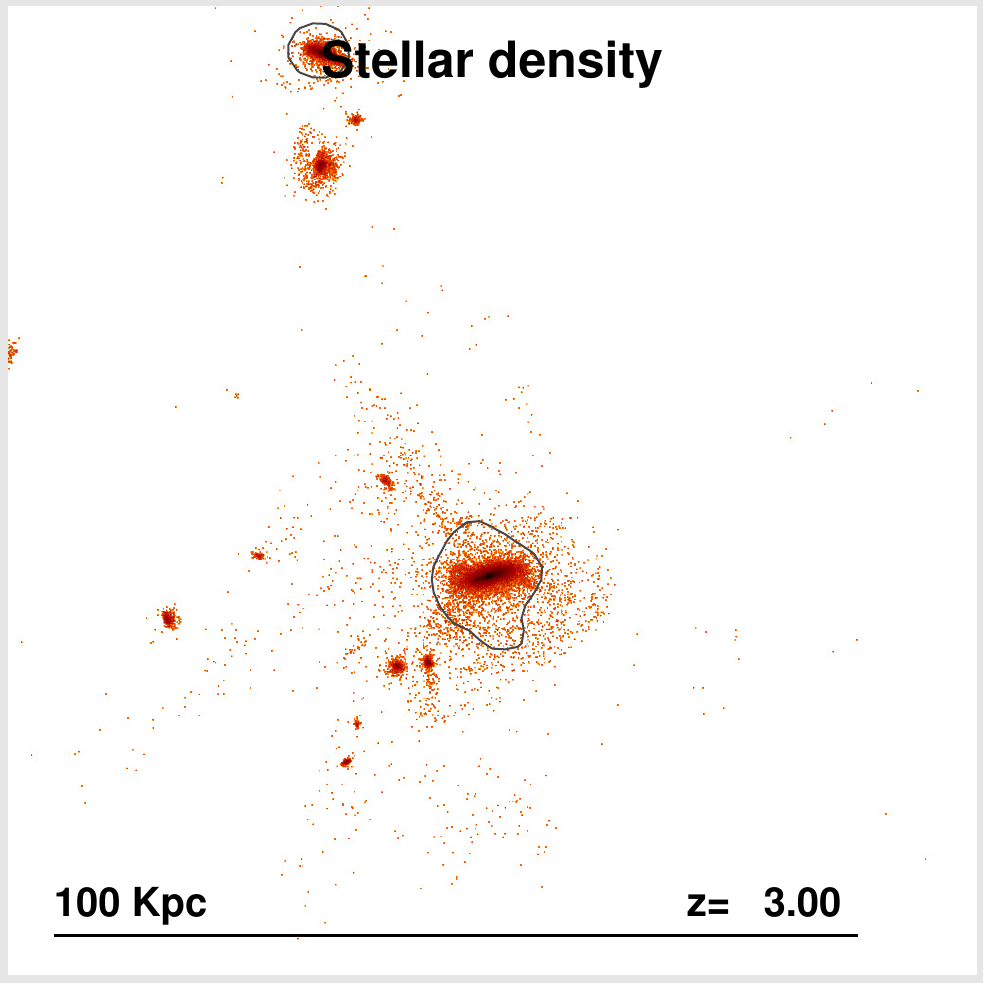}}\hspace{-1mm}
  \subfloat{\includegraphics[width=0.33\textwidth]
    {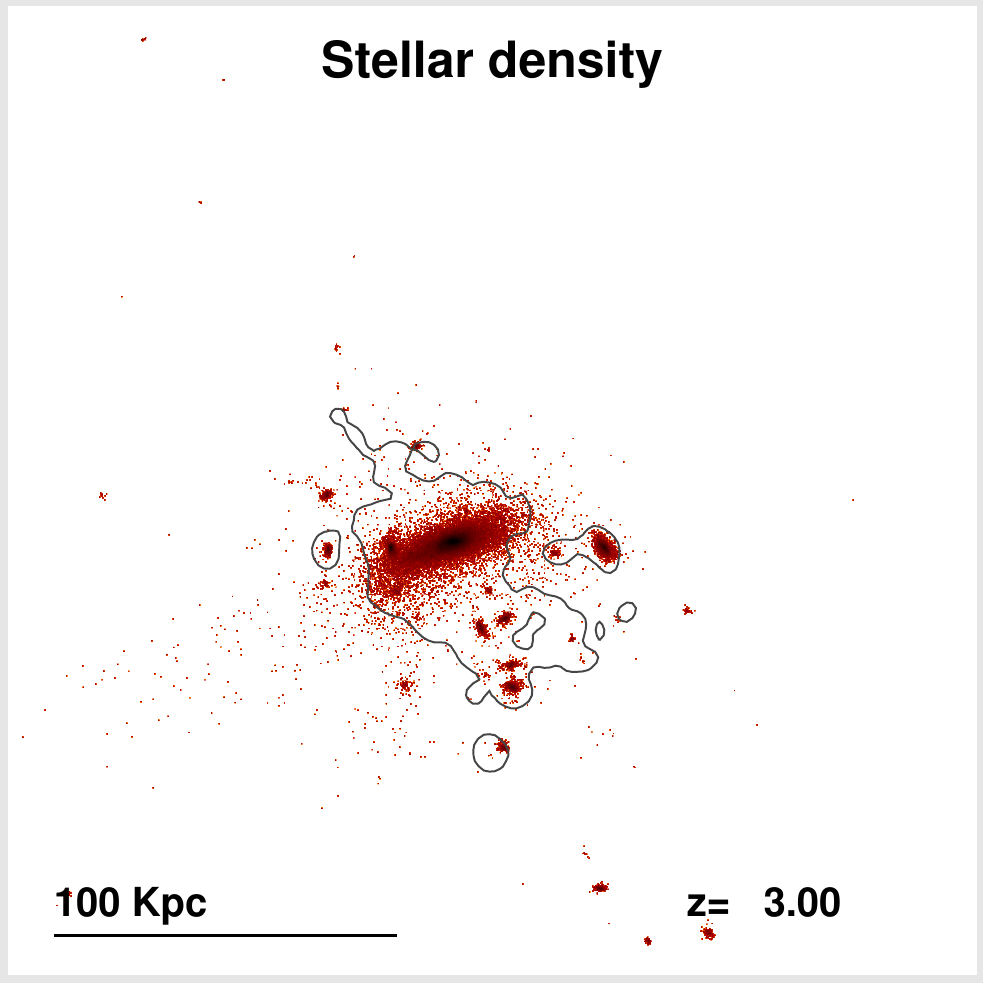}}\hspace{-1mm}
  \subfloat{\includegraphics[width=0.33\textwidth]
    {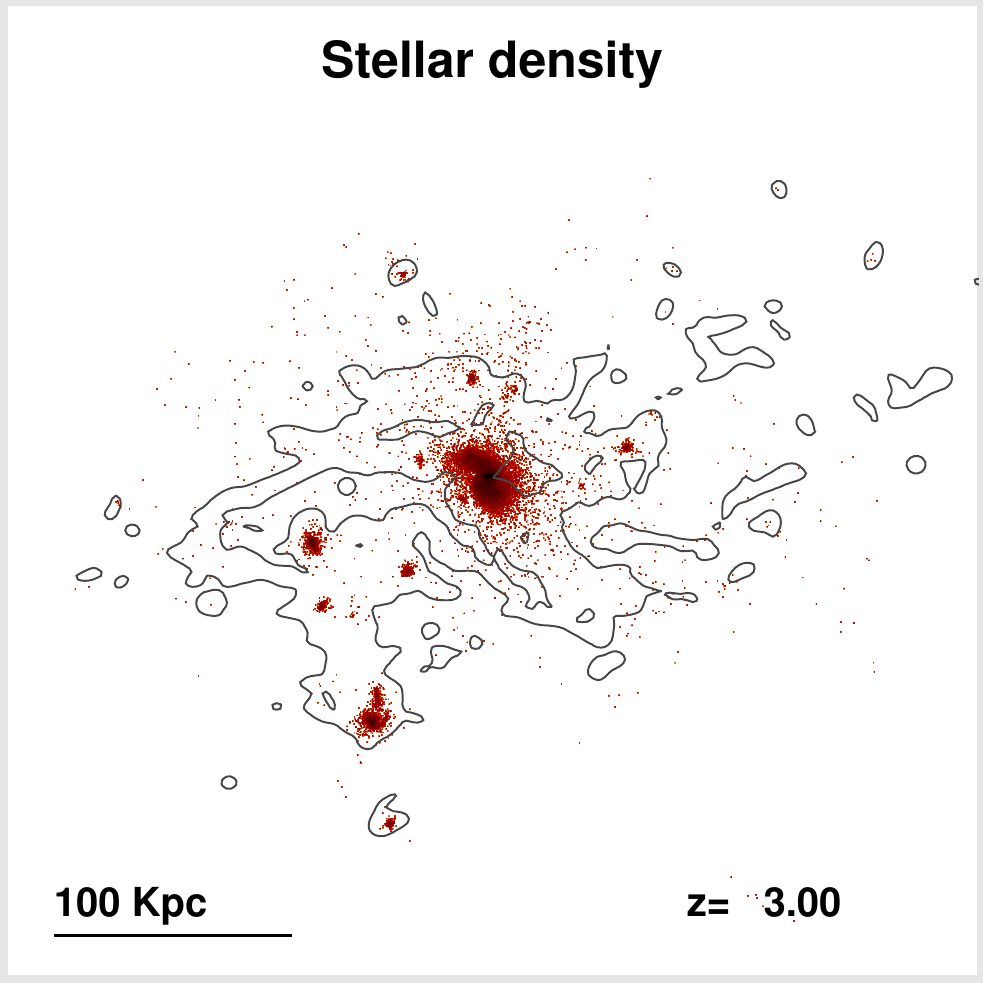}}\hspace{-1mm}
  \vspace{-0.3mm}

  \caption[]{\label{LyObs.fig}\textbf{Top row:} Mock images showing
    predicted observed surface brightness in the targeted halos of H1,
    H2 and H3, from left to right. The contours mark $10^{-17}$,
    $10^{-18}$ and $10^{-19}$ $\mathrm{erg \, s^{-1} \, cm^{-2} \,
      arcsec^{-2}}$. The images where computed using an optimistic PSF
    of FWHM 0.6 arcsec. \textbf{Bottom row:} Stellar density maps for
    the same halos, illustrating that the bright spots of \lya{}
    emission are centered on galaxies. Over-plotted are contours
    marking $10^{-18}\,\mathrm{erg \, s^{-1} \, cm^{-2} \,
      arcsec^{-2}}$ in observed surface brightness (the middle
    contours from the upper maps).}
\end{center}\end{figure*}

We now consider mock observations of our simulated halos. To produce
those we first convert the rest-frame surface brightness maps in
\Fig{SLE.fig} to \textit{observed} surface brightness $I$, using
\begin{equation}\label{mock.eq}
  I=\frac{S \fa}{4\pi(1+z)^4},
\end{equation}
where $\fa$ is a cosmological transmission factor that accounts for
absorption and scattering of \lya{} photons on the LOS from the object
to the observer. We adopt in this paper a value of $\fa=0.66$ based on
the work of \cite{FaucherGiguere:2008p3910}. ($\fa$ is only applied to
mock observations and not to the intrinsic emissivity and luminosity,
Figures \ref{SLE.fig} and \ref{HLL.fig}). To the result of
\Eq{mock.eq}, we then apply a Gaussian point spread function (PSF)
with a $0.6$ arcsec full width at half maximum (FWHM) to mimic
atmospheric and instrumental distortion, and assume a camera pixel
size of $0.2$ arcsec (Fig. \ref{LyObs.fig}). This corresponds to very
good seeing conditions in state-of-the-art instruments. We present
maps made with a PSF about twice as broad in Appendix
\ref{Sec:moremaps}. These are directly comparable to the observations
of M11.

Unlike \FG{}, we don't model the scattering of \lya{} photons in this
work. These authors show that \lya{} transfer dominates the spectral
shape of extended \lya{} emission, but hint that it has little effect
on the morphology and extent.  Their Fig. 8 shows this to be the case
for a halo corresponding in mass to our H1 halo, if only the \lya{}
emissivity of gas is considered -- though their Fig. 9 also
illustrates that strong point-like sources can produce extended \lya{}
structures via scattering. We will assume here that scattering has
little overall effect on our predicted morphologies and extents, in
the case that these structures are already well extended, though we do
expect that it will likely produce subtle changes in observable LAB
areas -- indeed Fig. 8 in \FG{} shows that the inclusion of scattering
can make some observable \lya{} structures narrow down and others
widen out. We will include and investigate the effect of \lya{}
scattering on spectral shapes, luminosities, and morphologies of our
objects in a future paper.

\subsubsection{Observed blobiness of cold accretion streams}
Mock observations of our three targeted halos are shown in the top row
of \Fig{LyObs.fig}. The middle contour in the maps is set at
$I_{-18}=10^{-18} \; \IS$, roughly corresponding to current
observation limits \citep[e.g. M11,][see Appendix \ref{Sec:moremaps}
for a more accurate comparison]{Erb:2011p5386}, and the inner and
outer contours correspond to ten times brighter and ten times dimmer,
i.e.  $10^{-17}$ and $10^{-19} \; \IS$.

Assuming $I_{-18}$ as our instrumental sensitivity limit, the H1 halo
(top left) is a \lya{} emitter that is centered on a galaxy,
circularly symmetric in shape, about $20$ kpc in diameter and doesn't
trace streams. Thus the H1 halo is not a LAB. The total observed
luminosity, i.e. $I$ integrated over the area within $I_{-18}$, is
$L_{obs}=6\ 10^{41}$ $\mathrm{erg \, s^{-1}}$.

The H2 halo observation (top middle) differs dramatically from that of
H1. At $I_{-18}$ we do see a borderline giant LAB, asymmetric and
about $100$ kpc in length, and we can see the end of an accretion
stream poking out to the north-west. The observed luminosity
integrated above $I_{-18}$ is $L_{obs}=2\ 10^{43}$ $\mathrm{erg \,
  s^{-1}}$.  The H3 halo (top right) has observable \lya{} emission
all over the place, is about $200$ kpc in diameter and very
asymmetric. Its observable luminosity is $L_{obs}= 10^{44}$
$\mathrm{erg \, s^{-1}}$.

Provided there is nothing special about these halos, we can conclude
that in general the {\it cooling emission from halos with masses
  greater than a few times $10^{12} \, \mathrm{M_{\sun}}$ can produce
  giant LABs ($\ga 100$ kpc) at redshift 3, assuming current
  instrument sensitivity limits}. Qualitatively this compares well
with \cite{Yang:2010p3447}, who find that at redshift 2.3, LABs should
occupy halos $\ga 10^{13} \, \mathrm{M_{\sun}}$.  Qualitatively again,
the maps presented in Sec. \ref{Sec:moremaps}, which mimic the
observational conditions of M11, show that the morphologies of our
simulated LABs are very similar to those observed.

Interestingly, we note that the LABs produced by cold accretion
streams are naturally extended in the direction of the main
large-scale filaments that they are connected to. This is particularly
visible for H2 and H3 in Fig. \ref{LyObs.fig} (see also
Fig. \ref{fig:ohmygod}), and lends support to the observational
findings of \citet{Erb:2011p5386}.

\begin{figure}\begin{center}
  \includegraphics[width=.5\textwidth]{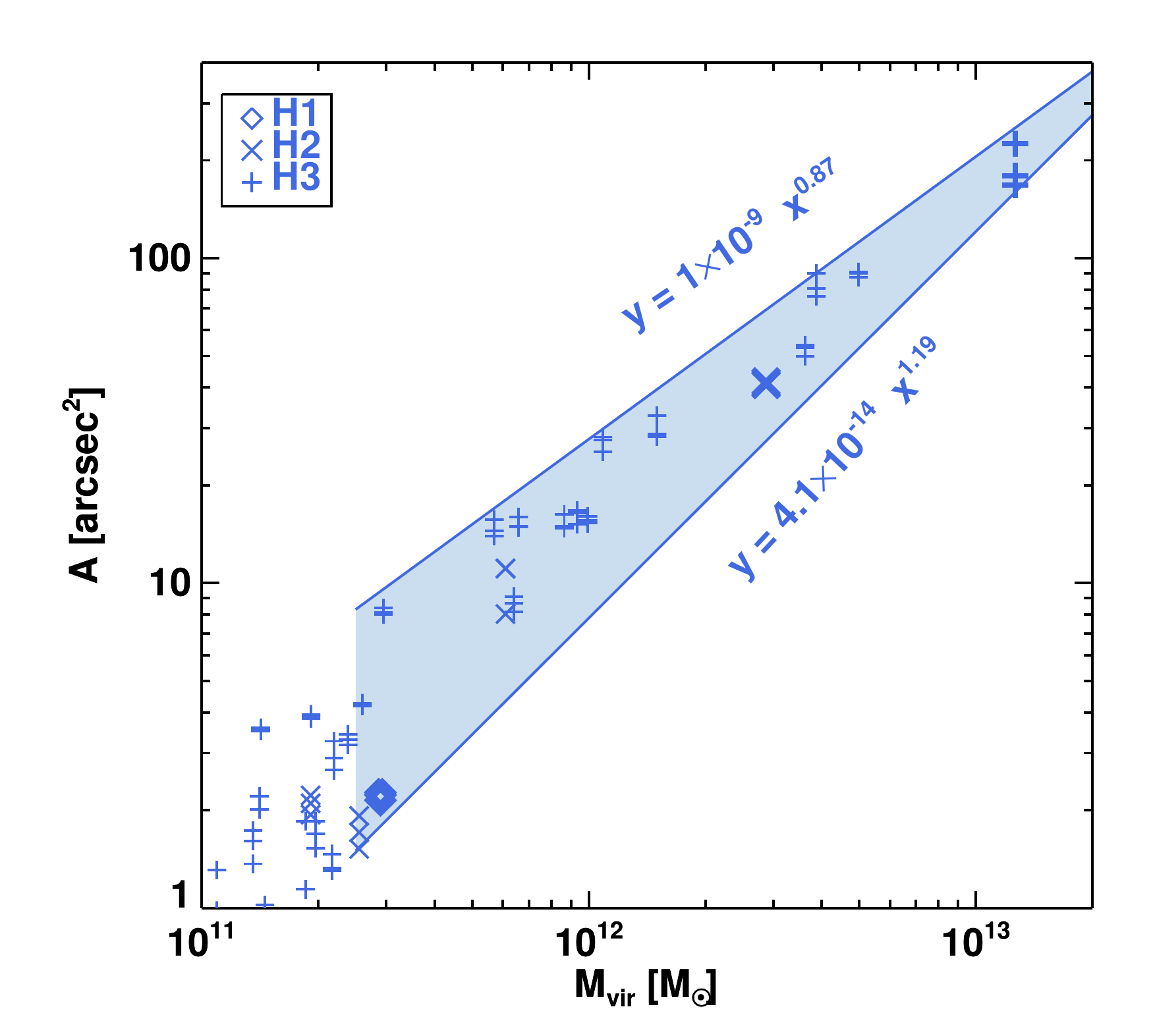}
  \caption[]{\label{P_A_mass.fig}Plot of mock observed areas within
    contours of $1.4\ 10^{-18}\; \IS$ versus halo mass for zoomed
    halos in the three simulations. Each halo is observed from three
    directions. The targeted halos are marked with thick symbols. The
    points are enveloped by two power-laws shown in the plot.}
\end{center}\end{figure}

\vsk Another matter are those mysterious LABs which do not seem to be
centered on observed galactic counterparts
\citep[e.g.][]{Steidel:2000p2213, Weijmans:2010p5527,
  Prescott:2011p5034}. We are not able to reproduce this phenomenon in
our simulations.  The bottom row of \Fig{LyObs.fig} shows stellar
densities in our targeted halos, with the $I_{-18}$ sensitivity
contour over-plotted.  Clearly all the peaks of \lya{} brightness
would have continuum counterparts in observations, unless these
counterparts would for some reason be hidden from view.  Such LABs are
rare among rare events, though, and our three simulations have little
statistical chance of reproducing such oddities. A larger sample of
simulations would be required to investigate this issue further.

\subsubsection{Size distribution of simulated
  LABs} \label{obsComp.sec}
We shall now statistically compare our results with a catalogue of
202 observed LABs from the surveys described in M11 (courtesy of
Yuichi Matsuda and team). The aim here is to derive a cumulative LAB
area function from our results and see how it compares with real data.

We follow M11 by assuming $z=3.1$ in Eq. \ref{mock.eq}, and applying a
PSF with FWHM=1.4 arcsec. We calculate the observed LAB area $A$ of
each halo within the zoom regions of our simulations by integrating
its total area above the surface brightness limit $I=1.4\ 10^{-18} \;
\IS$.  We `observe' each halo in three directions ($x$, $y$, and
$z$). In Fig. \ref{P_A_mass.fig}, we plot the LAB areas as a function
of halo mass. The large thick symbols correspond to our targeted halos
H1, H2, and H3. The observed LAB area is a reasonably well-behaved
function of halo mass, with more massive halos producing larger LABs,
and the points can be bracketed by a couple of power laws of indexes
0.87 and 1.19 (see Fig. \ref{P_A_mass.fig}).

\begin{figure}\begin{center}
  \includegraphics[width=.5\textwidth]{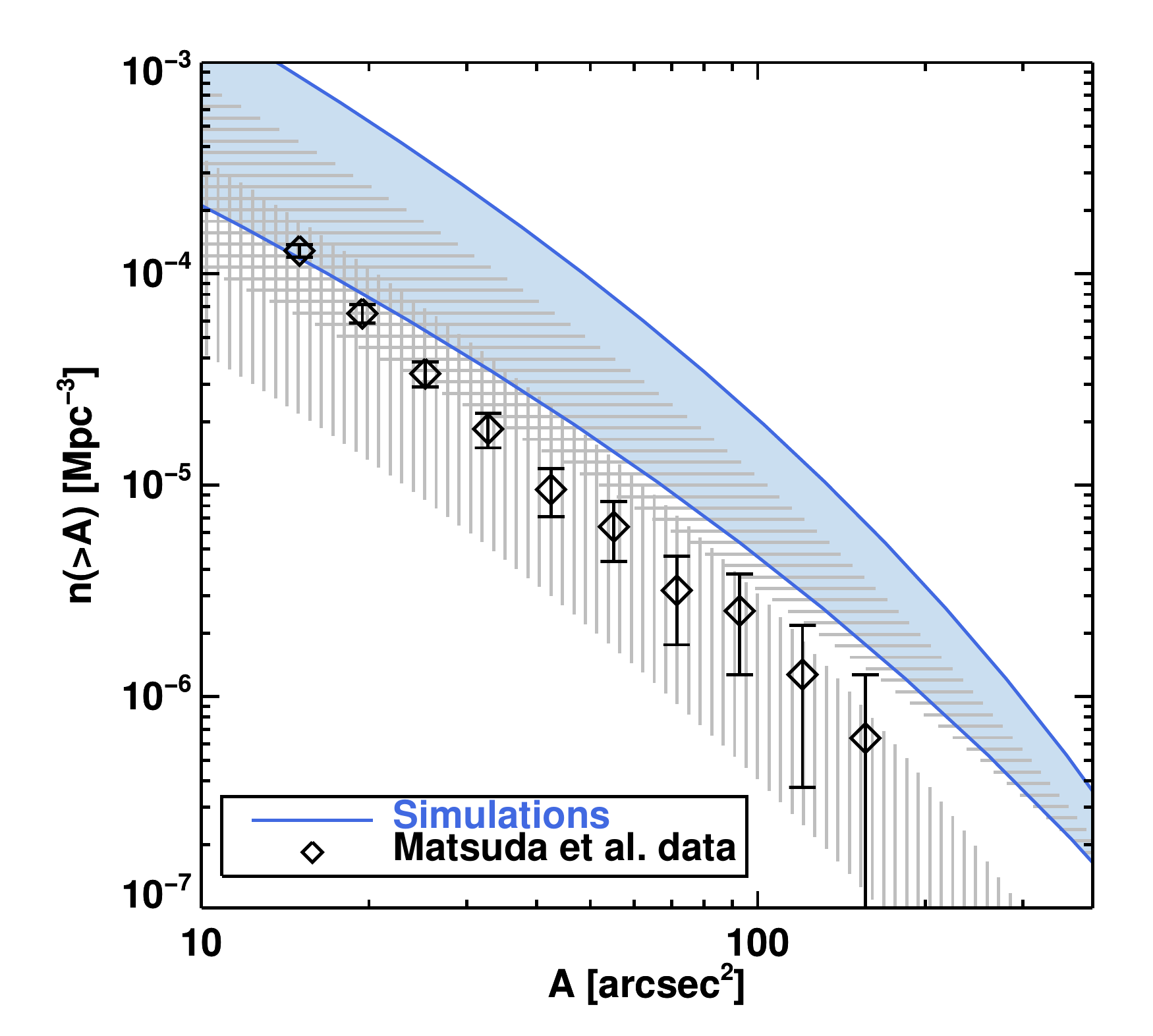}
  \caption[]{\label{f_area.fig} The shaded region represents the
    boundaries of our predicted LAB area function, derived from the
    power laws in \Fig{P_A_mass.fig}. The black symbols mark a rough
    area function derived from a sample of 202 observed LABs from the
    survey of M11. The horizontally and vertically line-filled regions
    represent similarly predicted area functions, but with gas
    densities of $\nh \ge 0.3$~$\cci$ and $\nh \ge 0.1$~$\cci$
    excluded, respectively.} \end{center}\end{figure}

We now make the assumption that extended \lya{} emission is an
inherent property of dark matter halos and that the observed LAB
properties are direct functions of halo mass. This assumption is
substantiated by our results (Figs. \ref{HLL.fig} and
\ref{P_A_mass.fig}).  We thus convolve the power laws of
Fig.~\ref{P_A_mass.fig} with the halo mass function at redshift $3.1$
\citep[taken from][]{Sheth:1999p4123} in order to produce the
cumulative area function envelope shown in \Fig{f_area.fig}.  There,
the black diamonds represent actual observations for comparison: they
are a rough estimate of the area function at redshift 3.1 based on the
202 LABs of the M11 survey, derived by binning the LABs by area and
dividing the count by the total survey volume of $1.57\ 10^6$ Mpc$^3$
(the error bars are Poissonian). The comparison between our predicted
area function and the observationally derived one is very
satisfactory, although we systematically over-predict the function by
a factor of 2-3.

There may be several causes to this over-prediction. First, our
derivation of the observed LAB area function is too simplified. For
example, we do not take into account the shape of the narrow band
filter, or any $1/V_{\rm max}$ corrections. This introduces systematic
errors that could well be of about a factor two.  Second, our mock
observations are also simplified, and do not include noise, which
could possibly affect the measured area in a systematic way. Third,
perhaps we have overshot in our choice of $\fa=0.66$. As noted by G10,
cosmic extinction may be stronger than average for sources that reside
in over-dense regions, as LABs tend to do. Fourth, our prediction is
based on only a few objects and to a lesser degree the same applies to
the observation-derived function.  Fifth, the predicted LAB areas are
sensitive to the applied PSF smoothing, which may not be entirely
consistent in all the 202 observed LABs. And finally, we may lack
physics in our simulations that would drive down the LAB areas. For
example, \lya{} scattering, if applied, could induce a slight spread
in the predicted rest-frame \lya{} surface brightness, which could in
some cases bring down both the observed area and luminosity within
sensitivity ordained brightness contours. Also, metal-line cooling may
drive down the \lya{} emissivity of gas by cooling it below $10^4$
K. Furthermore, as shown by \cite{vandeVoort:2011p5673},
\cite{FaucherGiguere:2011p5611}, and \cite{vandeVoort:2011p5669},
feedback driven winds can destroy cold accretion streams in the
vicinity of galaxies, hence terminating their \lya{} emissivities.

Our predicted area function is not very sensitive to the density
threshold of gas applied throughout this paper, where we have excluded
ISM densities ($\nh>1$~$\cci$) in our analysis.  To illustrate this,
\Fig{f_area.fig} also shows, with line filled regions, predicted area
function envelopes that have been derived from our simulations via a
convolution with the Sheth-Thormen halo mass function, but including
only more diffuse gas, $\nh<0.3$~$\cci$ (i.e. sub CGM densities) and
$\nh<0.1$~$\cci$, for the horizontal and vertical line-fillings
respectively. The prediction using the sub-CGM densities is close to
the original prediction, which can be expected since these densities
account for $\sim60\%$ of the total luminosities of all three targeted
halos (see Figs.  \ref{SLE.fig} and \ref{HLL.fig}). Even using
$\nh<0.1$~$\cci$ gas only (which is comparable to the more
conservative prescriptions used in \FG) still produces giant LABs
hosted by massive halos and gives an area function that is compatible
to the observational data. This confirms that the extent of our
simulated LABs is largely driven by low density cold streams.

We have also compared our results to observations via a LAB luminosity
function (rather than the area function just discussed). However,
since LAB emissivity typically peaks around compact sources, and since
we neither model the emission nor absorption coming from the compact
ISM regions, such a comparison is less robust than using the area
function which should be more or less dictated by the state of more
diffuse gas on much larger scales. The luminosity comparison, which is
discussed in detail in Appendix \ref{lumfunc.app}, is actually
surprisingly good, but it is problematic to draw any conclusions from
it because of the lack of modelling of compact regions.

\begin{figure}\begin{center}
  \includegraphics[width=.5\textwidth]{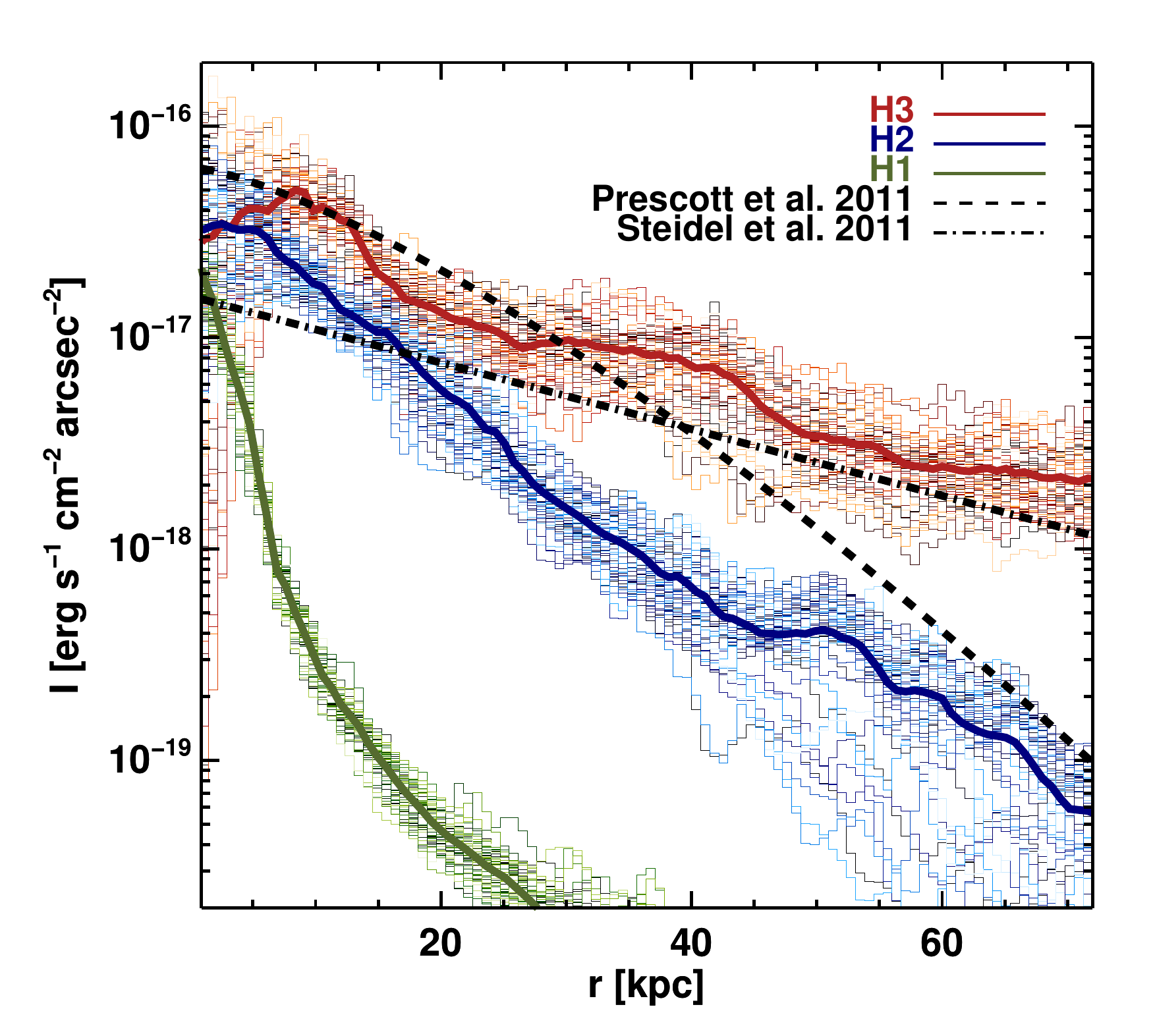}
  \caption[]{\label{LyStack.fig}Surface brightness profiles of our
    three targeted halos (solid) compared to the profile of LABd05
    from \cite{Prescott:2011p5034} (dashed) and an average surface
    brightness profile of 11 LABs from \cite{Steidel:2011p5455}
    (dot-dashed). The thin solid lines represent different random
    orientations (50 for each halo) and the thick solid ones are
    averages of the thin ones.}
\end{center}\end{figure}

\subsubsection{Surface brightness profiles}

\Fig{LyStack.fig} shows observed surface brightness profiles of our
three targeted halos. Profiles are taken for each halo in 50 planes of
random orientation, represented by thin coloured lines, and then these
are averaged into the thick coloured lines. The profiles are
transformed from rest-frame to observed surface brightness via
\Eq{mock.eq}, assuming redshift 3, but no smoothing or cosmic
extinction is applied, and as before \lya{} scattering is
neglected. Over-plotted are surface brightness profiles from
observations: The black dashed curve represents a Sersic fit to the
observed profile of the giant LABd05 at redshift $2.656$ from
\cite{Prescott:2011p5034}, which we have scaled to z=3. Notably,
LABd05 doesn't have a galactic counterpart at, or even close to the
peak of \lya{} emission, though it has 17 small galaxies substantially
offset from the peak (by $\ga 20$ kpc). The black dot-dashed curve is
an exponential disk fit to an average of 11 LAB profiles observed at
$z\approx2-3$, reported in \cite{Steidel:2011p5455}, with no scaling
applied.

The profiles of H2 and H3 are similar in shape and magnitude to the
observed profiles. Interestingly, each of those compares favourably to
different observations, with the H2 profile being similar to LABd05
and the H3 profile similar to the 11 LABs from
\cite{Steidel:2011p5455}. The comparison indicates that these
observations fit well within the model of cold accretion powered LABs,
but due to the very limited statistics of our simulations (i.e. one
halo per mass bin of three), and different redshifts of the observed
LABs, it is problematic to make quantitative deductions, e.g. about
masses of the host halos of observed LABs. Rather than representing
different halo masses, the different profile shapes (and to some
degree their magnitudes) may just as well reflect the different
morphologies one may find in galactic groups and clusters.

\cite{Prescott:2011p5034} compare the LABd05 surface brightness
profile with simulated profiles from G10 and \FG{}, and find that the
simulations appear to fit very badly with reality, with the G10
profile being both too peaky at the center and too shallow at large
radius, and the \FG{} profile being too weak and steep. Our H1 profile
actually agrees with the simulated profile from \FG{} (their model 7,
see Fig. 9 in \citealt{Prescott:2011p5034}), and it seems to us that
\FG{} in fact don't pose any mismatch with the LABd05 observation: The
fault lies in \cite{Prescott:2011p5034} assuming that the surface
brightness profile scales linearly with halo mass, which is not at all
the case judging from our simulations (and to be fair, these authors
admit that their assumption is probably not accurate).

We admittedly don't provide large statistics here, but we can conclude
that the surface brightness profiles produced by our simulations do
not disagree with LAB observations, and at the same time we can argue
that neither do the simulations of \FG{}.

\begin{figure}\begin{center}
  \includegraphics[width=.5\textwidth]{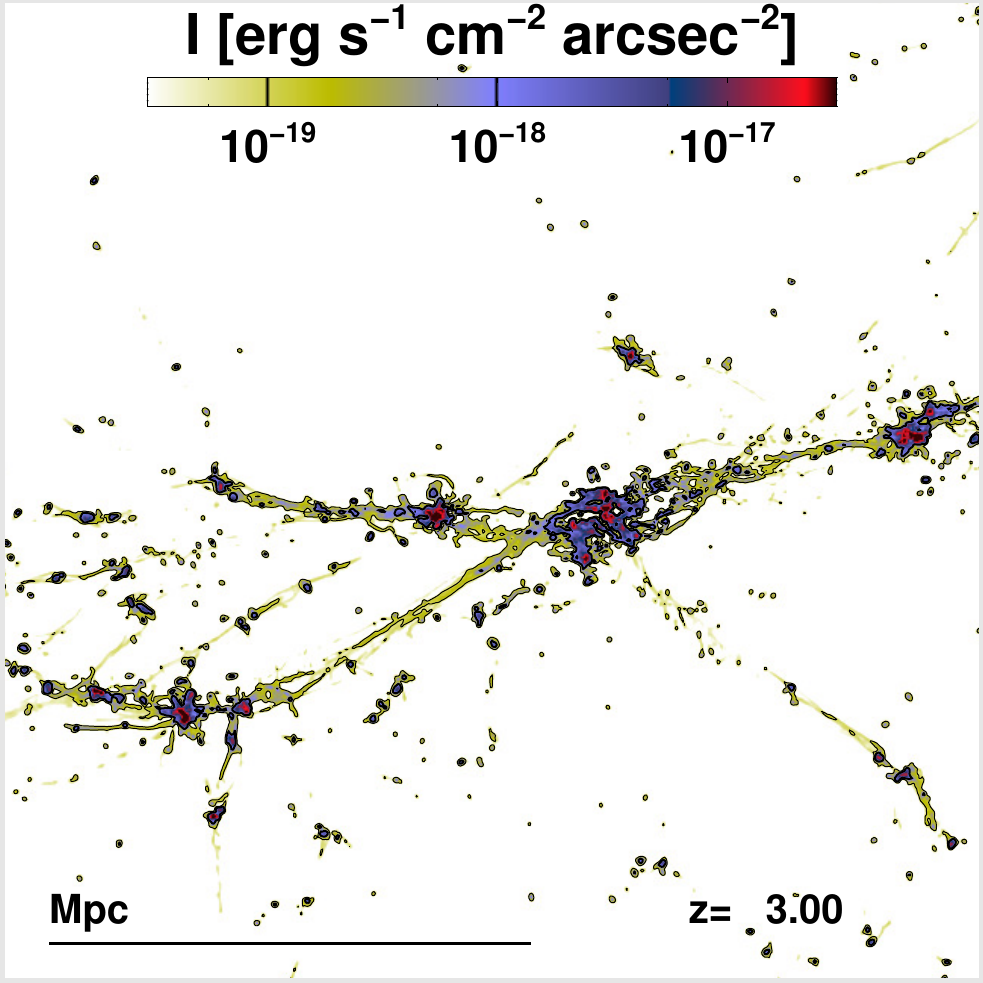}
  \caption[]{\label{fig:ohmygod} Same as Fig. \ref{LyObs.fig}, but
    showing the large-scale \lya{} map of the H3 halo and its
    environment. Thick inner (thin outer) contours mark $I=10^{-18}
    \;(10^{-19}) \, \IS$. Large-scale streams connecting massive halos
    and extending over several Mpc would be visible at $10^{-19}\,
    \IS$.}
\end{center}\end{figure}

\subsubsection{Implications for future observations}
Having demonstrated reasonable agreement between our simulations and
LAB observations, we now wish to highlight a prediction from our work
which is particularly relevant in the context of direct searches for
IGM emission at high redshifts.  The outermost contours in the upper
row of \Fig{LyObs.fig} mark \lya{} brightness at $10^{-19} \; \IS$. At
this limit, accretion streams start to show up even in the least
massive halo, and in the more massive halos we would detect them
unambiguously. The deepest observations to date are not quite there
yet, but almost, and this is an exciting perspective. Perhaps even
more exciting is the map shown on Fig. \ref{fig:ohmygod}, where the
thin (resp. thick) contours again mark the limit at $10^{-19}$
(resp. $10^{-18}$) $\IS$. This zoomed out view of our H3 halo shows
that deep \lya{} observations around massive halos may even reveal the
large-scale filamentary structure of the IGM on scales of a few Mpc !

\begin{figure*}\begin{center}
    \hspace{-2mm}
  \subfloat{\includegraphics[width=0.38\textwidth]
    {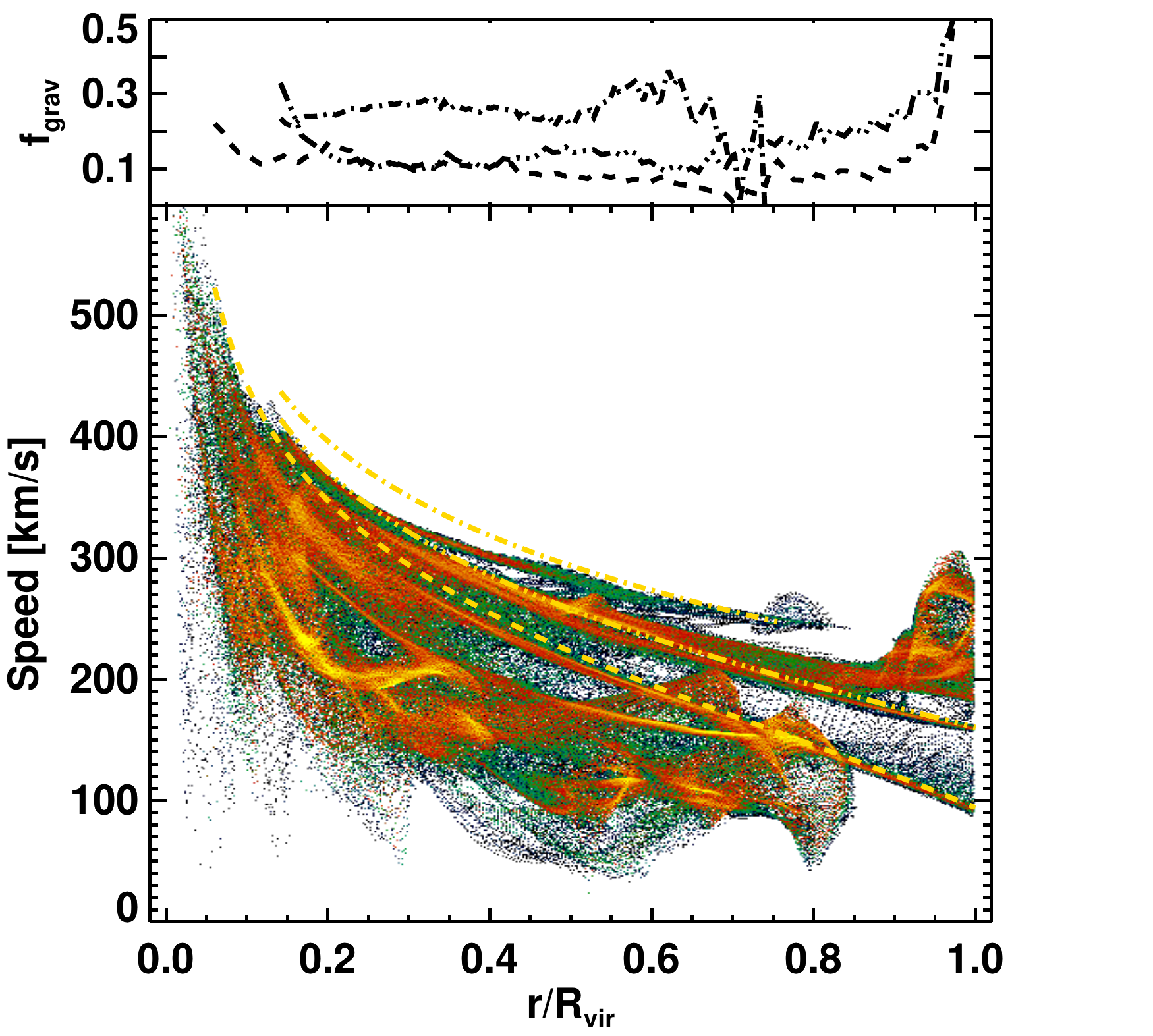}}\hspace{-13mm}
  \subfloat{\includegraphics[width=0.38\textwidth]
    {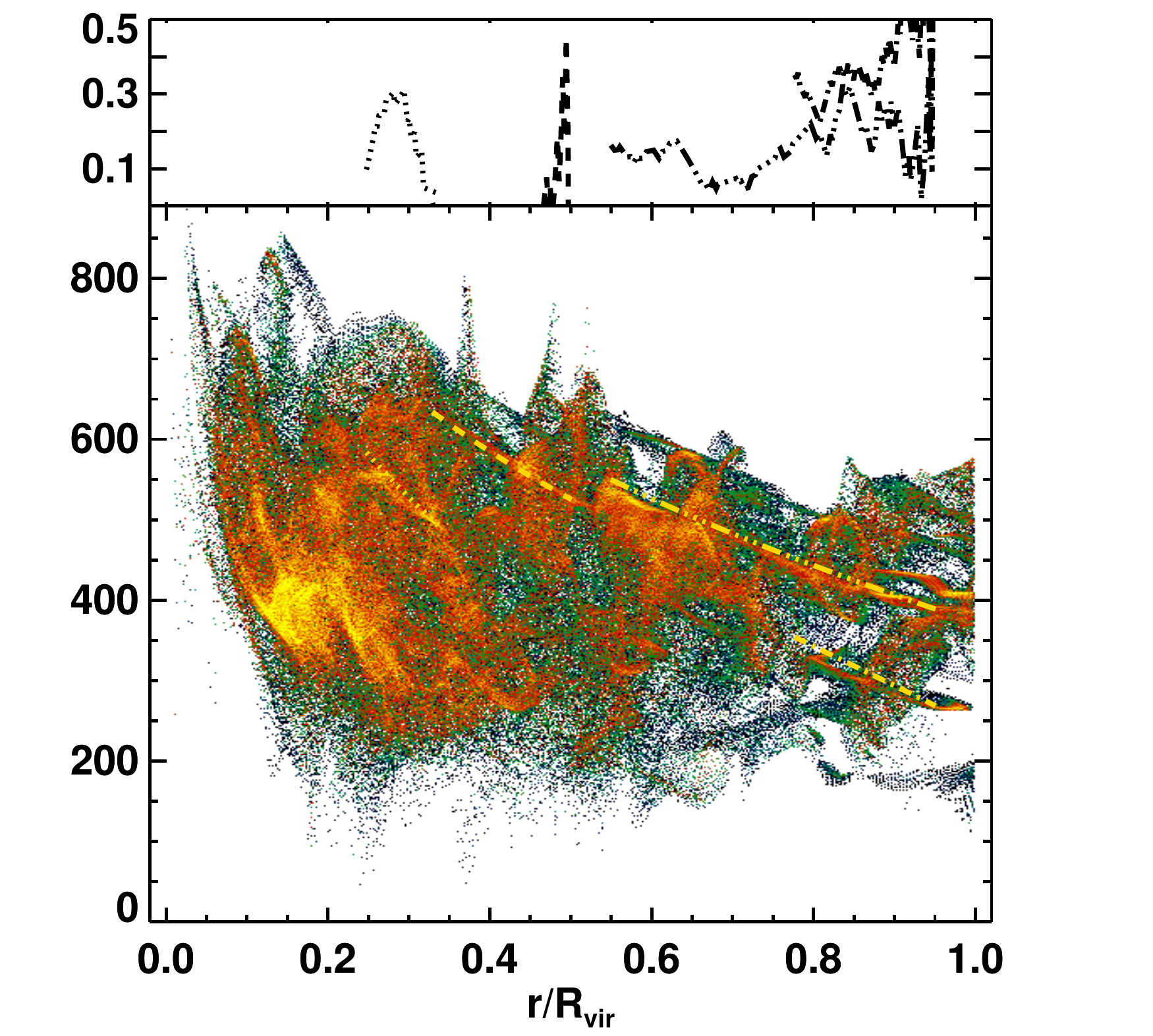}}\hspace{-13mm}
  \subfloat{\includegraphics[width=0.38\textwidth]
    {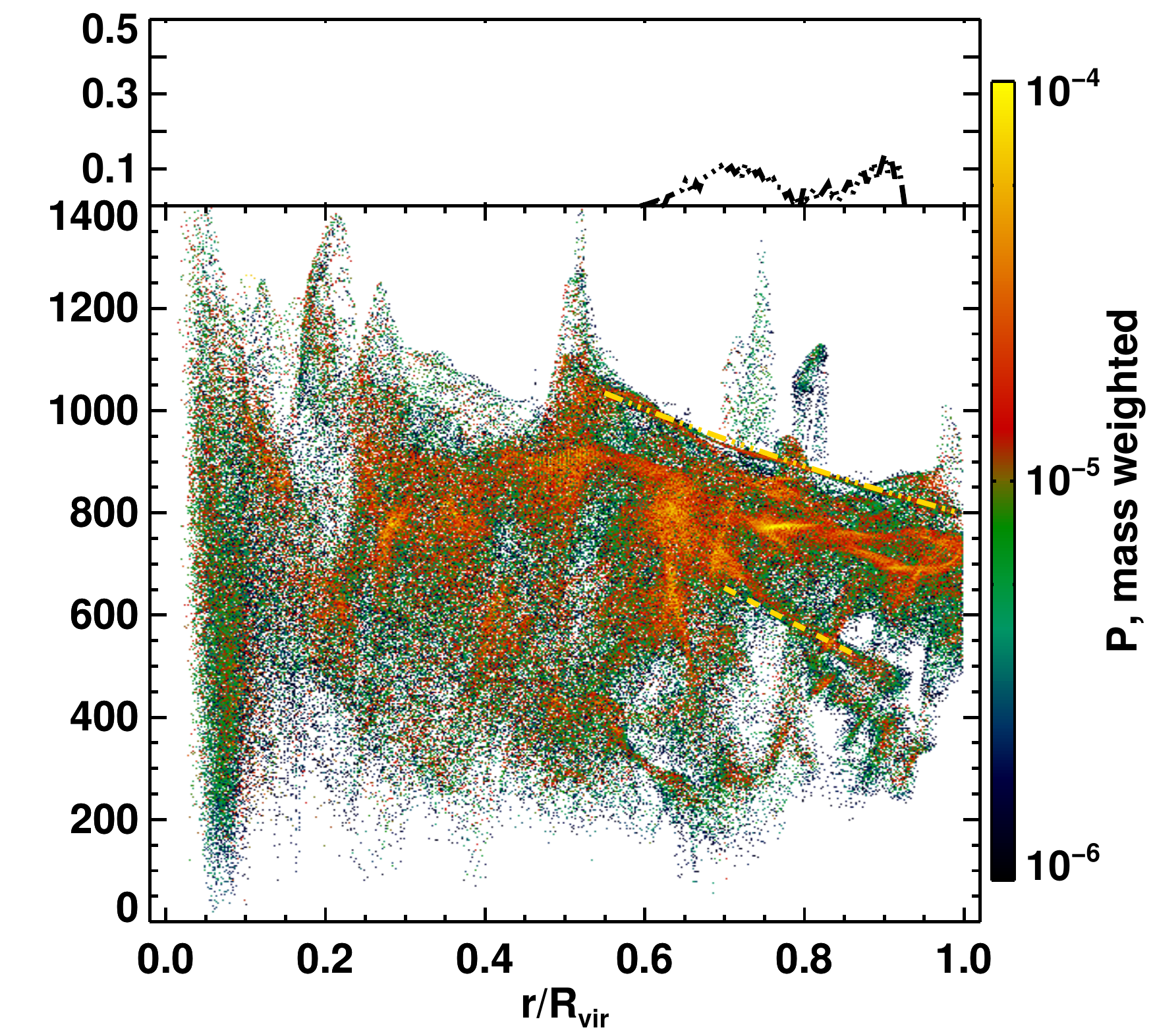}}\hspace{-1mm}
  \vspace{-0.3mm}
  \caption[]{\label{speed.fig}Phase diagrams showing speed of stream
    gas as a function of radius within targeted halos H1, H2 and H3
    from left to right. The color scale represents mass weighted
    probability per speed-radius bin. Only stream densities are
    included here so that the streams stand out in the diagrams and
    are not drowned in the more diffuse gas. Over-plotted in yellow
    are free-falling speed profiles. Plots above each phase diagram
    show gravitational heating efficiency in the clearest streams,
    compared to the free-fall profiles. }
\end{center}\end{figure*}

Although such observations are extremely challenging (if not plain
impossible) today, upcoming instruments, such as MUSE
\citep{Bacon:2006p5315} or K-CWI \citep{Martin:2010p5354}, should
greatly increase our chances of observing directly this source term of
galaxy formation in a very near future.

\section{What drives the \lya{} emission?} \label{Sec:What} Since
\lya{} scattering and stellar feedback are not included in our
simulations, the only possible power sources of \lya{} emission in our
results are gravity and the UV background.  We now look at how
gravitational heating contributes to the \lya{} emission along streams
and attempt to quantify its efficiency. We consider the contribution
of UV fluorescence and show how it is sub-dominant for typical values
of the UV background. We conclude this section by discussing to what
extent locally enhanced UV fluxes could boost \lya{} emission from
cold streams.

\subsection{Gravitational efficiency}
Gravitational heating is generally viewed as a progressive release of
gravitational potential energy that heats the gas along the cold
streams \citep{Dijkstra:2009p3780}. As long as this heating is not too
fast it can be balanced by radiative cooling, and as long as the gas
remains metal-poor and at temperatures $\sim 10^4$ K, \lya{} emission
is the dominant cooling mechanism, meaning that the thermal energy is
efficiently converted into \lya{} photons.

Gravitational heating in cold streams can be parametrized by the
\textit{gravitational efficiency} $f_{grav}$, the fraction of the
change in gravitational potential that dissipates into thermal energy
during in-fall. The rest is converted into bulk kinetic energy,
increasing the speed of the gas. A value of $f_{grav}=1$ thus means
perfect conversion of potential into heating, implying constant
in-fall speed, and $f_{grav}=0$ means that there is no conversion into
thermal energy and the stream should be in
free-fall. \cite{Dijkstra:2009p3780} derive an analytic model in which
$f_{grav} \ga 20\%$ is required if \lya{} blobs are to be driven by
gravitational heating in cool flows.

For each stream we can distinguish in our simulated halos in a given
output, we can extract the stream speed profile $v_{str}(r)$ by
following its core from end to end. Using this and a corresponding
free-fall profile $v_{ff}(r)$ for a body starting at a position and
speed identical to the outer end of the stream, we can estimate
$f_{grav}$ with
\begin{equation}\label{fgrav.eq}
  f_{grav}(r) = \frac{v_{ff}^2(r)-v_{str}^2(r)}{v_{ff}^2(r)-v_{init}^2},
\end{equation}
where $v_{init}$ is the speed at the outer starting position.

We calculate an approximate free-fall profile for the stream by
assuming static state and spherical symmetry, integrating the
free-fall speed from the starting position towards the halo center
using \begin{equation}\label{speed.eq}
  dv_{ff}(r) = \frac{1}{v_{ff}(r)} \frac{GM(<r)}{r^{2}} \; dr,
\end{equation}
where $r$ is radius, $G$ is the gravitational constant and $M(<r)$ is
the total halo mass within $r$.

 In practice we divide the halo mass into radial bins $r_i$, where
 increasing $i$ corresponds to decreasing radius, and solve
 \Eq{speed.eq} by recursively computing
\begin{equation}\label{speed2.eq}
  v_{ff}(r_{i+1}) = v_{ff}(r_i)+\frac{1}{v(r_i)} 
  \frac{GM(<r_{i+1})}{r_{i+1}^{2}} (r_{i+1}-r_{i}),
\end{equation}
where $M(<r_{i+1})$ is the mass measured within $r_{i+1}$ in the
simulation, and the initial condition is the stream speed at the outer
end, $v_{ff}(r_0)=v_{init}$.

In \Fig{speed.fig} we show phase diagrams for the three targeted halos
of gas speed versus radius (normalized to $R_{vir}$), where we exclude
all but gas at stream densities, so that the streams can stand out
more clearly. In the smallest halo (H1) the streams pop out nicely,
smooth and undisturbed basically over the whole radius range, though
they do dilute a bit at the central 10\% of $R_{vir}$. In H2 we can
still see streams, but they are much more disrupted, and not
distinguishable within the central 20\% of $R_{vir}$. In H3 only a few
streams can be distinguished in the outer $40\%$ of $R_{vir}$, and in
the central $\sim 20\%$ they are completely destroyed.

For those streams we can clearly distinguish in the diagrams, we have
plotted in yellow the corresponding free-fall profiles, using
\Eq{speed2.eq}, which show approximately the speeds that the streams
would follow were they in free-fall. Qualitatively it can be seen that
the streams are close to free-fall, though usually they lag a
little behind the free-fall profile, and conversely on some occasions
we even see streams that seem to accelerate faster than free-fall (due
to sub-halos and/or the inaccuracy of assuming static state and
spherical symmetry in our free-fall calculation).

We plot our estimates of $f_{grav}$ using \Eq{fgrav.eq} directly above
each phase diagram. For two of the three streams we have extracted in
the H1 halo we get a fairly consistent estimate of $f_{grav} \sim 0.1$
from the halo outskirts towards the central $\sim 15\%$ of $R_{vir}$,
whereas for the third (and more diffuse) stream we get a value which
is two to three times higher. In the H2 halo things are much messier,
and for those fragments of streams that we can extract we find a large
scatter in $f_{grav}$, going from negative values to about $0.3$ (the
initial large values are a numerical noise due to resolution in the
phase-space). Finally, in the H3 halo, we can only extract two streams
at the outer edges of the halo, one of them showing $f_{grav}\sim 0.1$
and the other accelerating faster than our free-fall approximation.

It appears that gravitational heating is inefficient if seen only as a
smooth and steady process along unperturbed streams as in H1. However,
heating and subsequent release of \lya{} photons seems to be more
efficient when it involves disrupted and wiggly streams. This also
appears reasonable, since the gas at the core of a straight and
unperturbed stream can flow virtually unopposed towards the central
galaxy whereas if the streams are wiggly and disrupted there should be
greater opposition from the surrounding hot and diffuse gas.

It remains to be seen how much photo-fluorescence from the UV
background is contributing the \lya{} emissivities compared to
gravitational processes, both smooth and messy. Before comparing these
factors we describe how they're derived from the simulation output.

\subsection{Computing the \lya{} contributions}\label{contr.sec}
Since we store the photon flux in each cell we can easily keep track
of the photo-heating and photoionization rates in the gas. If we
assume that every photoionization leads to a recombination and that
all the energy provided by photo-heating is released via collisional
excitations,\footnote{The timescale for recombinations in streams is
  on the order of $10^5$ - $10^6$ years, which is short compared to
  the timescale for the in-fall of streams in these halos, $\sim 100 $
  million years. The cooling timescale in streams is typically on the
  order of $10^4$ to $10^5$ years.} we can estimate the UV
contribution to the \lya{} emissivity in each cell as:
\begin{equation}\label{LyEmUV.eq}
\eps_{UV} = 0.7 \; \mathcal{H}_{\gamma} + 0.68 \; \Gamma
\; \nhi \; \epsilon_{\lyam},
\end{equation}
where $\mathcal{H}_{\gamma}$ is the photo-heating rate, the
$0.7$-factor is the conversion efficiency of cooling into \lya{}
photons\footnote{This factor (roughly) represents the ratio of $C_{Ly
    \alpha}(T)$ to the hydrogen collisional excitation cooling rate
  discussed in \Sec{emission.sec}. It means that we assume $70\%$ of
  the energy dissipated via cooling to go into \lya{} photons.}, and
the second term on the right is akin to \Eq{LyEmRec.eq}. We refer to
Appendix \ref{App_PHrate.sec} for how to calculate the photo-heating
rate. The UV contribution tends to be overestimated and can in fact be
estimated higher than $\eps$ in hot regions where collisional
excitation is not the dominant cooling channel, but since these
regions are \lya{} dim anyway this isn't a concern.

The only other driver of \lya{} emission in our simulations are
hydrodynamical processes which we can coin gravitational heating.
Thus the approximate gravitational contribution to \lya{} emissivity
can be calculated in each gas cell as:
\begin{equation}\label{LyEmgrav.eq}
\eps_{grav} = \rm{max}\;(0, \; \eps - \eps_{UV}).
\end{equation}

\begin{figure*}\begin{center}
  \subfloat{\includegraphics[width=0.33\textwidth]
    {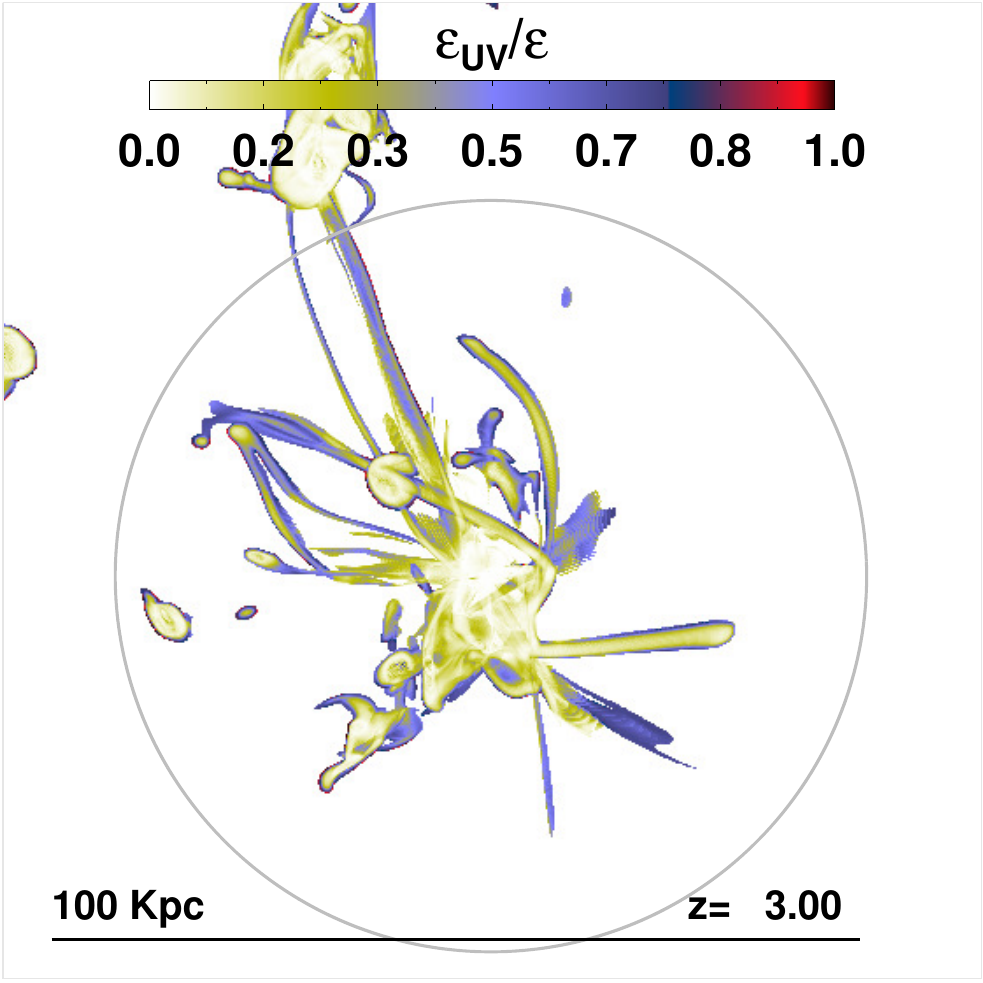}}\hspace{-1mm}
  \subfloat{\includegraphics[width=0.33\textwidth]
    {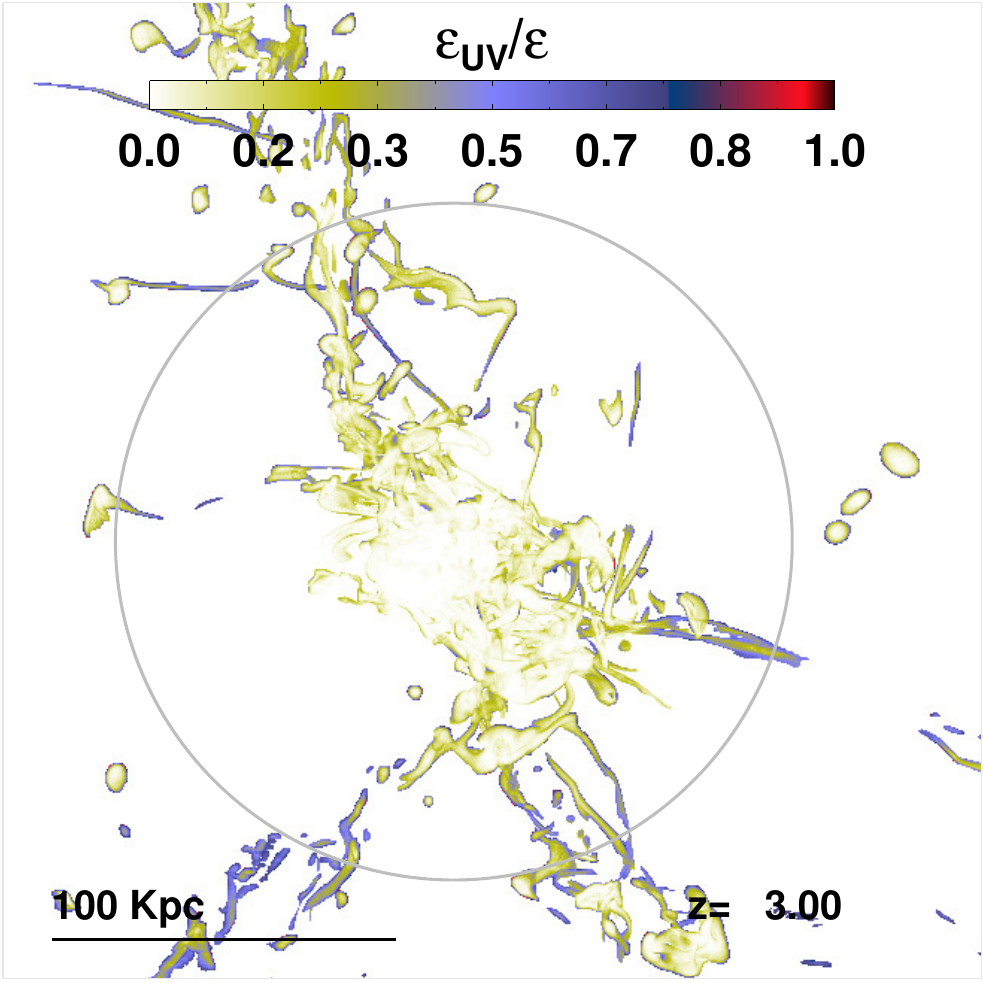}}\hspace{-1mm}
  \subfloat{\includegraphics[width=0.33\textwidth]
    {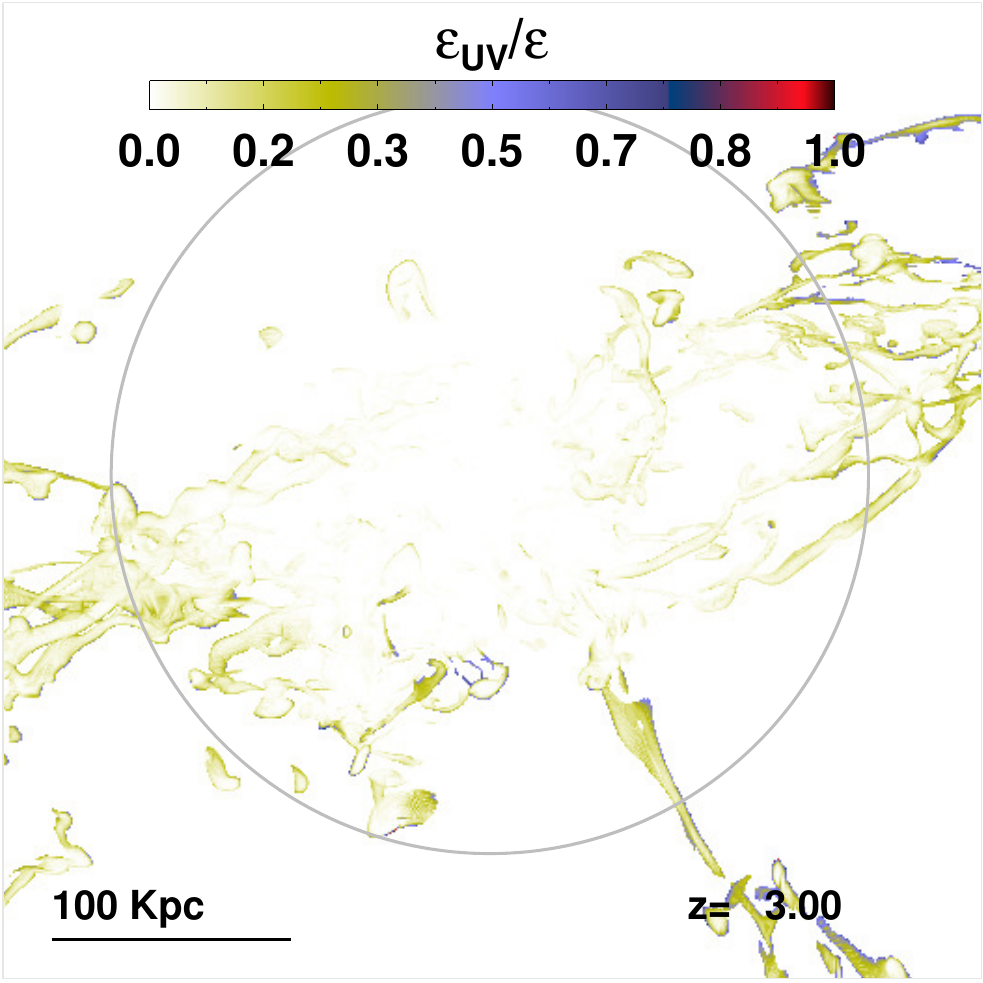}}\hspace{-1mm}
  \vspace{-3.3mm}

  \hspace{-7mm}
  \subfloat{\includegraphics[width=0.387\textwidth]
    {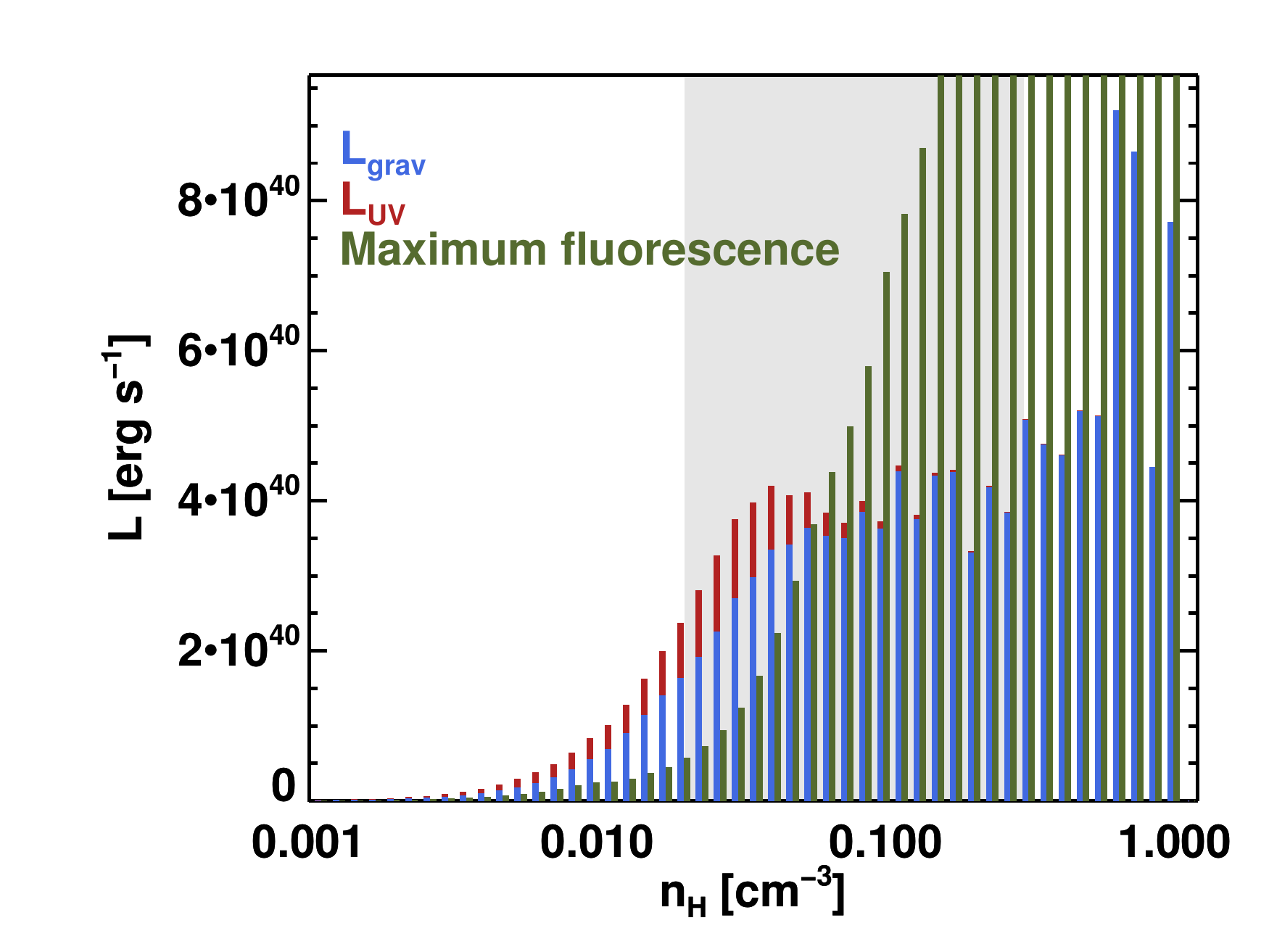}}\hspace{-11.5mm}
  \subfloat{\includegraphics[width=0.387\textwidth]
    {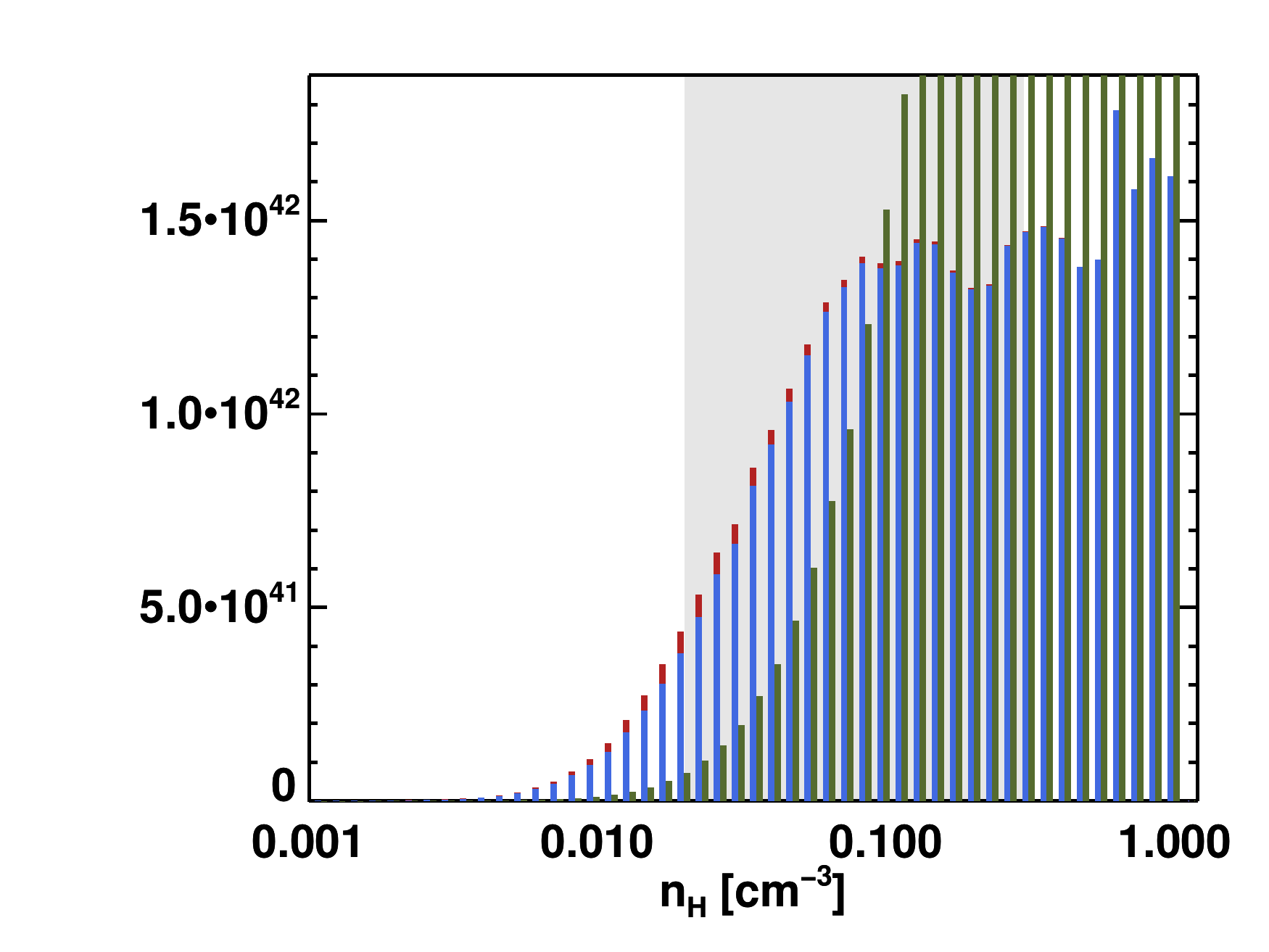}}\hspace{-13mm}
  \subfloat{\includegraphics[width=0.387\textwidth]
    {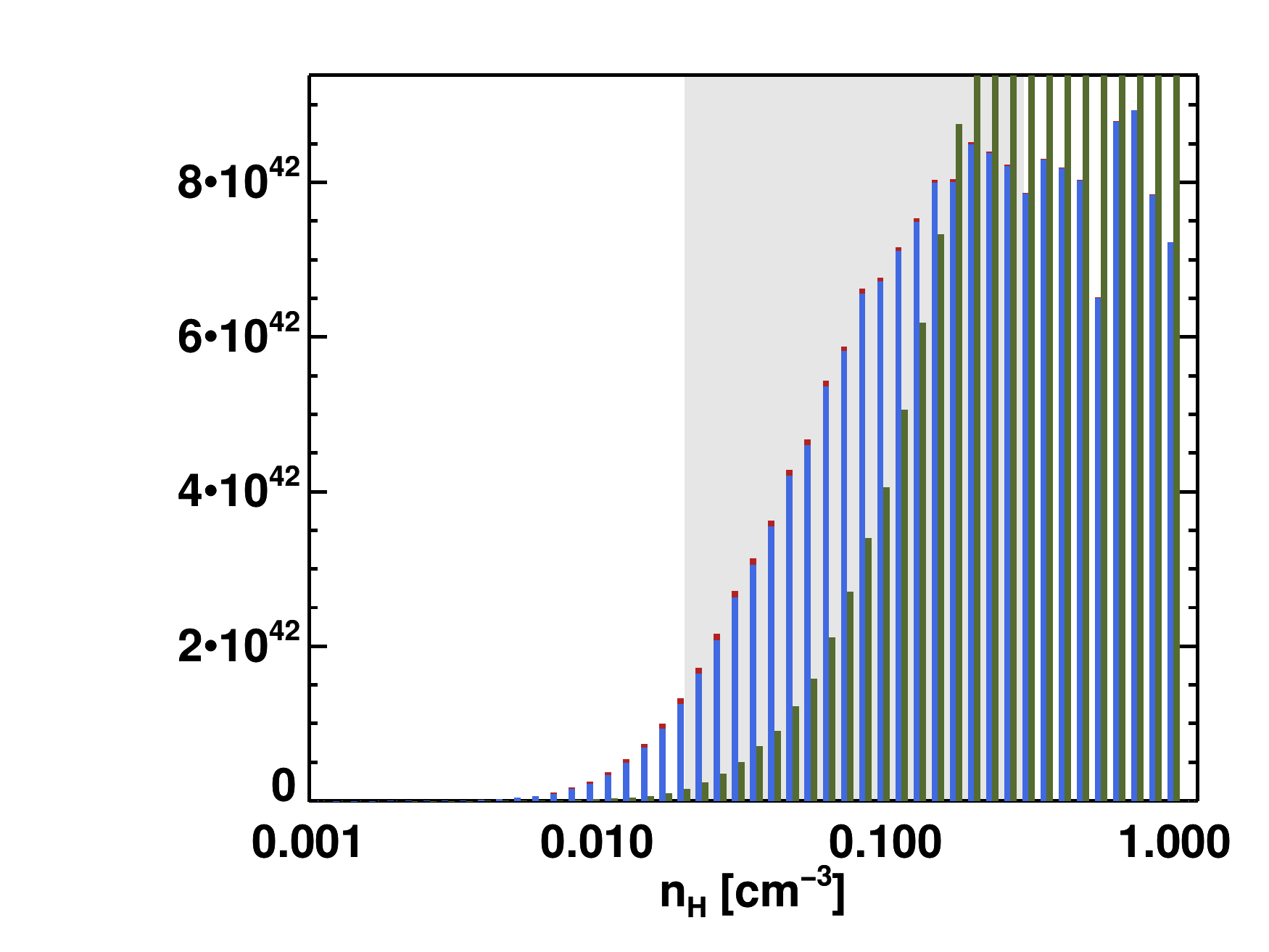}}\hspace{-3mm}
  \vspace{-1.3mm}
  \caption[]{\label{SLE_UV.fig}\textbf{Top row:} Fractional UV
    background contribution to the gas \lya{} emissivity in the three
    main halos, H1, H2 and H3, from left to right. Shown are mass
    weighted averages along the LOS, and everything below stream
    densities ($0.02 \mathrm \; \cci$) is ignored. \textbf{Bottom
      row:} Density distribution of the total \lya{} luminosity of the
    same halos, split into the UV (red) and the gravitational (blue)
    contributions. Green columns denote the \lya{} luminosity from
    maximum fluorescence. The shaded area represents stream
    densities. Note that the histograms have different scales on the
    y-axes.}
\end{center}\end{figure*}

\begin{figure*}\begin{center}
  \subfloat{\includegraphics[width=0.33\textwidth]
    {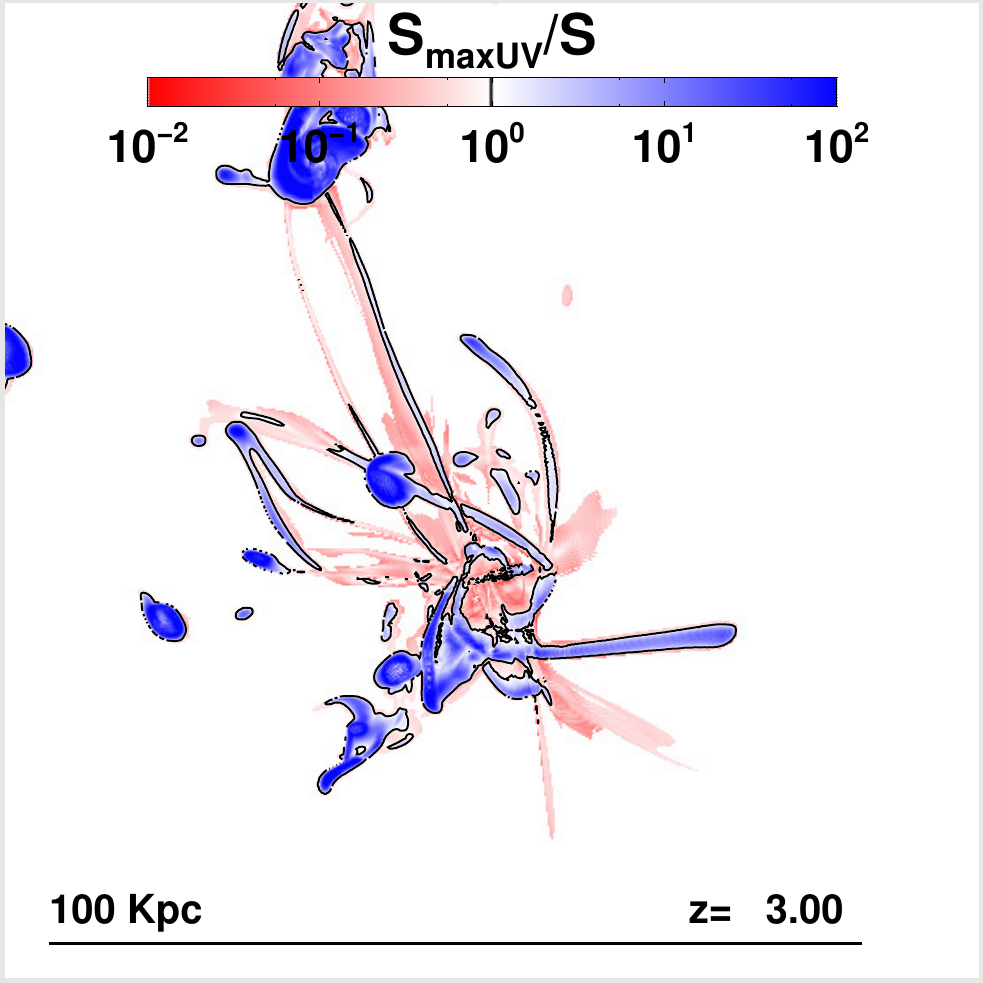}}\hspace{-1mm}
  \subfloat{\includegraphics[width=0.33\textwidth]
    {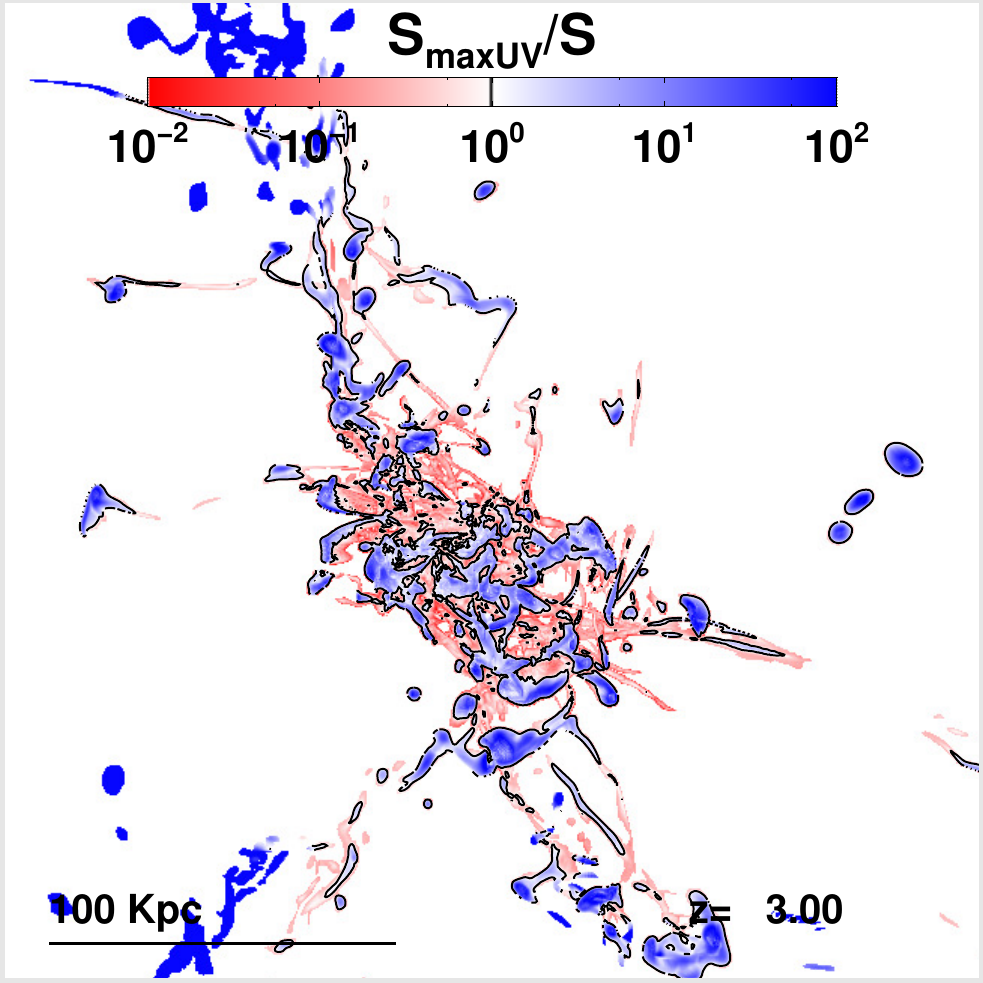}}\hspace{-1mm}
  \subfloat{\includegraphics[width=0.33\textwidth]
    {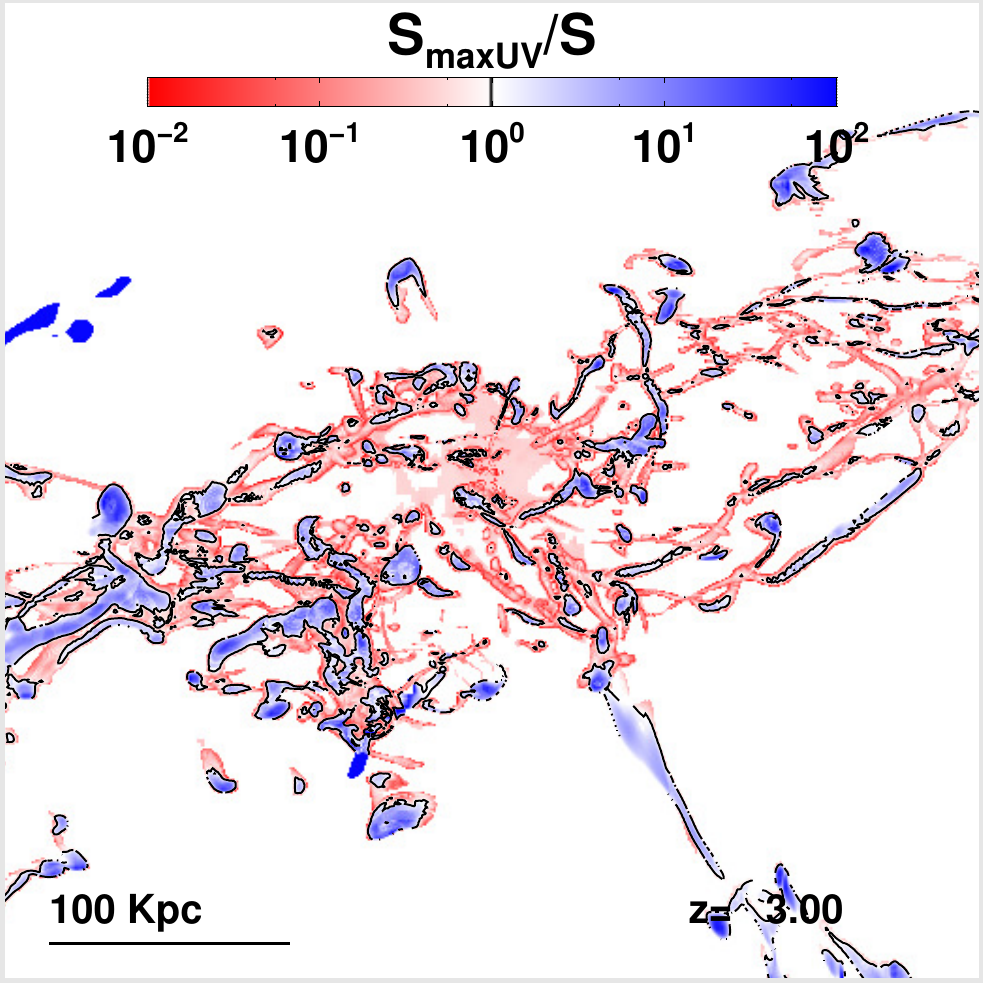}}\hspace{-1mm}
  \caption[]{\label{SLE_Ion.fig}Ratio of the rest-frame \lya{} surface
    brightness assuming maximum fluorescence (i.e. infinite UV flux,
    $S_{\rm maxUV}$) to our fiducial model, for halos H1, H2 and H3,
    from left to right. Fluorescence considerably boosts the densest
    clumps at the price of dimming the most diffuse streams.}
\end{center}\end{figure*}

\subsection{Gravitational heating vs. UV fluorescence}\label{fluor.sec}
Applying \Eq{LyEmUV.eq} we calculate the UV contribution to \lya{}
luminosity in each gas cell. We find that the relative UV contribution
becomes weaker with increasing halo mass, with the ratio to the total
halo \lya{} luminosity going from $8\%$ in H1 to $2\%$ in H2 and $1\%$
in H3 (see bottom row of \Fig{SLE.fig}). However, the relative UV
contribution is generally stronger on the edges of the halos than near
their centers.

In the top row of \Fig{SLE_UV.fig} we map the fractional UV
contribution to the \lya{} emissivity in streams and CGM gas, and in
the bottom row of the same figure we plot the density distribution of
the total luminosity, split into the UV (red) and gravitational (blue)
contributions.  As the maps and histograms show, the UV contribution
is negligible everywhere except for the smooth and diffuse streams
with $n_H \la 0.05 \; \cci$ in the H1 halo and on the outskirts of the
H2 halo. A comparison with the mock observations in \Fig{LyObs.fig}
reveals that these diffuse streams where the UV background
contribution is non-negligible are nowhere close to being observable
and all the observable emission is completely dominated by the
gravitational contribution. The UV background contribution to extended
\lya{} emission can thus safely be ignored, at least until the
observational sensitivity increases by two orders of magnitude or so.

Inclusion of local stellar UV radiation in our simulations may boost
the UV contribution, and thus both the total luminosity of the halos
and the extent of observable emission. Alternatively, the presence of
a luminous quasar nearby may also significantly enhance the \lya{}
luminosity through fluorescence, as demonstrated by
\citet[][]{Cantalupo:2005p4317} and \citet{Kollmeier:2010p3256}.

\subsection{Maximum fluorescence}\label{Sec:maxflu}
Even though we lack in this work the inclusion of local sources of UV
radiation, we can still evaluate the upper limit to the fluorescent
\lya{} luminosity that we can get from our simulated structures. This
gives us a idea of the relative luminosity increase a local UV
enhancement would provide, both in terms of the global luminosities of
our halos and in terms of where the \lya{} emissivity is boosted and
where it is dimmed, compared to the gravitationally driven emission we
have calculated.

To show the maximum fluorescent \lya{} luminosities we can obtain, we
re-calculate the \lya{} emissivity of the simulated gas in the limit
that $\xhii=1$ everywhere, corresponding to an infinite flux of UV
photons. Note that in this limit $\eps_{coll}$ is zero everywhere and
the \lya{} emissivity is purely recombinative.

The green columns in the histograms in the bottom row of
\Fig{SLE_UV.fig} show the \lya{} luminosity of gas in different
density bins in this limit. In all three halos at $\nh \ga 0.1
\mathrm{cm^{-3}}$, maximum fluorescence outshines the normal gas due
to the increase in HII abundance, while at lower densities it is
dimmer than what we predict with gravitational heating, because
collisional emission goes to zero.

In Fig. \ref{SLE_Ion.fig}, we show the effect of maximum fluorescence
on the \lya{} emissivity of the gas within our three simulated halos.
This is displayed as the ratio of $S_{\rm maxUV}$ -- the luminosity
computed assuming an infinite amount of ionizing photons everywhere
--, to $S$ -- the luminosity used in the rest of the paper, which
assumes a standard (though inhomogeneous) background value. Clearly, a
strongly enhanced UV fluorescence will boost the \lya{} emission in a
significant fraction of the gas (the blue part), and may contribute
significantly to LABs, as demonstrated by \cite{Cantalupo:2005p4317}
and \cite{Kollmeier:2010p3256}. However, from the perspective of
observing accretion streams, the price to pay is the strong dimming of
lower density structures (red).

This maximum fluorescence scenario is obviously optimistic, and only a
tiny volume fraction of the Universe will likely come close to it, in
the vicinity of rare and bright quasars in over-dense regions. Most of
the IGM will more likely be in a regime comparable to our fiducial
description, and its \lya{} luminosity will be powered by collisional
excitation.

\section{Summary and conclusions}
We have in this work addressed the questions of whether gravitational
heating may be the main driver of LABs, and how close we are to making
direct and unambiguous detections of cold accretion streams via their
\lya{} emission.

To this purpose we have run and analyzed cosmological RHD simulations
specifically tailored to accurately predict \lya{} emission from
extended structures. These simulations are idealized in the sense that
the effects of stellar feedback and \lya{} scattering are ignored such
as to isolate the efficiency of gravitational heating in generating
\lya{} photons. Our analysis is focused on redshift 3, which
corresponds to most LAB observations.

Our approach improves upon previous works in the following ways: (a)
Using \ramsesrt{}, our newly developed RHD version of the \ramses{}
code, we include on-the-fly propagation of UV photons, which allows us
to consistently and accurately model the self-shielding state in
accretion streams and their resulting temperatures and ionization
fractions, which are all very important to accurately predict their
\lya{} emissivity.  (b) We apply a novel refinement strategy that
allows us to optimally resolve accretion streams to an unprecedented
degree and on much larger scales than previously. This allows us to
spatially resolve the competition between gravitational heating and
radiative cooling in those streams.  (c) We post-process our
simulation outputs with very small timesteps to ensure we also resolve
said competition temporally. Failing to do this leads to a dramatic
underestimate of \lya{} emissivity of gas due to the commonly utilized
numerical method of operator splitting, and previous works may have
been marked by this problem.  (d) We simulate more massive halos than
hitherto done, based on the growing consensus that LABs are hosted by
the most massive halos in the Universe.

\vsk There are nontheless issues regarding uncertainties that
potentially affect our results. One is the likely presence of
artificial overcooling in shocks. As pointed out by
\cite{Creasey:2011p5926}, shocks -- or the mean-free paths of
particles inside them -- are almost exclusively under-resolved in
cosmological simulations. The artificially broadened shocks can
prevent the creation of hot and diffuse gas phase and instead allow it
to efficiently cool and remain at temperatures where \lya{} emission
is the most effective cooling channel. We may thus over-predict the
\lya{} emissivities of shock regions in our simulations. However, this
effect should be most severe in regions where gas is shocking on to
galactic disks, and should thus be mainly constrained to CGM regions,
and to the densest gas under consideration, i.e. $n_H\sim 1 \
\cci{}$. Weaker shocks may also exist at the boundaries of the
disrupted streams in and around our more massive halos, but it seems
unlikely that numerical overcooling is a big issue here, due to the
high resolution, large volumes, low densities, and the fact that the
\lya{} emissivity is not particularly concentrated at the stream
boundaries.

It is an unavoidable fact that the denser the gas in our simulations,
or in any simulations for that matter, the larger the uncertainty in
its \lya{} emissivity. In particular, at densities $\ga 0.1 \ \cci$,
gas may cool down to $<< 10^4$ K via molecular or metal-line cooling,
neither of which is included in our simulations.  The densest gas is
also in general the most \lya{} luminous in our simulations: What we
term CGM gas ($n_H>0.3\ \cci$) consistently accounts for $40\%$ of the
total \lya{} luminosities of our halos, so we can estimate the total
\lya{} luminosities to be uncertain by (very) roughly $50\%$, and even
more if we exclude still more diffuse gas than the CGM.  We have
however shown that our results and conclusions regarding LAB areas are
not sensitive to the density threshold applied (i.e. above which
densities we ignore \lya{} emissivity).

\vsk Our main results are the following:
\begin{itemize}
\item{ Cold accretion streams in halos more massive than $\sim
    10^{12}$~M$_\odot$ produces extended and luminous \lya{} nebulae
    which are by large compatible with LABs observed at $z\sim 3$, in
    terms of morphology, luminosity and extent. Gravity alone provides
    most of the energy, and we find that extra sources such as UV
    fluorescence, \lya{} scattering or superwinds are not
    necessary. This clearly doesn't rule out these other processes
    though, as they are likely all significant in the case of LABs,
    and further work is needed to study their complex interplay.}

\item{In our simulations, LAB area and luminosity are reasonably
    well-behaved functions of halo mass. We use these relations to
    compute the cumulative luminosity and area distributions, and find
    that they are in reasonable agreement with observations given the
    relatively large uncertainties. This comparison however suggests
    that the combined effects of SN feedback, \lya{} scattering and an
    enhanced local UV field may possibly have a negative impact on the
    luminosity and extent of simulated LABs, when conjoined with cold
    accretion. }

\item{ The model of gravitational heating as a driver of extended
    \lya{} emission works according to our results, but we need to
    alter our notion of \textit{how} it works: It is inefficient in
    the classic sense where gas accretion is smooth. Rather the
    accretion is messy and disrupted in massive halos and probably
    involves some mass loss to the surrounding hot diffuse medium.  }

\item{Our examination of maximum photofluorescence hints that in
    extreme cases local UV enhancement, e.g. near quasars, can boost
    the \lya{} luminosity of LABs and to a lesser degree their
    extent. As demonstrated by \citet[][]{Cantalupo:2005p4317} and
    \citet{Kollmeier:2010p3256}, this means that large accretion flows
    may be more easily observed in the proximity of quasars than
    elsewhere.}

\item{We find that cold accretion streams should be unambiguously
    observable via direct \lya{} emission for the first time in the
    near future, on upcoming instruments such as MUSE and K-CWI which
    will allow to probe emission at surface brightnesses as low as
    $\sim 10^{-19}\IS$.}

\end{itemize}

Although we have significantly improved on previous work, a large
number of theoretical issues remain to be addressed. In forthcoming
papers, we plan to investigate the effects of \lya{} scattering
SNe-driven winds and local UV enhancement from star formation.

\section*{Acknowledgements}
We thank Dominique Aubert and Romain Teyssier for helping us implement
radiative transfer in \ramses{}. We are grateful to Yuichi Matsuda for
kindly providing observational data, and we acknowledge valuable help
and discussion on this work from Sebastiano Cantalupo, Stephanie
Courty, Julien Devriendt, Tobias Goerdt, Joop Schaye and Romain
Teyssier. Last but not least, we thank the referee, Claude-Andr\'{e}
Faucher-Gigu\`{e}re, for his careful and constructive review of this
paper.

This work was funded in part by the Marie Curie Initial Training
Network ELIXIR of the European Commission under contract
PITN-GA-2008-214227. The simulations were performed using the HPC
resources of CINES under the allocation 2011-c2011046642 made by GENCI
(Grand Equipement National de Calcul Intensif). We also acknowledge
computing resources at the CC-IN2P3 Computing Center
(Lyon/Villeurbanne - France), a partnership between CNRS/IN2P3 and
CEA/DSM/Irfu. JB acknowledges support from the ANR BINGO project
(ANR-08-BLAN-0316-01).


\clearpage
\appendix

\section{The quasi-homogeneous UV background}\label{App_UV.sec}
\begin{figure}\begin{center}
  \includegraphics[width=0.5\textwidth]{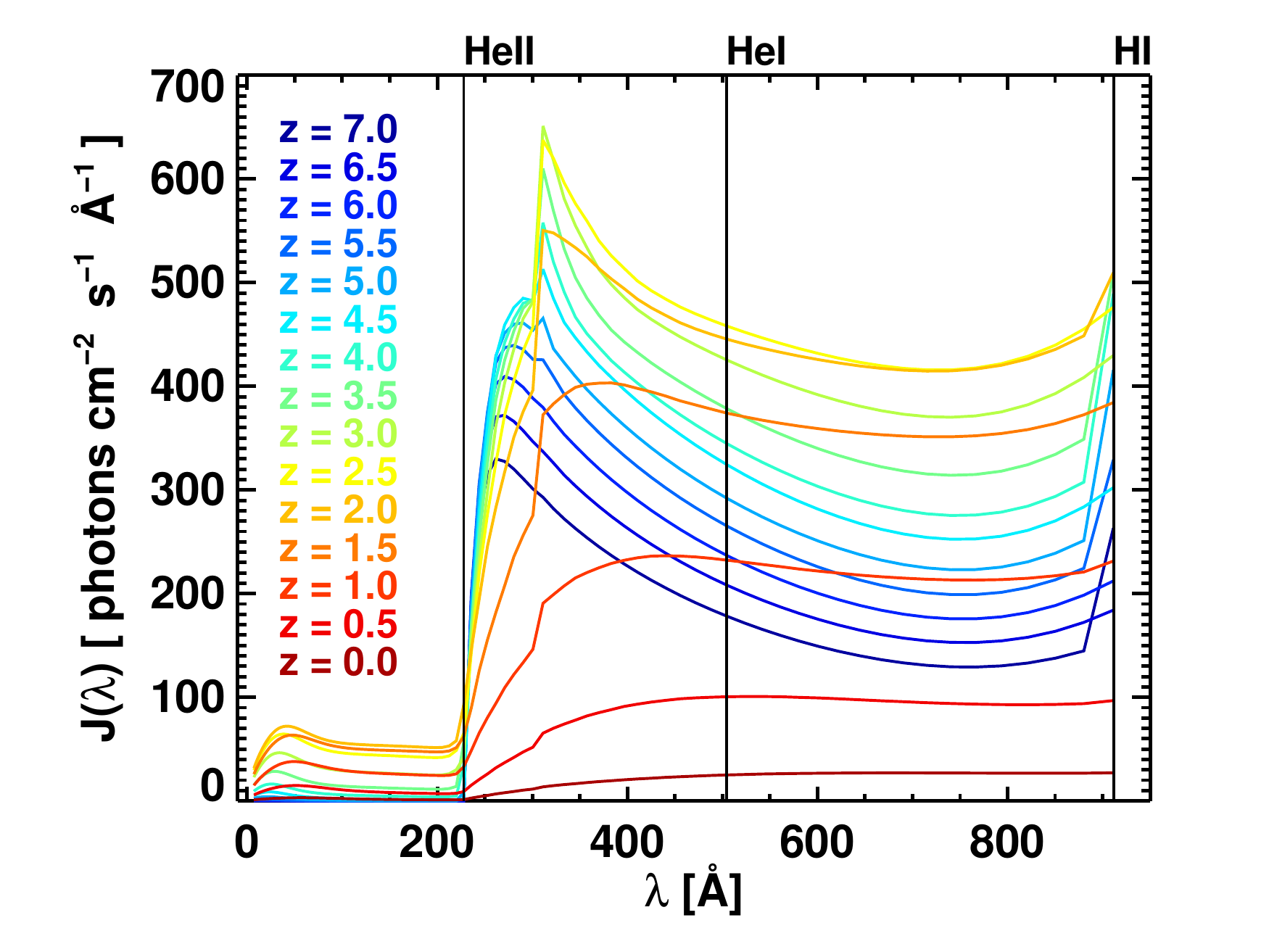}
  \caption[]{\label{FG1.fig}Evolution with redshift of the
    Lyman-continuum part of the UV spectrum of
    \cite{FaucherGiguere:2009p1685} which we use in our
    simulations. The plot shows photon flux versus wavelength
    ($\lambda$) for selected redshifts. The vertical lines indicate
    how we split the spectrum into three (\hi{}-, \hei{}-,
    \heii{}-ionizing) photon packages.}
\end{center}\end{figure}

\begin{figure*}\begin{center}
  \subfloat{\includegraphics[width=0.45\textwidth]
    {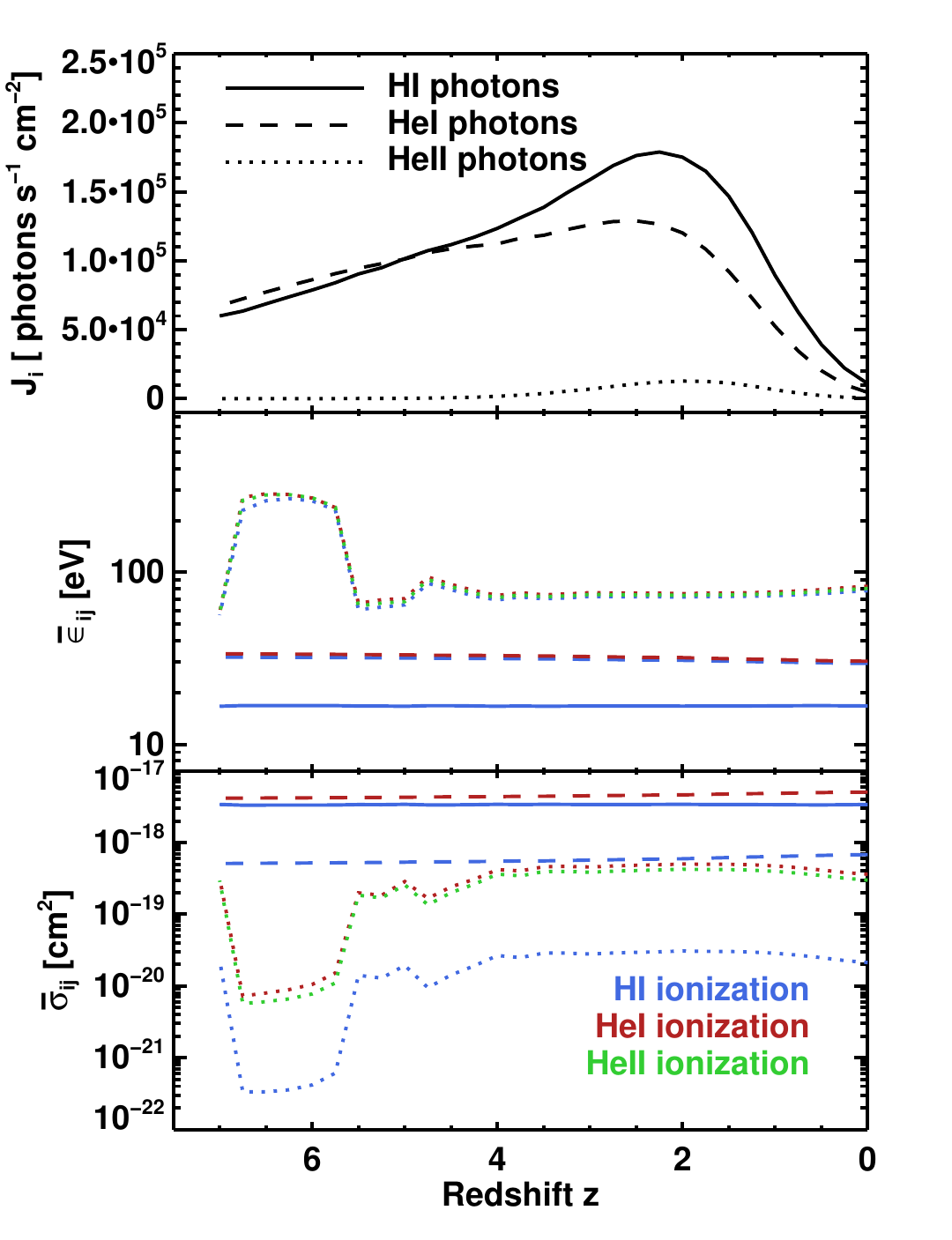}}
  \subfloat{\includegraphics[width=0.45\textwidth]
    {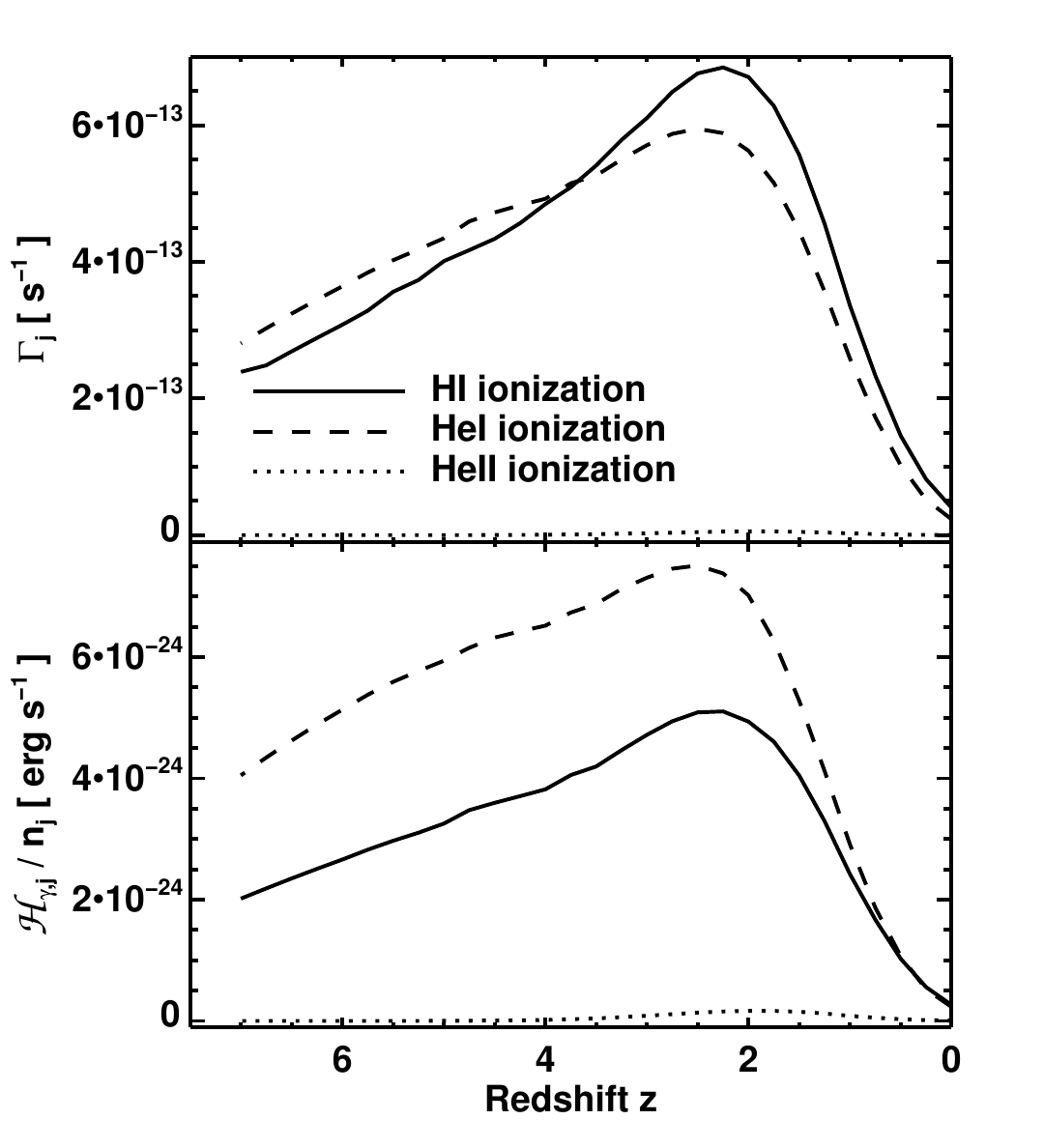}}
  \caption[]{\label{FG2.fig}\textbf{Left.}Redshift-dependent
    properties of the three photon packages, integrated from the
    spectra in Fig. \ref{FG1.fig}. \textbf{Right:} The redshift
    evolution of the photoionization rates against each species (upper
    plot) and per-species photo-heating rates (lower).}
\end{center}\end{figure*}

We use the UV background model of \cite{FaucherGiguere:2009p1685},
which is available on the web, and consists of the redshift-evolving
spectrum shown in \Fig{FG1.fig}. As indicated by vertical lines in the
plot, we discretize the spectrum into three photon \textit{packages};
\hi{} ionizing with frequencies in the range
($\nu_{\hisub},\nu_{\heisub}$); \hei{} ionizing in the range
($\nu_{\heisub},\nu_{\heiisub}$); and \heii{} ionizing in the range
($\nu_{\heiisub},\infty$). All photons belonging to a package $i$
share the common properties of flux $J_i$, average cross sections
$\bar{\sigma}_{ij}$, where $j$ stands for the three ionizable
primordial species \hi{}, \hei{} and \heii{}, and average energies
$\bar{\epsilon}_{ij}$ per photoionization (again versus the three
species). These properties are integrated from the redshift-dependent
UV spectrum and updated every coarse timestep in the simulation. They
are derived in the following way:

For each package $i$ that is defined for the frequency interval
($\nui,\nuf$) and given the UV spectrum $J(\nu) \; \rm{[photons \;
  cm^{-2} \; s^{-1}} \; Hz^{-1}]$ (Fig. \ref{FG1.fig}), we assign an
average photoionization cross section against each species $j$ (\hi{},
\hei{}, \heii{}) as
\begin{equation} 
  \bar{\sigma}_{ij} = \frac{ \int_{\nu_{i0}}^{\nu_{i1}} \sigma_j(\nu)
  J(\nu) \; d\nu } { \int_{\nu_{i0}}^{\nu_{i1}} J(\nu) \; d\nu},
\end{equation}
where we use the expressions for $\sigma_j(\nu)$ from
\cite{Hui:1997p2465}. Similarly, we assign to each photon package
average photon energies per photoionization event against each species:  
\begin{equation} 
  \bar{\epsilon}_{ij} = 
  \frac{ \int_{\nui}^{\nuf} h \nu \, \sigma_j(\nu) J(\nu) \;
  d\nu } { \int_{\nui}^{\nuf} \sigma_j(\nu) J(\nu) \; d\nu},
\end{equation}
where $h$ is Planck's constant. The flux injected isotropically into
each diffuse gas cell is derived for each package as

\begin{equation}
  J_{i} = \int_{\nui}^{\nuf} J(\nu) \; d\nu.
\end{equation}

When injected this way, the photons flow into adjacent cells which are
above the UV density threshold and thus evolve into \textit{local}
photon fluxes $F^{\gamma}_i$ representing the quasi-homogeneous UV
field. \Fig{FG2.fig} shows how the package properties evolve with
redshift.

\section{Calculating the photoionization and photo-heating rates}
\label{App_PHrate.sec}
The hydrogen photoionization rate $\Gamma$ for the Lyman-continuum, in
units of ionization events per hydrogen atom per unit time, is given
by
\begin{equation}\label{phrateA.eq}
  \Gamma=\int_{0}^{\infty} \sighi(\nu) J(\nu) \; d\nu,
\end{equation}
where $\sighi$ is the hydrogen ionization cross section and $J$ is the
local photon flux, integrated over all directions. Since the UV
spectrum in our RHD simulations is discretized into three photon
packages, the photoionization rate is extracted from each gas cell in
the simulation output as
\begin{equation}
  \Gamma= \sum_{i=1}^{3} \bar{\sigma}_{i \rm{H \scriptscriptstyle I}}
   F^{\gamma}_i,
\end{equation}
where $\bar{\sigma}_{i \rm{H \scriptscriptstyle I}}$ is the average
hydrogen ionization cross section for package $i$ and $F^{\gamma}_i$
is the \textit{local} flux of package $i$ photons (see Appendix
\ref{App_UV.sec}).

\begin{figure*}\begin{center}
  \renewcommand*{\thesubfigure}{}
  \rule{\textwidth}{2pt}
  \subfloat[H2]{\includegraphics[width=1\textwidth]
    {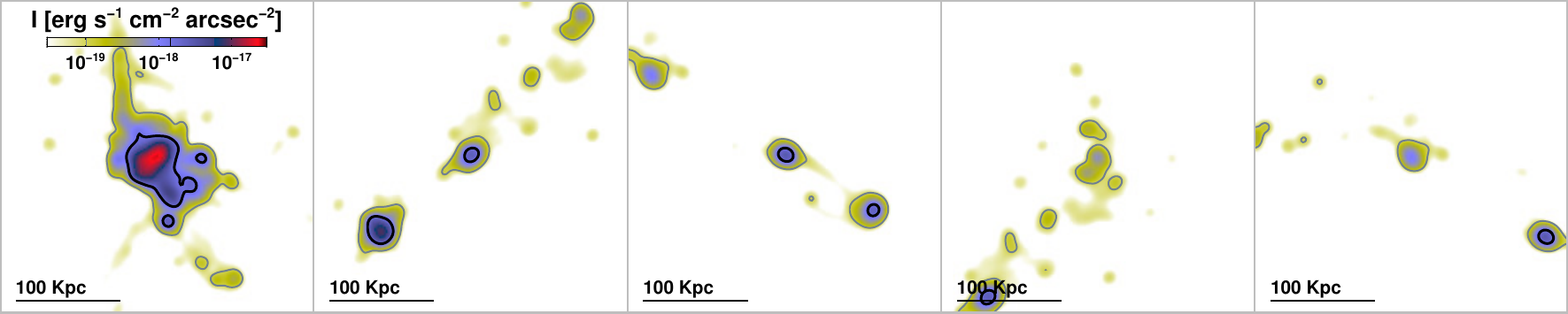}}\\ \hspace{-1mm}
  \rule{\textwidth}{2pt}
  \subfloat[H3]{\includegraphics[width=1\textwidth]
    {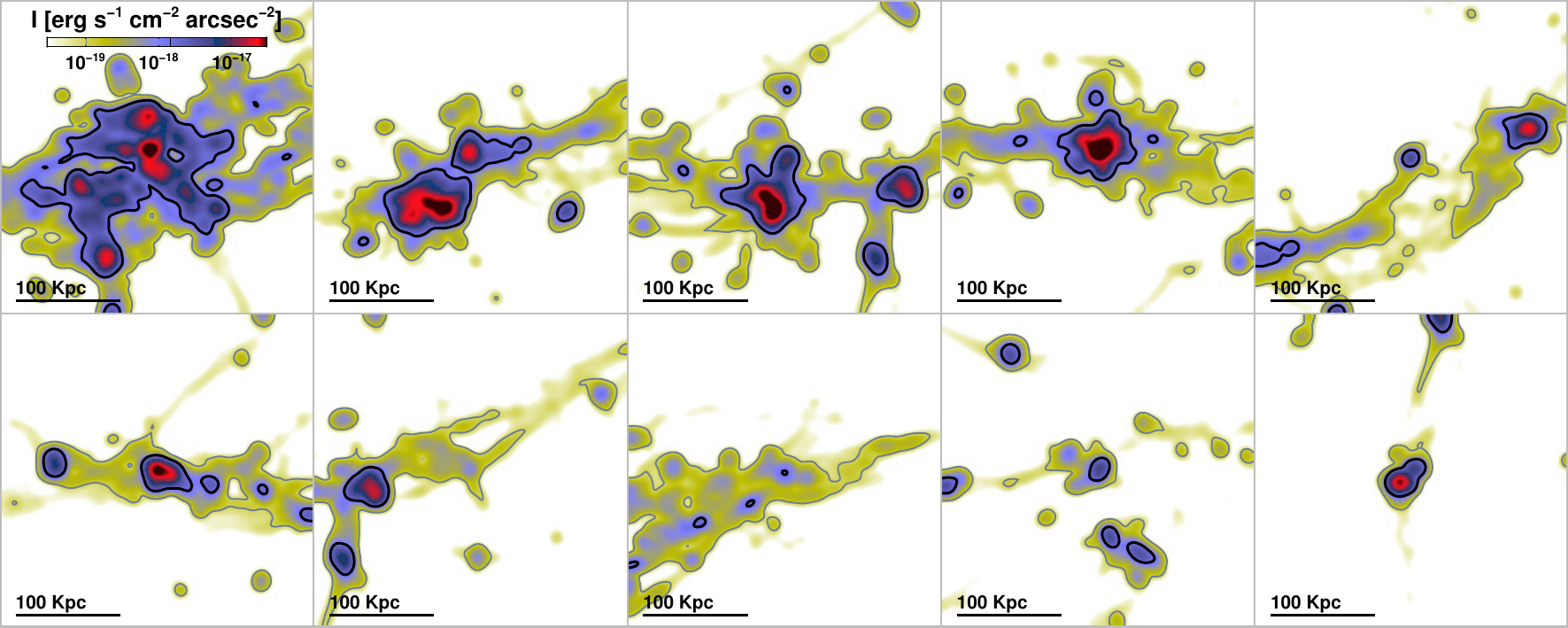}}\\ \hspace{0mm}
  \caption[]{\label{ObsPanels.fig}Mock observations showing $I$ for
    the largest objects in the H2 and H3 simulations at redshift 3,
    smoothed with a Gaussian PSF of FWHM=1.4 arcsec to match recent
    observations. The physical widths of these thumbnail squares are
    all identical at $300$ kpc ($\approx 40$ arcsec). The thick inner
    contours correspond to $I=1.4\ 10^{-18} \; \IS$ and the thin outer
    ones to $10^{-19} \; \IS$.}
\end{center}\end{figure*}

\vsk
The photo-heating rate $ \mathcal{H}_{\gamma}$ for the Lyman-continuum,
in units of energy per time per volume, is given by
\begin{equation}\label{PHeat1.eq}
  \mathcal{H}_{\gamma} = \sum_j^{\rm{\hi{}, \hei{}, \heii{}}} n_j
  \int_{0}^{\infty} \sigma_j(\nu) F^{\gamma}(\nu) \left[
    h \nu - \epsilon_j \right] d\nu,
\end{equation}
where we sum the photo-heating rates over the primordial ion species
\hi{}, \hei{} and \heii{}.  Here $n_j$ and is the number density of ion
species $j$, $\sigma_j(\nu)$ is the species' cross-section,
$F^{\gamma}(\nu)$ is local photon flux, $h$ is Planck's constant and
$\epsilon_{j}$ the photoionization-threshold energies for species $j$.

With the discretization of the UV spectrum into three photon packages
(see Appendix \ref{App_UV.sec}) the integral in eq. (\ref{PHeat1.eq})
becomes a sum:
\begin{equation}\label{PHeat2.eq}
  \mathcal{H}_{\gamma} = \sum_j^{\rm{\hi, \hei, \heii}} n_j
  \sum_{i=1}^{3} \bar{\sigma}_{ij} F^{\gamma}_{i} 
  \left( \bar{\epsilon}_{ij} - \epsilon_j  \right),
\end{equation}
where $F^{\gamma}_{i}$ is the local flux of photons in package $i$,
and $\bar{\sigma}_{j i}$ and $\bar{\epsilon}_{ij}$ are the photon
package properties defined in appendix \ref{App_UV.sec}. We plot the
redshift evolution of the per-species ionization- and heating rates in
\Fig{FG2.fig} (right).

\section{Mock LAB maps}\label{Sec:moremaps}
\Fig{ObsPanels.fig} shows mock observation thumbnails of the most
luminous halos in the H2 and H3 simulations, produced in the same way
as the ones in \Fig{LyObs.fig}, but applying observational parameters
to match the surveys of M11 for direct comparison (these images can
also be compared with thumbnails in e.g. \citealt{Yang:2010p3447} and
\citealt{Erb:2011p5386}). In practice, this means that we assume our
objects are at redshift 3.1, smooth the images with a PSF with
FWHM=1.4 arcsec, and put the thick inner surface brightness contours
at $I=1.4\ 10^{-18} \; \IS$ (the thin outer ones are at $I=10^{-19} \;
\IS$).  Morphologically our mocks resemble real LABs, asymmetric with
a slight tendency to be filamentary and often having short
sub-filaments that poke out of the main structure. In other words, our
simulated LABs look like real ones.

\section{Comparing the LAB luminosity function with observations}
\label{lumfunc.app}
\begin{figure*}\begin{center}
    \hspace{-2mm}
  \subfloat{\includegraphics[width=0.45\textwidth]
    {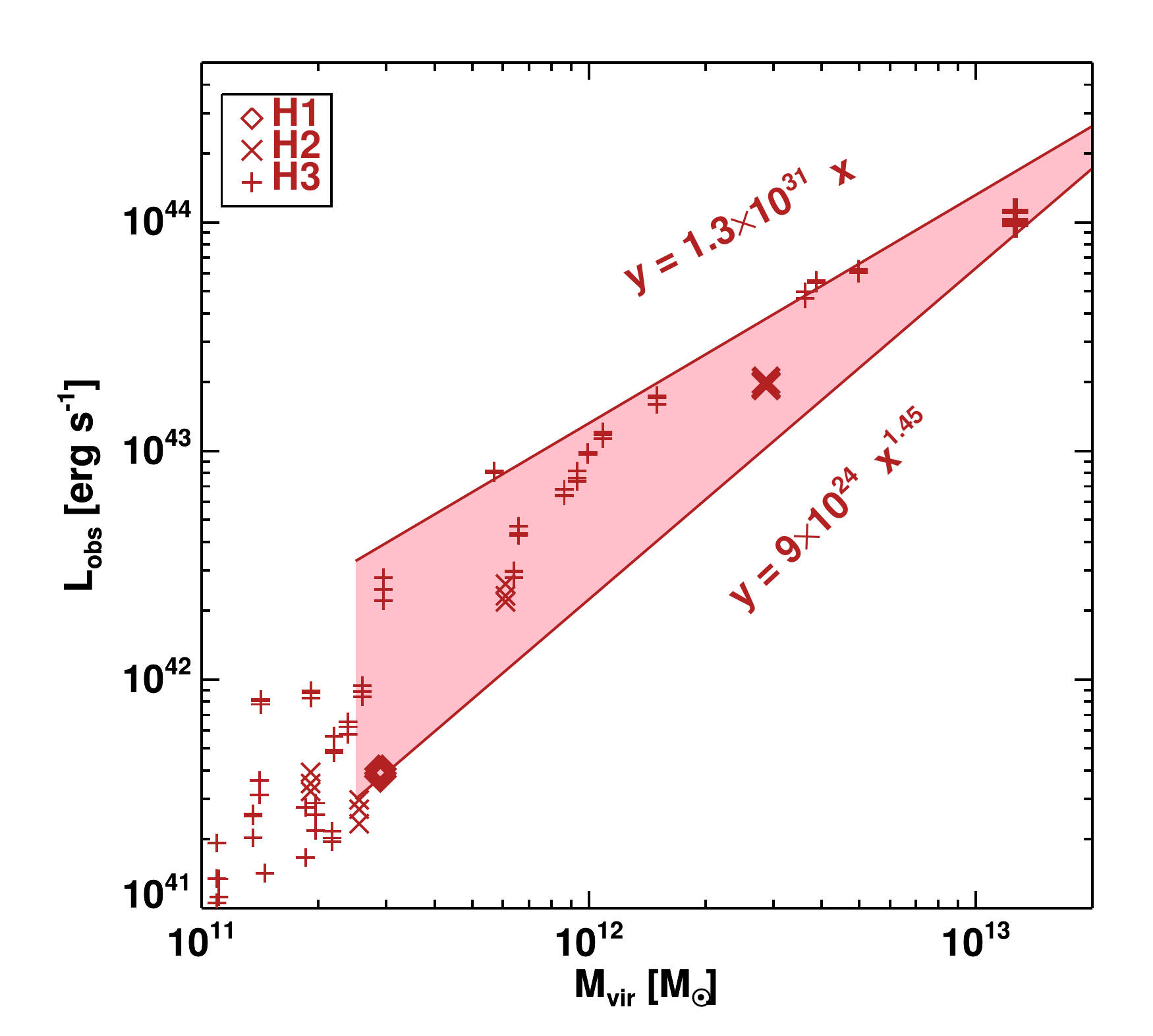}\label{P_ALL_mass.fig}}
  \subfloat{\includegraphics[width=0.45\textwidth]
    {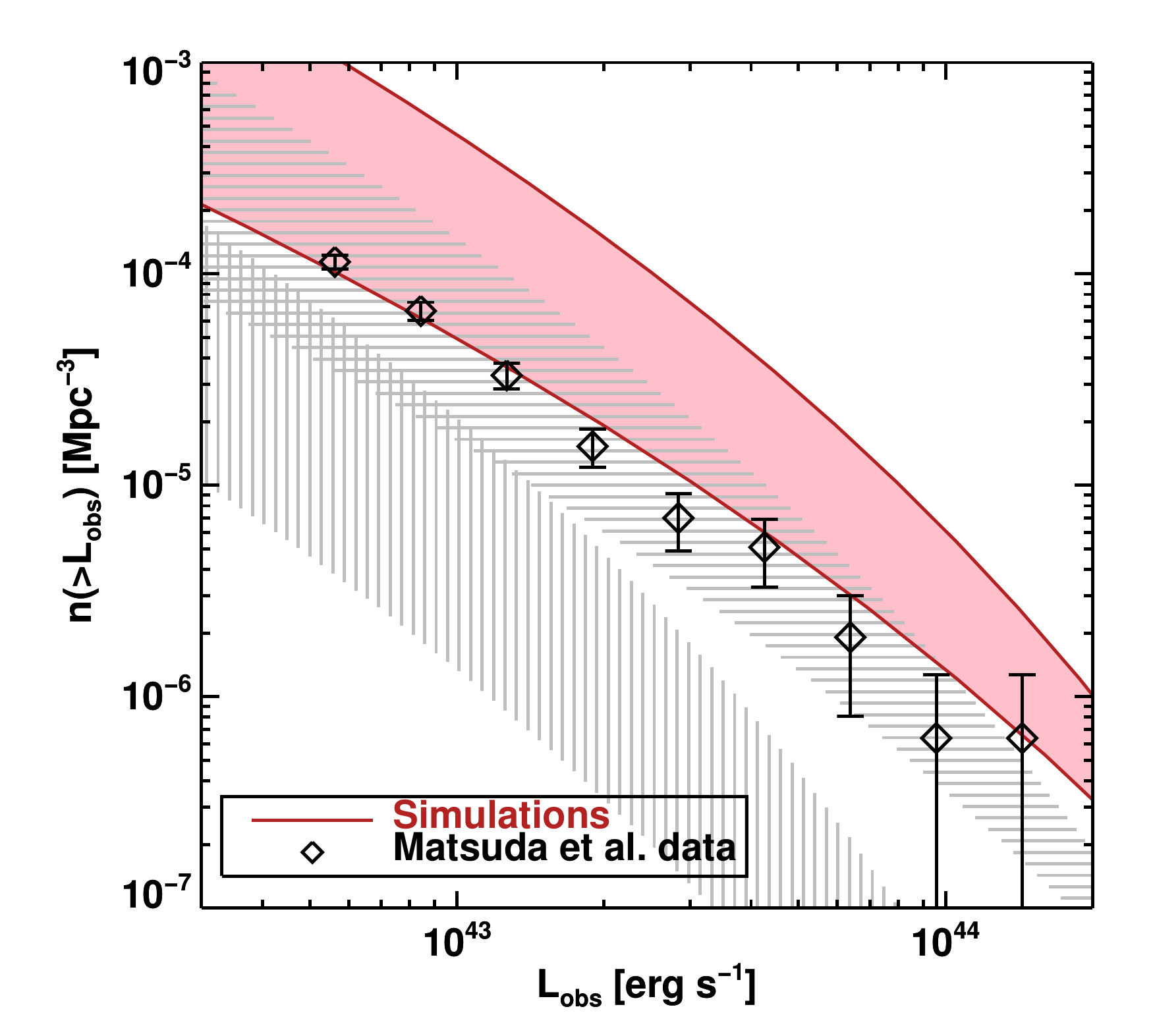}\label{f_LL.fig}}
  \caption[]{\label{P_ALL.fig}\textbf{Left:} Mock observed \lya{}
    luminosities within contours of $1.4\ 10^{-18}\; \IS$ versus halo
    mass for all zoomed halos in the three simulations. Each halo is
    observed along three simulation box axes and thus is represented
    by three points in the plot. The targeted halos are marked with
    thick symbols. The points are enveloped by two power laws shown in
    the plot. \textbf{Right:} Predicted luminosity function of LABs
    (shaded region), according to the power-law envelope from the left
    plot. The black symbols mark a rough luminosity function derived
    from a sample of 202 observed LABs from the survey of M11.  The
    horizontally and vertically line-filled regions represent
    predicted luminosity functions with gas densities of $\nh \ge
    0.3$~$\cci$ and $\nh \ge 0.1$~$\cci$ excluded, respectively.}
\end{center}\end{figure*}
In \Sec{obsComp.sec}, we derive a cumulative area function from our
mock LAB observations and compare to observations. We have done the
same comparison for a cumulative luminosity function, though our
prediction should be less robust than the area function due to our
lack of ISM modelling. Following M11, we assume $z=3.1$ and apply a
PSF with FWHM=1.4. We calculate the observed \lya{} luminosities
$L_{obs}$ of all halos within the zoom regions of our simulations by
integrating the surface brightness within $I=1.4\ 10^{-18} \; \IS$
contours. These luminosities are plotted against halo mass in
\Fig{P_ALL.fig} (left). Note the difference between $L_{obs}$ in this
plot and $L$ in \Fig{HLL.fig}: There we plot intrinsic luminosities of
halos whereas here we plot observable luminosities, assuming
instrument sensitivity and cosmological extinction
($f_{\alpha}=0.66$). The plot points are enveloped by a shaded region
bordered by two power laws, with indexes $1$ and $1.45$ as indicated
in the plot. These power laws are convolved with a Sheth-Tormen halo
mass function \citep{Sheth:1999p4123} at redshift $3.1$ in order to
produce the cumulative luminosity function envelope in \Fig{P_ALL.fig}
(right).

The black diamonds in \Fig{P_ALL.fig} (right) represent the
observations from M11, derived by binning the LABs by luminosity and
dividing the count by the total survey volume, with Poissonian error
bars. The comparison of our results to the observations is good,
somewhat surprisingly so considering the lack of modelling of the
emission and absorption in the compact peaks of \lya{} emission, which
contribute substantially to the total luminosity. As with the area
function, we over-predict LAB abundances, though the prediction here
is slightly closer to observations than in the case of areas.

The plot also shows, with line filled regions, predicted luminosity
function envelopes, where gas of densities $\nh \ge 0.3$~$\cci$ and
$\nh \ge 0.1$~$\cci$ is excluded from the analysis, for the horizontal
and vertical line-fillings respectively. Much as with the area
function (\Fig{f_area.fig}), excluding CGM densities and above ($\nh
\ge 0.3$~$\cci$) has relatively small impact on the luminosity
function. However using the lower density threshold of excluding gas
with $\nh \ge 0.1$~$\cci$ gives a very abrupt change in the function,
illustrating that the simulated LAB luminosities are more sensitive to
the applied density cut than their areas.

\section{Comparison to previous theoretical work}
Our work is similar in nature to the work of \FG{} and G10 (see
\Sec{Intro.sec}), and for comparison to those we have chosen the mass
of the H1 halo to be similar to the halos on which they focus their
analysis.

\vsk Fig. 12 in G10 shows a plot of halo luminosities versus virial
masses at redshift $3.1$, much like our \Fig{HLL.fig}. Their
mass-luminosity power-law exponent is $\sim 0.8$, which is
considerably shallower than our value of $\sim 1.25$. Their less
massive halos are more luminous than ours, with our ones catching up
around $10^{12} \mathrm{M_{\sun}}$. Their targeted halos of masses
$\approx 4\ 10^{11} \, \mathrm{M_{\sun}}$ are typically a few times
more luminous than ours, and much more extended in \lya{}
emission. Their Fig. 10 shows mock observations of two of their
targeted halos.  Also using $I_{-18}$ as a sensitivity limit, they
have observable \lya{} emission which is very asymmetric, clearly
traces accretion streams and extends to about $100$ kpc in length.

Their prediction is probably a bit over the top, since a giant LAB in
a halo of this size implies that LABs should be very common, and this
contradicts the generally accepted view that they are uncommon and
associated with unusually over-dense regions in the Universe
\citep[e.g.][]{Steidel:2000p2213,Prescott:2008p3849,Yang:2010p3447}.
The cause of their over-prediction can probably be traced to an
overestimate in gas temperatures due to their self-shielding
approximation being applied in post-processing, as pointed out by FG10
(see also our \Sec{emission.sec}).

Fig. 2 in \FG{} shows a plot of halo luminosities versus halo masses
at redshift $3$ for the various numerical approaches. Their estimate
which is most comparable to ours (their prescription 7, that sums all
gas) has a power law with exponent $\sim 1.1$, steeper than that of
G10, but still a bit shallower than our exponent, and their halos are
slightly \lya{} dimmer than ours, typically around half the luminosity
for a given mass, though this varies quite a lot due to scatter. The
luminosity difference may be partly explained by the overdensity of
our simulated regions, which tends to increase the brighness of halos
of similar mass, going from the least massive to the most massive
simulation. Their more conservative prescription excludes star-forming
gas from their analysis, which in their simulations is gas with
$\nh>0.13 \ \cci$, likely a more realistic threshold than our $\nh>1 \
\cci$. On this exclusion, the luminosity drops by 1-2 orders of
magnitude. This is a bit more dimming than we find in our results: If
the left plot of \Fig{P_ALL.fig} is considered, where the region
filled with vertical grey lines corresponds to our luminosities where
gas with $\nh>0.1 \ \cci$ is excluded from our analysis, it can be
seen that the luminosity drops by $\la 1$ order of magnitude compared
to our normal prescription of including all gas with $\nh>1 \ \cci$.

They also show mock observations of a $2.5\ 10^{11} \,
\mathrm{M_{\sun}}$ halo at redshift 3, that includes \lya{}
scattering. Their Fig. 7, middle left, can be compared to ours (again,
their prescription 7, that sums all gas). A contour at $I_{-18}$ marks
a very circular source centered on a galaxy, about $15$ kpc in
diameter, which is similar to our H1 halo observation. The \lya{}
luminosity of their halo is $8\ 10^{41}$ $\mathrm{erg \, s^{-1}}$,
close to the `observed' luminosity of our H1 halo of $6 \ 10^{41}$
$\mathrm{erg \, s^{-1}}$.

In terms of the emission coming from $\sim 10^{11} \,
\mathrm{M_{\sun}}$ halos we thus seem to be in fair agreement with
\FG{}, though the LABs produced by our simulations appear to be
somewhat more luminous, even when matching their more conservative
prescriptions. In terms of LAB extent it is harder to tell, since they
don't consider mock observations of halos more massive than $3 \
10^{11} \mathrm{M_{\sun}}$, and it is thus hard to tell whether or not
their massive halos produce LAB-like objects.


\section{Table of symbols}\label{App_Symbols.sec}
\begin{table}
  \caption{Table of symbols}
  \begin{tabular}{l p{6.5cm}}
    \toprule
    $A$ & Area \\
    $\recB(T)$ & Case-B recombination rate for hydrogen \\
    $C_{\lyam}(T)$ & Rate of collisional excitations \\
    $\eps$ & \lya{} emissivity \\
    $\epsilon_{\lyam}$ & Energy of a \lya{} photon ($10.2 \; 
                        \mathrm{eV}$) \\
    $\fa$ & Cosmological transmission factor for mock observations (we
    use $\fa=0.66$). \\
    $f_c$ & Light-speed fraction \\
    $f_{grav}$ & Gravitational efficiency \\
    $G$ & Gravitational constant \\
    $\Gamma$ & Hydrogen photoionization rate \\
    $\mathcal{H}_{\gamma}$ & Photo-heating rate \\
    $I$ & Observed \lya{} surface brightness \\
    $I_{-18}$ & Fiducial observational sensitivity limit, set to 
    $10^{-18}$ $\mathrm{erg \, s^{-1} \, cm^{-2} \, arcsec^{-2}}$ \\
    $L$ & \lya{} luminosity \\
    $M_{vir}$ & Virial mass \\
    $\nhuv$ & Density threshold for UV background emitting gas \\
    $n_i$ & Number density of species $i$ \\
    $N_i$ & Column density of species $i$ \\
    $r$ & Radius from halo center \\
    $R_{vir}$ & Virial radius \\
    $S$ & Rest-frame \lya{} surface brightness \\
    $T$ & Temperature \\
    $v$ & Speed \\
    $x_i$ & Ionization fraction of ion species $i$ \\
    $z$ & Cosmological redshift \\
    \bottomrule 
  \end{tabular}
\end{table}


\bibliography{ref,jeje} 

\begin{thebibliography}{81}
\expandafter\ifx\csname natexlab\endcsname\relax\def\natexlab#1{#1}\fi

\bibitem[{Aubert {et~al}\mbox{.}(2004)Aubert, Pichon, \&
  Colombi}]{Aubert:2004p4312}
Aubert D., Pichon C., Colombi S., 2004, MNRAS, 352, 376

\bibitem[{Aubert \& Teyssier(2008)}]{Aubert:2008p1439}
Aubert D., Teyssier R., 2008, MNRAS, 387, 295

\bibitem[{{Aubert} \& {Teyssier}(2010)}]{AubertTeyssier10}
{Aubert} D., {Teyssier} R., 2010, \apj, 724, 244

\bibitem[{Bacon {et~al}\mbox{.}(2006)Bacon, Bauer, B{\"o}hm, Boudon,
  Brau-Nogue, Caillier, Capoani, Carollo, Champavert, Contini, Daguise, Dalle,
  Delabre, Devriendt, Dreizler, Dubois, Dupieux, Dupin, Emsellem, Ferruit,
  Franx, Gallou, Gerssen, Guiderdoni, Hahn, Hofmann, Jarno, Kelz, Koehler,
  Kollatschny, Kosmalski, Laurent, Lilly, Lizon, Loupias, Lynn, Manescau,
  McDermid, Monstein, Nicklas, Per{\`e}s, Pasquini, P{\'e}contal,
  P{\'e}contal-Rousset, Pello, Petit, Picat, Popow, Quirrenbach, Reiss,
  Renault, Roth, Schaye, Soucail, Steinmetz, Str{\"o}bele, Stuik, Weilbacher,
  Wozniak, \& de~Zeeuw}]{Bacon:2006p5315}
Bacon R. {et~al.}, 2006, The Messenger, 124, 5

\bibitem[{Birnboim \& Dekel(2003)}]{Birnboim:2003p3602}
Birnboim Y., Dekel A., 2003, MNRAS, 345, 349

\bibitem[{Birnboim {et~al}\mbox{.}(2007)Birnboim, Dekel, \&
  Neistein}]{Birnboim:2007p4448}
Birnboim Y., Dekel A., Neistein E., 2007, MNRAS, 380, 339

\bibitem[{Brooks {et~al}\mbox{.}(2009)Brooks, Governato, Quinn, Brook, \&
  Wadsley}]{Brooks:2009p3604}
Brooks A.~M., Governato F., Quinn T., Brook C.~B., Wadsley J., 2009, ApJ, 694,
  396

\bibitem[{Callaway {et~al}\mbox{.}(1987)Callaway, Unnikrishnan, \&
  Oza}]{Callaway:1987p2948}
Callaway J., Unnikrishnan K., Oza D.~H., 1987, Phys. Rev. A, 36, 2576

\bibitem[{Cantalupo {et~al}\mbox{.}(2005)Cantalupo, Porciani, Lilly, \&
  Miniati}]{Cantalupo:2005p4317}
Cantalupo S., Porciani C., Lilly S.~J., Miniati F., 2005, ApJ, 628, 61

\bibitem[{Cen(1992)}]{Cen:1992p2824}
Cen R., 1992, ApJS, 78, 341

\bibitem[{Cen(2011)}]{Cen:2011p5588}
Cen R., 2011, AJ, 742, L33

\bibitem[{Creasey {et~al}\mbox{.}(2011)Creasey, Theuns, Bower, \&
  Lacey}]{Creasey:2011p5926}
Creasey P., Theuns T., Bower R.~G., Lacey C.~G., 2011, MNRAS, 415, 3706

\bibitem[{Dekel \& Birnboim(2006)}]{Dekel:2006p4450}
Dekel A., Birnboim Y., 2006, MNRAS, 368, 2

\bibitem[{Dekel {et~al}\mbox{.}(2009)Dekel, Birnboim, Engel, Freundlich,
  Goerdt, Mumcuoglu, Neistein, Pichon, Teyssier, \& Zinger}]{Dekel:2009p1318}
Dekel A. {et~al.}, 2009, Nature, 457, 451

\bibitem[{Dijkstra {et~al}\mbox{.}(2006)Dijkstra, Haiman, \&
  Spaans}]{Dijkstra:2006p4697}
Dijkstra M., Haiman Z., Spaans M., 2006, ApJ, 649, 37

\bibitem[{Dijkstra \& Loeb(2009)}]{Dijkstra:2009p3780}
Dijkstra M., Loeb A., 2009, MNRAS, 400, 1109

\bibitem[{Dubois \& Teyssier(2008)}]{Dubois:2008p1848}
Dubois Y., Teyssier R., 2008, A{\&}A, 477, 79

\bibitem[{Erb {et~al}\mbox{.}(2011)Erb, Bogosavljevi{\'c}, \&
  Steidel}]{Erb:2011p5386}
Erb D.~K., Bogosavljevi{\'c} M., Steidel C.~C., 2011, AJ, 740, L31

\bibitem[{Fardal {et~al}\mbox{.}(2001)Fardal, Katz, Gardner, Hernquist,
  Weinberg, \& Dav{\'e}}]{Fardal:2001p3736}
Fardal M.~A., Katz N., Gardner J.~P., Hernquist L., Weinberg D.~H., Dav{\'e}
  R., 2001, ApJ, 562, 605

\bibitem[{Faucher-Gigu{\`e}re \& Kere{\v s}(2011)}]{FaucherGiguere:2011p3606}
Faucher-Gigu{\`e}re C.-A., Kere{\v s} D., 2011, MNRAS, 412, L118

\bibitem[{Faucher-Gigu{\`e}re {et~al}\mbox{.}(2010)Faucher-Gigu{\`e}re, Kere{\v
  s}, Dijkstra, Hernquist, \& Zaldarriaga}]{FaucherGiguere:2010p5372}
Faucher-Gigu{\`e}re C.-A., Kere{\v s} D., Dijkstra M., Hernquist L.,
  Zaldarriaga M., 2010, ApJ, 725, 633

\bibitem[{Faucher-Gigu{\`e}re {et~al}\mbox{.}(2011)Faucher-Gigu{\`e}re, Kere{\v
  s}, \& Ma}]{FaucherGiguere:2011p5611}
Faucher-Gigu{\`e}re C.-A., Kere{\v s} D., Ma C.-P., 2011, MNRAS, 417, 2982

\bibitem[{Faucher-Gigu{\`e}re {et~al}\mbox{.}(2009)Faucher-Gigu{\`e}re, Lidz,
  Zaldarriaga, \& Hernquist}]{FaucherGiguere:2009p1685}
Faucher-Gigu{\`e}re C.-A., Lidz A., Zaldarriaga M., Hernquist L., 2009, ApJ,
  703, 1416

\bibitem[{Faucher-Gigu{\`e}re {et~al}\mbox{.}(2008)Faucher-Gigu{\`e}re,
  Prochaska, Lidz, Hernquist, \& Zaldarriaga}]{FaucherGiguere:2008p3910}
Faucher-Gigu{\`e}re C.-A., Prochaska J.~X., Lidz A., Hernquist L., Zaldarriaga
  M., 2008, ApJ, 681, 831

\bibitem[{Francis {et~al}\mbox{.}(1996)Francis, Woodgate, Warren, Moller,
  Mazzolini, Bunker, Lowenthal, Williams, Minezaki, Kobayashi, \&
  Yoshii}]{Francis:1996p4544}
Francis P.~J. {et~al.}, 1996, ApJ, 457, 490

\bibitem[{Fumagalli {et~al}\mbox{.}(2011)Fumagalli, Prochaska, Kasen, Dekel,
  Ceverino, \& Primack}]{Fumagalli:2011p2943}
Fumagalli M., Prochaska J.~X., Kasen D., Dekel A., Ceverino D., Primack J.~R.,
  2011, eprint arXiv, 1103, 2130

\bibitem[{Furlanetto {et~al}\mbox{.}(2005)Furlanetto, Schaye, Springel, \&
  Hernquist}]{Furlanetto:2005p3744}
Furlanetto S.~R., Schaye J., Springel V., Hernquist L., 2005, ApJ, 622, 7

\bibitem[{Gnedin \& Abel(2001)}]{Gnedin:2001p2858}
Gnedin N.~Y., Abel T., 2001, New Astronomy, 6, 437

\bibitem[{Goerdt {et~al}\mbox{.}(2010)Goerdt, Dekel, Sternberg, Ceverino,
  Teyssier, \& Primack}]{Goerdt:2010p1237}
Goerdt T., Dekel A., Sternberg A., Ceverino D., Teyssier R., Primack J.~R.,
  2010, MNRAS, 407, 613

\bibitem[{Haiman \& Rees(2001)}]{Haiman:2001p4742}
Haiman Z., Rees M.~J., 2001, ApJ, 556, 87

\bibitem[{Haiman {et~al}\mbox{.}(2000)Haiman, Spaans, \&
  Quataert}]{Haiman:2000p4632}
Haiman Z., Spaans M., Quataert E., 2000, ApJ, 537, L5

\bibitem[{Hui \& Gnedin(1997)}]{Hui:1997p2465}
Hui L., Gnedin N.~Y., 1997, MNRAS, 292, 27

\bibitem[{Iliev {et~al}\mbox{.}(2006)Iliev, Ciardi, Alvarez, Maselli, Ferrara,
  Gnedin, Mellema, Nakamoto, Norman, Razoumov, Rijkhorst, Ritzerveld, Shapiro,
  Susa, Umemura, \& Whalen}]{Iliev:2006p1394}
Iliev I.~T. {et~al.}, 2006, MNRAS, 371, 1057

\bibitem[{Iliev {et~al}\mbox{.}(2009)Iliev, Whalen, Mellema, Ahn, Baek, Gnedin,
  Kravtsov, Norman, Raicevic, Reynolds, Sato, Shapiro, Semelin, Smidt, Susa,
  Theuns, \& Umemura}]{Iliev:2009p1494}
Iliev I.~T. {et~al.}, 2009, MNRAS, 400, 1283

\bibitem[{Katz {et~al}\mbox{.}(1996)Katz, Weinberg, \&
  Hernquist}]{Katz:1996p2854}
Katz N., Weinberg D.~H., Hernquist L., 1996, ApJS, 105, 19

\bibitem[{Keel {et~al}\mbox{.}(1999)Keel, Cohen, Windhorst, \&
  Waddington}]{Keel:1999p4529}
Keel W.~C., Cohen S.~H., Windhorst R.~A., Waddington I., 1999, AJ, 118, 2547

\bibitem[{Keel {et~al}\mbox{.}(2009)Keel, White, Chapman, \&
  Windhorst}]{Keel:2009p4857}
Keel W.~C., White R.~E., Chapman S., Windhorst R.~A., 2009, AJ, 138, 986

\bibitem[{Kere{\v s} {et~al}\mbox{.}(2005)Kere{\v s}, Katz, Weinberg, \&
  Dav{\'e}}]{Keres:2005p3601}
Kere{\v s} D., Katz N., Weinberg D.~H., Dav{\'e} R., 2005, MNRAS, 363, 2

\bibitem[{Kimm {et~al}\mbox{.}(2011)Kimm, Slyz, Devriendt, \&
  Pichon}]{Kimm:2011p4491}
Kimm T., Slyz A., Devriendt J., Pichon C., 2011, MNRAS, 413, L51

\bibitem[{Kollmeier {et~al}\mbox{.}(2010)Kollmeier, Zheng, Dav{\'e}, Gould,
  Katz, Miralda-Escud{\'e}, \& Weinberg}]{Kollmeier:2010p3256}
Kollmeier J.~A., Zheng Z., Dav{\'e} R., Gould A., Katz N., Miralda-Escud{\'e}
  J., Weinberg D.~H., 2010, ApJ, 708, 1048

\bibitem[{{Komatsu} {et~al}\mbox{.}(2011){Komatsu}, {Smith}, {Dunkley},
  {Bennett}, \& et~al.}]{KomatsuEtal11}
{Komatsu} E., {Smith} K.~M., {Dunkley} J., {Bennett} C.~L., et~al., 2011,
  \apjs, 192, 18

\bibitem[{Laursen \& Sommer-Larsen(2007)}]{Laursen:2007p4741}
Laursen P., Sommer-Larsen J., 2007, ApJ, 657, L69

\bibitem[{Martin {et~al}\mbox{.}(2010)Martin, Moore, Morrissey, Matuszewski,
  Rahman, Adkins, \& Epps}]{Martin:2010p5354}
Martin C., Moore A., Morrissey P., Matuszewski M., Rahman S., Adkins S., Epps
  H., 2010, Ground-based and Airborne Instrumentation for Astronomy III. Edited
  by McLean, 7735, 21, (c) 2010: American Institute of Physics

\bibitem[{Maselli {et~al}\mbox{.}(2003)Maselli, Ferrara, \&
  Ciardi}]{Maselli:2003p4122}
Maselli A., Ferrara A., Ciardi B., 2003, MNRAS, 345, 379

\bibitem[{Matsuda {et~al}\mbox{.}(2004)Matsuda, Yamada, Hayashino, Tamura,
  Yamauchi, Ajiki, Fujita, Murayama, Nagao, Ohta, Okamura, Ouchi, Shimasaku,
  Shioya, \& Taniguchi}]{Matsuda:2004p3081}
Matsuda Y. {et~al.}, 2004, AJ, 128, 569

\bibitem[{Matsuda {et~al}\mbox{.}(2011)Matsuda, Yamada, Hayashino, Yamauchi,
  Nakamura, Morimoto, Ouchi, Ono, Kousai, Nakamura, Horie, Fujii, Umemura, \&
  Mori}]{Matsuda:2011p5426}
Matsuda Y. {et~al.}, 2011, MNRAS, 410, L13

\bibitem[{Mori {et~al}\mbox{.}(2004)Mori, Umemura, \& Ferrara}]{Mori:2004p4829}
Mori M., Umemura M., Ferrara A., 2004, ApJ, 613, L97

\bibitem[{Nilsson {et~al}\mbox{.}(2006)Nilsson, Fynbo, M{\o}ller,
  Sommer-Larsen, \& Ledoux}]{Nilsson:2006p3525}
Nilsson K.~K., Fynbo J. P.~U., M{\o}ller P., Sommer-Larsen J., Ledoux C., 2006,
  A{\&}A, 452, L23

\bibitem[{Ocvirk {et~al}\mbox{.}(2008)Ocvirk, Pichon, \&
  Teyssier}]{Ocvirk:2008p2688}
Ocvirk P., Pichon C., Teyssier R., 2008, MNRAS, 390, 1326

\bibitem[{Ohyama {et~al}\mbox{.}(2003)Ohyama, Taniguchi, Kawabata, Shioya,
  Murayama, Nagao, Takata, Iye, \& Yoshida}]{Ohyama:2003p4783}
Ohyama Y. {et~al.}, 2003, ApJ, 591, L9

\bibitem[{Osterbrock \& Ferland(2006)}]{2006agna.book.....O}
Osterbrock D.~E., Ferland G.~J., 2006, Astrophysics of gaseous nebulae and
  active galactic nuclei

\bibitem[{Palunas {et~al}\mbox{.}(2004)Palunas, Teplitz, Francis, Williger, \&
  Woodgate}]{Palunas:2004p4550}
Palunas P., Teplitz H.~I., Francis P.~J., Williger G.~M., Woodgate B.~E., 2004,
  ApJ, 602, 545

\bibitem[{Prescott {et~al}\mbox{.}(2011)Prescott, Dey, Brodwin, Chaffee, Desai,
  Eisenhardt, Floc'h, Jannuzi, Kashikawa, Matsuda, \&
  Soifer}]{Prescott:2011p5034}
Prescott M. K.~M. {et~al.}, 2011, eprint arXiv, 1111, 630

\bibitem[{Prescott {et~al}\mbox{.}(2009)Prescott, Dey, \&
  Jannuzi}]{Prescott:2009p3951}
Prescott M. K.~M., Dey A., Jannuzi B.~T., 2009, ApJ, 702, 554

\bibitem[{Prescott {et~al}\mbox{.}(2008)Prescott, Kashikawa, Dey, \&
  Matsuda}]{Prescott:2008p3849}
Prescott M. K.~M., Kashikawa N., Dey A., Matsuda Y., 2008, ApJ, 678, L77

\bibitem[{Press {et~al}\mbox{.}(1992)Press, Teukolsky, Vetterling, \&
  Flannery}]{1992nrfa.book.....P}
Press W.~H., Teukolsky S.~A., Vetterling W.~T., Flannery B.~P., 1992, Numerical
  recipes in FORTRAN. The art of scientific computing

\bibitem[{Prunet {et~al}\mbox{.}(2008)Prunet, Pichon, Aubert, Pogosyan,
  Teyssier, \& Gottloeber}]{Prunet:2008p5196}
Prunet S., Pichon C., Aubert D., Pogosyan D., Teyssier R., Gottloeber S., 2008,
  AJS, 178, 179

\bibitem[{Rasera \& Teyssier(2006)}]{Rasera:2006p2855}
Rasera Y., Teyssier R., 2006, A{\&}A, 445, 1

\bibitem[{Rauch {et~al}\mbox{.}(2011)Rauch, Becker, Haehnelt, Gauthier,
  Ravindranath, \& Sargent}]{Rauch:2011p5439}
Rauch M., Becker G.~D., Haehnelt M.~G., Gauthier J.-R., Ravindranath S.,
  Sargent W. L.~W., 2011, MNRAS, 418, 1115

\bibitem[{Rees \& Ostriker(1977)}]{Rees:1977p4388}
Rees M.~J., Ostriker J.~P., 1977, MNRAS, 179, 541

\bibitem[{Ribaudo {et~al}\mbox{.}(2011)Ribaudo, Lehner, Howk, Werk, Tripp,
  Prochaska, Meiring, \& Tumlinson}]{Ribaudo:2011p5454}
Ribaudo J., Lehner N., Howk J.~C., Werk J.~K., Tripp T.~M., Prochaska J.~X.,
  Meiring J.~D., Tumlinson J., 2011, ApJ, 743, 207

\bibitem[{{Schaye}(2001)}]{Schaye01}
{Schaye} J., 2001, \apjl, 562, L95

\bibitem[{Schaye(2004)}]{Schaye:2004p5757}
Schaye J., 2004, ApJ, 609, 667

\bibitem[{Sheth \& Tormen(1999)}]{Sheth:1999p4123}
Sheth R.~K., Tormen G., 1999, MNRAS, 308, 119

\bibitem[{Silk(1977)}]{Silk:1977p4383}
Silk J., 1977, ApJ, 211, 638

\bibitem[{Smith \& Jarvis(2007)}]{Smith:2007p4610}
Smith D. J.~B., Jarvis M.~J., 2007, MNRAS, 378, L49

\bibitem[{Steidel {et~al}\mbox{.}(2000)Steidel, Adelberger, Shapley, Pettini,
  Dickinson, \& Giavalisco}]{Steidel:2000p2213}
Steidel C.~C., Adelberger K.~L., Shapley A.~E., Pettini M., Dickinson M.,
  Giavalisco M., 2000, ApJ, 532, 170

\bibitem[{Steidel {et~al}\mbox{.}(2011)Steidel, Bogosavljevi{\'c}, Shapley,
  Kollmeier, Reddy, Erb, \& Pettini}]{Steidel:2011p5455}
Steidel C.~C., Bogosavljevi{\'c} M., Shapley A.~E., Kollmeier J.~A., Reddy
  N.~A., Erb D.~K., Pettini M., 2011, ApJ, 736, 160

\bibitem[{Taniguchi \& Shioya(2000)}]{Taniguchi:2000p4771}
Taniguchi Y., Shioya Y., 2000, ApJ, 532, L13

\bibitem[{Teyssier(2002)}]{Teyssier:2002p1740}
Teyssier R., 2002, A{\&}A, 385, 337

\bibitem[{Toro(1999)}]{Toro99}
Toro E.~F., 1999, Riemann Solvers and Numerical Methods for Fluid Dynamics: A
  Practical Introduction

\bibitem[{Truelove {et~al}\mbox{.}(1997)Truelove, Klein, McKee, Holliman,
  Howell, \& Greenough}]{Truelove:1997p3217}
Truelove J.~K., Klein R.~I., McKee C.~F., Holliman J.~H., Howell L.~H.,
  Greenough J.~A., 1997, ApJ, 489, L179

\bibitem[{Tweed {et~al}\mbox{.}(2009)Tweed, Devriendt, Blaizot, Colombi, \&
  Slyz}]{Tweed:2009p4217}
Tweed D., Devriendt J., Blaizot J., Colombi S., Slyz A., 2009, A{\&}A, 506, 647

\bibitem[{van~de Voort \& Schaye(2011)}]{vandeVoort:2011p5669}
van~de Voort F., Schaye J., 2011, eprint arXiv, 1111, 5039

\bibitem[{van~de Voort {et~al}\mbox{.}(2011{\natexlab{a}})van~de Voort, Schaye,
  Altay, \& Theuns}]{vandeVoort:2011p5667}
van~de Voort F., Schaye J., Altay G., Theuns T., 2011{\natexlab{a}}, eprint
  arXiv, 1109, 5700

\bibitem[{van~de Voort {et~al}\mbox{.}(2011{\natexlab{b}})van~de Voort, Schaye,
  Booth, Haas, \& Vecchia}]{vandeVoort:2011p5673}
van~de Voort F., Schaye J., Booth C.~M., Haas M.~R., Vecchia C.~D.,
  2011{\natexlab{b}}, MNRAS, 414, 2458

\bibitem[{Weijmans {et~al}\mbox{.}(2010)Weijmans, Bower, Geach, Swinbank,
  Wilman, de~Zeeuw, \& Morris}]{Weijmans:2010p5527}
Weijmans A.-M., Bower R.~G., Geach J.~E., Swinbank A.~M., Wilman R.~J.,
  de~Zeeuw P.~T., Morris S.~L., 2010, MNRAS, 402, 2245

\bibitem[{White \& Rees(1978)}]{White:1978p4389}
White S. D.~M., Rees M.~J., 1978, MNRAS, 183, 341

\bibitem[{{Wise} \& {Cen}(2009)}]{WiseCen09}
{Wise} J.~H., {Cen} R., 2009, \apj, 693, 984

\bibitem[{Yang {et~al}\mbox{.}(2010)Yang, Zabludoff, Eisenstein, \&
  Dav{\'e}}]{Yang:2010p3447}
Yang Y., Zabludoff A., Eisenstein D., Dav{\'e} R., 2010, ApJ, 719, 1654

\bibitem[{Zheng {et~al}\mbox{.}(2011)Zheng, Cen, Weinberg, Trac, \&
  Miralda-Escud{\'e}}]{Zheng:2011p5486}
Zheng Z., Cen R., Weinberg D., Trac H., Miralda-Escud{\'e} J., 2011, ApJ, 739,
  62

\end{thebibliography}

\end{document}